\documentclass[twocolumn,resetfootnote]{aastex701}

\usepackage{mathtools,amsmath} 
\usepackage{graphicx}
\usepackage{xcolor}
\usepackage{txfonts}
\usepackage{subcaption}
\usepackage{caption}
\usepackage{tabularx} 
\usepackage{array}    
\usepackage{comment}
\usepackage{soul}
\usepackage{xcolor}
\usepackage{chngcntr} 
\counterwithin{figure}{section}
\counterwithin{table}{section}
\usepackage{float}
\usepackage{soul} 
\usepackage{booktabs}  
\usepackage{csvsimple}
\usepackage{xspace}

\newcommand{\pgtw}{PG\,0026+129\xspace}
\newcommand{\pgfi}{PG\,0052+251\xspace}

\defcitealias{ricci_tight_2023}{R23} 
\defcitealias{gupta_bass_2024}{G24} 

\usepackage{chngcntr}
\counterwithout{figure}{section}

\usepackage{chngcntr}
\counterwithout{figure}{section}
\counterwithout{table}{section}
\usepackage[labelfont=bf]{caption}

\begin{document}

\title{The Millimeter/X-ray Relation in Rapidly Accreting Supermassive Black Holes at $z<$ 0.16}

\author[0009-0001-8342-7522]{Sophie M. Venselaar}
\affiliation{Instituto de Estudios Astrofísicos, Facultad de Ingeniería y Ciencias, Universidad Diego Portales, Av. Ejército Libertador 441, Santiago, Chile}
\affiliation{Department of Astronomy, University of Geneva, ch. d’Ecogia 16, 1290 Versoix, Switzerland}
\email[show]{\href{mailto:sophie.venselaar@unige.ch}{sophie.venselaar@unige.ch}}

\author[0000-0001-5231-2645]{Claudio Ricci}
\affiliation{Instituto de Estudios Astrofísicos, Facultad de Ingeniería y Ciencias, Universidad Diego Portales, Av. Ejército Libertador 441, Santiago, Chile}
\affiliation{Department of Astronomy, University of Geneva, ch. d’Ecogia 16, 1290 Versoix, Switzerland}
\affiliation{Kavli Institute for Astronomy and Astrophysics, Peking University, Beijing 100871, People’s Republic of China}
\email{..}

\author[0000-0002-5761-2417]{Santiago Del Palacio}
\affiliation{Department of Space, Earth and Environment, Chalmers University of Technology, 412 96 Gothenburg,
Sweden}
\email{..}

\author[0009-0007-9018-1077]{Kriti K. Gupta}
\affiliation{STAR Institute, Liège Université, Quartier Agora - Allée du six Août, 19c, B-4000 Liège, Belgium}
\affiliation{Sterrenkundig Observatorium, Universiteit Gent, Krijgslaan 281 S9, B-9000 Gent, Belgium}
\affiliation{Leibniz-Institut für Astrophysik Potsdam (AIP), An der Sternwarte 16, D-14482 Potsdam, Germany}
\email{..}

\author[0000-0001-9910-3234]{Chin-Shin Chang}
\affiliation{Department of Astronomy, University of Geneva, Chemin Pegasi 51, 1290 Versoix, Switzerland}
\affiliation{Joint ALMA Observatory, Avenida Alonso de Cordova 3107, Vitacura 7630355, Santiago, Chile}
\email{..}

\author[0000-0003-1200-5071]{Roberto Serafinelli}
\affiliation{Instituto de Estudios Astrofísicos, Facultad de Ingeniería y Ciencias, Universidad Diego Portales, Av. Ejército Libertador 441, Santiago, Chile}
\affiliation{INAF - Osservatorio Astronomico di Roma, Via Frascati 33, 00078, Monte Porzio Catone, Roma, Italy}
\email{..}

\author[0000-0002-1292-1451]{Macon A. Magno}
\affiliation{George P. and Cynthia Woods Mitchell Institute for Fundamental Physics and Astronomy, Texas A\&M University, College Station, TX,
77845, USA}
\affiliation{CSIRO Space and Astronomy, ATNF, PO Box 1130, Bentley WA 6102, Australia}
\email{..}

\author[0000-0002-7962-5446]{Richard Mushotzky}
\affiliation{Department of Astronomy, University of Maryland, College Park, MD 20742, USA}
\email{..}

\author[0000-0003-2914-2507]{Elena Shablovinskaya}
\affiliation{Max-Planck-Institut für Radioastronomie, Auf dem Hügel 69, Bonn D-53121, Germany}
\email{..}

\author[0000-0002-6808-2052]{Taiki Kawamuro}
\affiliation{Department of Earth and Space Science, Osaka University, 1-1 Machikaneyama, Toyonaka 560-0043, Osaka, Japan}
\email{..}

\author[0000-0001-7568-6412]{Ezequiel Treister}
\affiliation{Instituto de Alta Investigaci\'on, Universidad de Tarapac\'a, Casilla 7D, Arica, Chile}
\email{..}

\author[0000-0002-6139-2226]{Jacob S. Elford}
\affiliation{Instituto de Estudios Astrofísicos, Facultad de Ingeniería y Ciencias, Universidad Diego Portales, Av. Ejército Libertador 441, Santiago, Chile}
\email{..}

\author{Susanne Aalto}
\affiliation{Department of Space, Earth and Environment, Chalmers University of Technology, Onsala Space Observatory, 439 92, Onsala,
Sweden}
\email{..}

\author[0000-0003-3474-1125]{George C. Privon}
\affiliation{National Radio Astronomy Observatory, 520 Edgemont Road, Charlottesville, VA 22903, USA}
\affiliation{Department of Astronomy, University of Virginia, 530 McCormick Road, Charlottesville, VA 22904 USA}
\affiliation{Department of Astronomy, University of Florida, P.O. Box 112055, Gainesville, FL 32611, USA}
\email{..}

\author[0000-0002-7998-9581]{Michael J. Koss}
\affiliation{Eureka Scientific, 2452 Delmer Street, Suite 100, Oakland, CA 94602-3017, USA}
\email{..}


\date{Accepted XXX. Received YYY; in original form ZZZ}

\begin{abstract}

%

A tight correlation between nuclear millimeter and X-ray emission has recently been found in nearby ($z < 0.01$) and low-Eddington ratio ($\rm \lambda_{Edd} < 0.1$) radio-quiet Active Galactic Nuclei (AGN), suggesting a common origin in the hot X-ray corona. We test this relation in nine more distant RQ AGN ($z \sim 0.06$--0.16) with higher bolometric luminosities ($\log(L_{\rm bol}/\mathrm{erg\,s^{-1}})=45.3$--46.3), Eddington ratios ($\rm \lambda_{Edd} = 0.19$--0.85), and X-ray bolometric corrections ($\kappa_{2-10}=29$--194), selected from the Burst Alert Telescope (BAT) survey. 
We obtained quasi-simultaneous observations with \textit{Swift} at 2--10\,keV and the Atacama Large Millimeter/submillimeter Array (ALMA) at 100\,GHz and with high angular resolution ($<0.14$\arcsec). 
We find that these high-luminosity AGN lie above the millimeter/X-ray correlation defined by lower-luminosity sources. A joint fit to both samples yields a second-degree polynomial with an intrinsic scatter of 0.32\,dex. Furthermore, the millimeter emission correlates linearly with both the UV disk luminosity and $L_{\rm bol}$, with intrinsic scatters of 0.45 and 0.35\,dex, respectively. 
We propose that the deviation from the linear millimeter/X-ray relation arises from a two-component coronal electron population: thermal electrons that produce X-rays, but become less efficient at higher luminosities, and nonthermal electrons that produce millimeter emission and remain tied to $L_{\rm bol}$. Additional millimeter emission from outflow-driven shocks may also contribute, though spectral energy distribution modeling and spectral index studies favor a coronal origin. 

\end{abstract}
\keywords{
\href{https://vocabs.ardc.edu.au/repository/api/lda/aas/the-unified-astronomy-thesaurus/current/resource.html?uri=http://astrothesaurus.org/uat/16}{Active galactic nuclei (16)}; 
\href{https://vocabs.ardc.edu.au/repository/api/lda/aas/the-unified-astronomy-thesaurus/current/resource.html?uri=http://astrothesaurus.org/uat/2035}{X-ray active galactic nuclei (2035)};
\href{https://vocabs.ardc.edu.au/repository/api/lda/aas/the-unified-astronomy-thesaurus/current/resource.html?uri=http://astrothesaurus.org/uat/1663}{Supermassive black holes (1663)};
\href{https://vocabs.ardc.edu.au/repository/api/lda/aas/the-unified-astronomy-thesaurus/current/resource.html?uri=http://astrothesaurus.org/uat/1647}{Submillimeter astronomy (1647)}}

\section{Introduction}\label{sect:intro}
 
Active Galactic Nuclei (AGN) are found at the center of $\sim10\%$ of local galaxies, and are some of the most luminous persistent sources of radiation in the Universe. They are powered by accretion onto supermassive black holes (SMBHs) and emit radiation over the full electromagnetic spectrum \citep[e.g., review by][]{Hickox18}.
The key nuclear components of an AGN include the accretion disk surrounding the SMBH, which emits primarily in the optical and ultraviolet (UV) bands; the hot corona close to the SMBH, responsible for most of the X-ray emission; and the surrounding dusty torus, which radiates in the infrared (IR) \citep[e.g., review by][]{ramos_almeida_nuclear_2017,Ricci2026}.

AGN can be categorized as jetted/radio-loud (RL) or non-jetted/radio-quiet (RQ) based on the presence or absence of a strong, relativistic radio jet and/or extended radio emission, which leads to observational differences in the radio, X-ray, and $\gamma$-ray bands \citep{padovani_two_2017}.
RQ AGN do not show strong jets or radio lobes, making them typically $\sim$10$^3$ times fainter in the radio regime when normalized by optical emission \citep{panessa_origin_2019}. Additionally, RQ AGN do not emit as bright in $\gamma$-rays \citep{ackermann_search_2012,padovani_two_2017,Liu2025NatAs} and have a thermal cutoff in the X-ray band at energies $\sim$50--200\,keV \citep{malizia_integral_2014,ricci_bat_2018}. 
These RQ AGN make up the majority of the AGN population, accounting for approximately $\sim90\%$ of all AGN \citep[e.g.,][]{padovani_vla_2011}.
However, despite their faint radio emission, RQ AGN still ubiquitously emit unresolved nuclear millimeter (mm) emission on scales smaller than $\sim 10$--20\,pc \citep[e.g.,][]{panessa_origin_2019,kawamuro_bass_2022,ricci_tight_2023}.

One proposed origin of this unresolved millimeter emission is the compact ($R\sim5-10\, R_{\rm g}$, with $R_{\rm g}$ the gravitational radius\footnote{The gravitational radius is defined as $ R_\mathrm{g} = GM_{\rm BH}/c^2$.}) and hot\footnote{The temperature of the corona is typically 50--100\,keV \citep[e.g.,][]{ricci_bat_2017,Tortosa2018}} X-ray corona, located close to the SMBH \citep[e.g.,][]{laor_origin_2008}. 
Recently, the size of the millimeter continuum emitting source has been measured for a RQ AGN at $z\sim0.658$ through microlensing and was indeed found to be extremely compact with a size of $R<100R_{\rm g}$ \citep{rybak_detection_2025}. 
The corona is believed to be heated through magnetic reconnections \citep[e.g.,][]{di_matteo_magnetic_1998}, though this remains a matter of debate \citep[e.g.,][]{inoue_detection_2018,Inoue2024,NhatLy2026}.
In the corona, optical/UV photons from the accretion disk are Comptonized by hot electrons
, scattering to higher energies and producing X-ray emission \citep[e.g.,][]{katz_nonrelativistic_1976,Haardt1991}. 
In the presence of a strong magnetic field, the electrons in the corona are expected to produce synchrotron emission as well \citep[e.g.,][]{Field93,laor_origin_2008,inoue_unveiling_2014,panessa_origin_2019}.
Optically thin synchrotron emission produces a power-law spectrum, $S_{\nu} \propto \nu^{-\alpha}$, with a spectral index $\alpha \sim 0.5$--1 
\citep[e.g.,][]{behar_discovery_2015}.
However, for a compact emitter as the corona, the emission becomes self-absorbed (synchrotron self-absorption; SSA) at frequencies of $\nu_{\rm SSA}\sim 100$--300\,GHz, producing a spectral turnover that transitions from optically thin to optically thick emission, following $S_{\nu}\propto \nu ^{2.5}$ \citep[e.g.,][]{laor_origin_2008,inoue_unveiling_2014}. 
The exact location of this turnover peak in the spectrum depends on the properties of the SSA region, such as its size and the magnetic field strength \citep[e.g.,][]{del_palacio_millimeter_2025}.
Recent observations have detected this peak \citep[e.g.,][]{inoue_detection_2018,del_palacio_millimeter_2025}, together with rapid millimeter variability \citep[e.g.,][]{baldi_milimetre-band_2015,behar_simultaneous_2020,petrucci_simultaneous_2023,michiyama_alma_2024,shablovinskaya_joint_2024}, both supporting the idea that synchrotron emission indeed originates from a compact and dense region.
However, this coronal origin of the millimeter emission in RQ AGN has not yet been conclusively confirmed. Alternative explanations include, for example, emission from a very compact and low-power jet 
\citep[e.g.,][]{panessa_origin_2019} or an outflow from the X-ray corona \citep{Hankla2026}.

If both the millimeter and X-ray emission originate from the corona, a strong correlation between them would be expected. Initially, \cite{behar_discovery_2015} found a relation of $ L_{\mathrm{95GHz}} = 10^{-4}L_{\rm 2-10keV}$ for eight RQ AGN, although \cite{behar_mm-wave_2018} later reported significant scatter when expanding the sample.
\cite{kawamuro_bass_2022} analyzed high-resolution Atacama Large Millimeter/submillimeter Array (ALMA) observations of 98 low-redshift ($z<0.05$) AGN at 230\,GHz, resolving scales up to $\sim200$\,pc. They found a tight correlation between the 230\,GHz and 14--150\,keV luminosities with a scatter of 0.36\,dex.
Recently, \citet{ricci_tight_2023} (hereafter \citetalias{ricci_tight_2023}) used high-resolution ALMA observations, probing physical scales of $<23$\,pc, to report an even tighter correlation (scatter of $\sim0.2$\,dex) between 100\,GHz and X-ray emission in a sample of 26 nearby AGN ($z<0.01$). 
These results provided compelling evidence for a common physical origin of the millimeter and X-ray emission.

The observed millimeter/X-ray correlation could serve as a powerful tool for studying highly obscured AGN since millimeter emission remains unaffected by obscuration up to column densities of $N_{\rm H}\sim10^{27}\,\rm cm^{-2}$ (e.g., \citealp{hildebrand_determination_1983}). 
This is about 3\,dex better than what can be achieved by hard X-ray ($>$10\,keV) studies, which are unaffected by obscuration up to $N_{\rm H}\sim10^{24}\,\rm cm^{-2}$ \citep[e.g.,][]{ricci_compton-thick_2015}. 
Therefore, this correlation would enable the study of highly obscured AGN through millimeter emission and could also serve as a probe of AGN power, as it appears to be independent of $N_{\rm H}$, black hole mass ($M_{\rm BH}$), star-formation rate (SFR) and Eddington ratio ($\lambda_{\rm Edd}=L_{\rm bol}/L_{\rm Edd}$), following \citetalias{ricci_tight_2023}. 
Furthermore, it could be used to infer $N_{\rm H}$, providing valuable insights into the AGN's obscuration properties \citepalias{ricci_tight_2023}, and to identify dual AGN at subkiloparsec nuclear separations (e.g., \citealp{Koss23,Droguett-Callejas26}).

In addition to the millimeter/X-ray relation, the relation between optical/UV and X-ray emission, which together define the bolometric luminosity ($L_{\rm bol}$), has been extensively studied for decades to fully understand the energy output of AGN \citep[e.g.,][]{tananbaum_x-ray_1979,avni_cosmological_1982,avni_x-ray_1986,wilkes_einstein_1994}. Various studies have found that the optical/UV emission from the disk and the X-ray emission from the corona are tightly correlated among themselves and with $L_{\rm bol}$ \citep[e.g.,][]{vasudevan_piecing_2007,lusso_x-ray_2010,gupta_bass_2024}. 
One way to characterize the relative contributions of the optical/UV and X-ray emission to the total bolometric output is through bolometric corrections, such as the X-ray bolometric correction $\kappa_{\rm X}=L_{\rm bol}/L_{\rm X}$. 
The value of $\kappa_{\rm X}$ was found to remain approximately constant at $\kappa_{X}\sim10$ for bolometric luminosities up to $\log( L_{\rm bol}/L_{\odot}) > 11$ \citep{duras_universal_2020}. 
Beyond this range, $\kappa_{\rm X}$ increases with increasing $L_{\rm bol}$ \citep[e.g.,][]{duras_universal_2020,gupta_bass_2024,gupta_bass_2025}. This suggested that, as the accretion rate increases, driven by the increased production of optical and UV photons from the accretion flow, the relative contribution of the corona to the total luminosity decreases.
This deviation suggests that the properties of the corona change at higher accretion rates, which could be due to the saturation of the Comptonizing electrons or to UV-driven disk winds that deplete the corona \citep[e.g.,][]{martocchia_wissh_2017, zappacosta_wissh_2020}.
The correlations between X-ray and millimeter emission, as well as between X-ray and UV emission, are well established for low-$\lambda_{\rm Edd}$ AGN. In more luminous AGN, $\kappa_{\rm X}$ is known to increase, altering the optical/UV to X-ray relation. However, to date, it remains unclear how the millimeter emission behaves in these more luminous sources.
To determine whether the nuclear millimeter emission in luminous AGN is more strongly correlated with X-ray coronal emission or optical/UV disk emission, we need high spatial resolution observations of the innermost regions of high-luminosity AGN, which will exclude contamination from the host galaxy. 
Thanks to ALMA's unmatched resolution and sensitivity, it is now possible to detect nuclear millimeter emission on scales of $<400$\,pc for AGN at $z<0.16$.

In this manuscript, we investigate the relation between millimeter emission and X-ray, optical/UV, and bolometric output for SMBHs accreting more rapidly than those studied by \citetalias{ricci_tight_2023}. The AGN in our sample exhibit higher bolometric corrections ($\kappa_{\rm 2-10} = 29$--194), higher Eddington ratios ($\lambda_{\rm Edd} = 0.19$--0.85), and larger bolometric luminosities ($\log(L_{\rm bol}/\rm erg\,s^{-1}) = 45.3$--46.3). 
To study these sources, we obtained new quasi-simultaneous observations with ALMA at 100\,GHz and with \textit{Swift} at 2--10\,keV. 
This data will allow us to test whether the tight millimeter/X-ray correlation reported by \citetalias{ricci_tight_2023} extends to more luminous AGN, or if these sources deviate from the relation. While it is known that the X-ray fraction decreases in higher-luminosity AGN, reflected by the rising $\kappa_X$, it remains unclear how the millimeter emission behaves and how this relation might evolve. 
The correlation derived in this work could, in the future, be used to probe the innermost regions of luminous, heavily obscured AGN that are otherwise hidden from X-ray surveys.


This paper is structured as follows: Section\,\ref{Sample} presents the sample of AGN used in this work and describes the observations with ALMA and \textit{Swift.}
Section\,\ref{Data} describes the data reduction and imaging procedures, as well as the methods used to measure the flux densities for each source.
Section\,\ref{Results} presents the resulting relations between millimeter and X-ray, optical/UV, and bolometric emission.
Section\,\ref{Discussion} discusses what processes might explain the relations we observe.
Finally, our conclusions are presented in Section\,\ref{Conclusion}.

In this paper, we adopt the standard cosmological parameters ($ H_0=70\ \rm km\,s^{-1}\,Mpc^{-1}$, $\Omega_{\rm m}=0.3$, $\Omega_{\Lambda}=0.7$).
The correlations in this work were obtained using the \textsc{statistics} module\footnote{\url{https://docs.scipy.org/doc/scipy/reference/stats.html}} from the Python \textsc{scipy} library \citep{Virtanen2020}.
We present correlations in the luminosity domain rather than flux space, since \citet{isobe_statistical_1986} demonstrated that luminosity relations are more appropriate for recovering intrinsic physical correlations.
When evaluating the significance of a correlation, we adopt the Spearman--r correlation test where a $p$-value\,$<0.01$ indicates a significant correlation.
Finally, all uncertainties presented are 1$\sigma$ 
unless stated otherwise. 

\section{Sample and observations}\label{Sample}

\begin{table*}[th!]
\setlength{\tabcolsep}{4.5pt}
\centering
\caption{Source sample and main parameters}
\begin{tabularx}{\textwidth}{ll|cccccccc}
\hline
\hline
(1) & (2) & (3) & (4) & (5) & (6) & (7) & (8) & (9) & (10)\\ 
Source  & SWIFT ID & $z$  & $\rm \lambda_{Edd}$ & $\rm \kappa_{2-10}$ & $ \log ( L _{\rm bol})$  & $\log (L_{\rm 14-150keV})$ & $\log \rm (SFR)$ & $\log (M_{\rm BH})$ & $\log (N_{\rm H})$ \\ 
        &    &  &                     &                     & ($\rm erg\ s^{-1}$) & ($\rm erg\ s^{-1}$) & ($\rm M_{\odot}\, yr^{-1}$) & ($\rm M_{\odot}$) & ($\rm cm^{-2}$) \\
\hline
Q\,0119$-$286  & SWIFT\,J0122.0$-$2818       &  0.116 & 0.85  & 194 & 46.3 & 44.6 & $<$1.7 & 8.2 & $<$20.0\\
        
PG\,0026+129   & SWIFT\,J0029.2+1319         &  0.142 & 0.32  & 50  & 46.2 & 44.8 & 1.5    & 8.5 & $<$20.0\\

PG\,0052+251   & SWIFT\,J0054.9+2524         &  0.155 & 0.28  & 33  & 46.1 & 44.8 & 1.8    & 8.5 & $<$20.0\\

Mrk\,813       & SWIFT\,J1427.5+1949         &  0.110 & 0.19  & 40  & 46.0 & 44.6 & $<$1.1 & 8.5 & $<$20.9\\

RHS\,61        & SWIFT\,J2325.6+2157         &  0.120 & 0.25  & 54  & 46.0 & 44.8 & $<$1.5 & 8.4 & $<$21.1\\

LEDA\,126226   & SWIFT\,J1416.9$-$1158       &  0.098 & 0.26  & 29  & 45.7 & 44.6 & $<$1.1 & 8.1 & $<$20.0\\

2MASX\,J02223523+2508143 &SWIFT\,J0222.3+2509&  0.060 & 0.19  & 54  & 45.5 & 44.1 & $<$0.7 & 8.1 & $<$20.0\\

2MASX\,J17311341+1442561 &SWIFT\,J1731.3+1442&  0.080 & 0.42  & 60  & 45.4 & 44.2 & $<$1.9 & 7.6 & $<$20.0\\

LEDA\,12773    & SWIFT\,J0325.0$-$4154       &  0.058 & 0.30  & 70  & 45.3 & 43.7 & 0.7    & 7.7 & 20.0   \\     
\hline
\hline
\end{tabularx}
\par\vspace{1ex}
\parbox{0.95\textwidth}{\small
    \textbf{Note:} (1) Source name, (2) Swift ID, (3) Redshift, (4) Eddington ratio, (5) 2-10\,keV bolometric correction, (6) Bolometric luminosity (all from \citetalias{gupta_bass_2024}), (7) BAT luminosity at 14--150\,keV \citep{ricci_bat_2017}, (8) Star-formation rate \citep{ichikawa_complete_2017,ichikawa_bat_2019}, (9) Black hole mass \citep{koss_bass_2022}, and (10) Column density \citep{ricci_bat_2017}.}
\label{tab:Sources}
\end{table*}

We obtained quasi-simultaneous observations with ALMA at 100\,GHz and with \textit{Swift} in the X-ray and UV bands of nine nearby ($z=0.058$--0.155) and unobscured AGN. Details on the sample selection are presented in Section\,\ref{sampleselection} and we provide a detailed discussion of the ALMA and \textit{Swift} observations in Sections\,\ref{almadata} and \ref{swiftdata}, respectively.

\subsection{Sample selection}\label{sampleselection}

Our nine targets were selected to have the highest bolometric luminosities in the all-sky Burst Alert Telescope (BAT) 70-month hard-X-ray survey\footnote{\url{https://swift.gsfc.nasa.gov/results/bs70mon/}} \citep{baumgartner_70_2013} that are observable with ALMA\footnote{ALMA can only observe sources within source Declinations of $-$70 and $+$40 \url{https://almascience.eso.org/documents-and-tools/cycle10/alma-technical-handbook}}. 
BAT, which is on board the Neil Gehrels \textit{Swift} Observatory, operates in the hard X-ray band, covering energies between 14\,keV and 195\,keV. 
Throughout the BAT survey, more than 1000 nearby AGN ($z < 0.1$) have been observed for 70 months \citep{baumgartner_70_2013}, including highly obscured AGN that had not been detected before \citep{ricci_compton-thick_2015}.
The BAT luminosities ($L_{\rm 14-150keV}$) of the sources in our sample are presented in Table\,\ref{tab:Sources} and were taken from \cite{ricci_bat_2017}.

Additionally, these sources are RQ AGN, thus their radio-loudness values ($R_{\rm X}$) are all below the radio-quietness threshold of $R_{\rm X} = L_{1.4\mathrm{GHz}} / L_{14\text{–}195\mathrm{keV}} \le -4.7$ \citep{teng_fermilat_2011}. Their radio-loudness is also consistent with that of the lower-luminosity sample from \citetalias{ricci_tight_2023}, which has an average value of $\log R_{\rm X}=-5.31$, compared to $\log R_{\rm X}=-5.37$ for our sample.

Furthermore, our sources were selected to have higher bolometric luminosities, Eddington ratios and bolometric corrections than the sample by \citetalias{ricci_tight_2023}.
Recently, \cite{gupta_bass_2024} (hereafter \citetalias{gupta_bass_2024}) performed a thorough analysis on 236 hard-X-ray selected and nearby ($z < 0.3$) unobscured AGN from the BAT AGN Spectroscopic Survey \citep[BASS\footnote{\url{https://www.bass-survey.com}};][]{koss_bat_2017,ricci_bat_2017,koss_bass_2022} based on simultaneous optical to X-ray observations.
BASS provided multi-wavelength observations of the BAT AGN and accurately measured AGN parameters such as $M_{\rm BH}$, $ N_{\rm H}$, and intrinsic X-ray luminosities. Additionally, SFRs were obtained by fitting the IR spectral energy distribution (SED) of these AGN \citep{ichikawa_complete_2017,ichikawa_bat_2019}. 
\citetalias{gupta_bass_2024} constructed broadband SEDs that were corrected for host galaxy contamination and, thereafter, obtained accurate values for important AGN parameters such as $L_{\rm bol}$, $\lambda_{\rm Edd}$, and $ \kappa_{\rm X}$, upon which the sample selection criterion in this work is based.

A total of ten sources from BASS met our criteria. 
However, one source (SWIFT\,J1255.0--2657) 
was not visible during the observing period of October, 2023. 
Therefore, nine targets were ultimately observed with ALMA.
These nine sources were all detected with both ALMA and \textit{Swift}; no targets were excluded on the basis of non-detections.

Our nine selected sources have, as aforementioned, higher bolometric luminosities ($\log (L_{\text{bol}}/\rm erg\,s^{-1})= 45.3$--46.3), higher bolometric corrections ($\kappa_{\text{2-10}}= 29$--194), and higher Eddington ratios ($\lambda_{\text{Edd}}= 0.19$--0.85) compared to the AGN sample by \citetalias{ricci_tight_2023}, which we use as a comparison sample throughout this work. 
The specific parameter values for each source, as well as BAT IDs and values for $M_{\rm BH}$, SFR and $N_{\rm H}$, are listed in Table\,\ref{tab:Sources}. 
The values for $M_{\rm BH}$ were taken from \citet{koss_bass_2022}. The SFR from \citet{ichikawa_complete_2017,ichikawa_bat_2019} and Y.\,D\'{i}az et al. (2026, in preparation). 
Finally, the $N_{\rm H}$ were taken from \citet{ricci_bat_2017}; all sources have $\log (N_{\rm H}\rm /cm^{-2}) < 22$, confirming they are unobscured.

\subsection{The ALMA observations}\label{almadata}

\begin{table*}[ht!]
\setlength{\tabcolsep}{4.5pt}
\centering
\caption{Details on the ALMA observations and data}
\begin{tabularx}{\textwidth}{ll|cccccccccc}
\hline
\hline
(1) & (2) & (3) & (4) & (5) & (6) & (7) & (8) & (9) & (10) \\ 
Source                &  Date         & Exposure & rms         &  $\theta$ & $\rm \theta_{pc}$ &   $ S_{\rm 100GHz}^{\rm peak}$ & $ \log( \nu F_{\rm 100GHz})$  & $ \log(\nu L_{\rm 100GHz})$ & $ \alpha_{\rm mm}$\\  
                      &  (yyyy-mm-dd) & (min)    & ($\mu$Jy beam$^{-1}$)  & (\arcsec)  & (pc) &  (mJy beam$^{-1}$)                    &(erg  s$^{-1}$ cm$^{-2}$)& (erg s$^{-1}$)           \\

\hline
Q\,0119$-$286           &  2023-10-02   & 85.9 & 6.3 &  0.112 & 270  & 0.08 $\pm$ 0.01   & $-$16.1 & 39.4 & 0.16$\pm$0.85 \\
        
PG\,0026+129            &  2023-10-26   & 49.9 & 8.5 &  0.129 & 381 &  1.32 $\pm$ 0.07    & $-$14.9 & 40.9 & 0.22$\pm$
0.35 \\

PG\,0052+251            & 2023-10-01    & 47.8 & 10.9 &  0.128 & 412 &  0.41 $\pm$ 0.02    & $-$15.4 & 40.4 & 0.92$\pm$0.35   \\

Mrk\,813                &  2023-10-10   & 26.3 & 13.4 &  0.123 & 281 &  0.38 $\pm$ 0.02    & $-$15.4 & 40.1 & 0.89$\pm$0.50 \\

RHS\,61                 &  2023-10-02   & 64.1 & 8.6 &  0.139 & 347 &  0.59 $\pm$ 0.03    & $-$15.2 & 40.3 & 0.28$\pm$
0.35  \\

LEDA\,126226            &  2023-10-09   & 25.7 & 13.8 &  0.105 & 214 &  2.74 $\pm$ 0.14    & $-$14.6 & 40.8 & 0.48$\pm$
0.35  \\

2MASX\,J022\\\,\,\,23523+2508143 & 2023-10-01 & 31.8 & 12.3 & 0.127 & 158 & 0.71 $\pm$ 0.04 & $-$15.1 & 39.8 & 1.41$\pm$
0.35 \\


2MASX\,J173\\\,\,\,11341+1442561&  2023-10-10   & 111.3 & 6.3 &  0.135 & 224 &  0.12 $\pm$ 0.01    & $-$15.9 & 39.3 & 0.15$\pm$
0.69  \\

LEDA\,12773             &  2023-10-01   & 82.9 & 6.7 &  0.106 & 128 &  0.19 $\pm$ 0.01    & $-$15.7 & 39.2 & 0.37$\pm$
1.15   \\         
\hline
\hline
\end{tabularx}
\par\vspace{1ex}
\parbox{0.95\textwidth}{\small
\textbf{Note:} (1) Source names, (2) Date of the observation, (3) Exposure time of the ALMA observations, (4) The rms of the observation, (5) The geometric mean resolution in arcsec, (6) The obtained geometric mean resolution in pc, (7) Peak flux density, (8) Flux density at 100\,GHz, (9) Luminosity at 100\,GHz, and (10) Spectral index obtained from the four SPWs. For further details, see Section\,\ref{Datamm}.}
\label{tab:ObsALMA}
\end{table*}

ALMA observations (Proposal ID 2023.1.01046.S; PI: C. Ricci) of our sample were carried out in Band\,3, which covers frequencies from 84\,GHz to 116\,GHz, and were taken between October\,1 and October\,26, 2023, with configuration C--8. 
The spectral configuration of the ALMA observations consisted of four spectral windows (SPWs) divided into 128 channels (0.01563 GHz wide), with central frequencies of 90.52\,GHz, 92.48\,GHz, 102.52\,GHz, and 104.48\,GHz, respectively. These four SPWs were combined to map the continuum. 
Furthermore, a standard calibration strategy was adopted: a single bright quasar was used as both the flux and bandpass calibrator, while a second quasar was used as a phase calibrator. Details about these observations are reported in Table\,\ref{tab:ObsALMA}.

\subsection{The Swift observations}\label{swiftdata}
The \textit{Swift}/X-Ray Telescope (XRT; \citealp{burrows_swift_2005}) operates in the X-ray band, ranging between energies of 0.3\,keV to 10\,keV. From the XRT observations of our sources (PI: C. Ricci), seven out of nine observations were obtained between October 6 and October 17, 2023, to be quasi-simultaneous with the ALMA observations. However, observations of Mrk\,813 and LEDA\,126226 were obtained on December 12 and December 26, 2023, respectively, since they could not be observed before due to Sun constraints.
All XRT observations were performed in Photon Counting mode. Observation dates and exposure times are listed in Table\,\ref{tab:ObsSwift}. 
Observation IDs are presented in Table\,\ref{tab:AdditionalSwiftData} (Appendix\,\ref{AppendixXray}).

In addition to X-ray observations from the XRT, optical/UV observations were obtained with the \textit{Swift} Ultraviolet/Optical Telescope (UVOT; \citealp{poole_photometric_2008,breeveld_further_2010}). UVOT can observe in six different filters, ranging between central wavelengths of 1928\,\AA\ (filter UVW2) to 5468\,\AA\ (filter V; \citealp{poole_photometric_2008}).
Each source was observed using the \textit{Swift}/UVOT filter that was available on the day of observation. The specific filters and the corresponding exposure times are listed in Table\,\ref{tab:ObsSwift}.

\begin{table*}[ht!]
\centering
\caption{Details on the \textit{Swift} observations and data}
\begin{tabularx}{\textwidth}{ll|ccc|ccccccccc}
\hline
&& XRT & & &UVOT\\
\hline
\hline
(1) & (2) & (3) & (4) & (5) & (6) & (7) \\
Source   & Date &   Exposure &$ \log(F_{\rm 2-10keV})$ &$ \log(L_{\rm 2-10keV})$&
Exposure & Filter and  & $ \log(\nu F_{\rm UV})$ & $ \log (\nu L_{\rm UV})$  \\ 

         & (yyyy-mm-dd)    & (s) & $\rm (erg\,s^{-1}cm^{-2})$ & $\rm (erg\,s^{-1})$&
         time (s) & $\rm \lambda_{center}$ (\AA)  & $\rm (erg\,s^{-1}cm^{-2})$ &$\rm (erg\,s^{-1})$  \\ 
\hline
Q\,0119$-$286            &2023-10-06 & 1452 &  $-$12.0   &    43.5& 1451          & W1 / 2600 & $-$11.0 & 44.6
    \\ 
PG\,0026+129             & 2023-10-11 & 712 &  $-$11.2    &    44.5& 710           & U / 3465  & $-$11.1 & 44.6
     \\

PG\,0052+251             &2023-10-06 & 1519 &   $-$11.1   &    44.7& 1517          & W1 / 2600 & $-$11.0 & 44.8
     \\

Mrk\,813                 &2023-12-01 & 1968 &  $-$11.4    &   44.1  & 1556, 394     & M2 / 2246 & $-$11.2 & 44.2
    \\

RHS\,61                  &2023-10-07 & 1617 &  $-$11.1    &   44.4  & \nodata             & \nodata         & \nodata             &  \nodata
    \\

LEDA\,126226             &2023-12-16 & 2023 &  $-$11.0    &   44.4 & 1497, 524     & M2 / 2246 & $-$10.9 & 44.5
      \\

2MASX\,J022\\\,\,\,23523+2508143 &2023-10-07 & 1380 &  $-$11.4    &    43.6& 1382          & U / 3465  & $-$10.8 & 44.2
     \\

2MASX\,J173\\\,\,\,11341+1442561 &2023-10-11 & 1667 &  $-$11.6    &    43.6 & 1666          & U / 3465  & $-$11.3 & 43.9
    \\

LEDA\,12773              &2023-10-12 & 1499 &  $-$11.4    &    43.5 & 932, 198, 152 & W2 / 1928 & $-$11.2 & 43.8
     \\     
\hline
\hline
\end{tabularx}
\par\vspace{1ex}
\parbox{0.95\textwidth}{\small
\textbf{Note:} (1) Source names, (2) Date of the observations, (3) Exposure time of the XRT observations, (4) Intrinsic flux at 2--10\,keV, (5) Intrinsic luminosity at 2--10\,keV,
(6) Exposure time of UVOT observations, (7) Filter and central wavelength of the UVOT filter used, (8) Intrinsic UV flux of the AGN, and (9) Intrinsic UV luminosity of the AGN. For further details, see Section\,\ref{sect:XrayDataReduction}.}
\label{tab:ObsSwift}
\end{table*}



\section{Data reduction, imaging and spectral analysis}\label{Data}

\subsection{The millimeter data}\label{Datamm}

The imaging and calibration of the ALMA data were performed with Common Astronomy Software Applications (\sc CASA\rm), version 6.5.4.9 \citep{mcmullin_casa_2007,casa_team_casa_2022}, and ALMA pipeline version 2023.1.0.124.\\
First, we created dirty images for the nine sources. The images of sources with multiple observations were combined, and from these images, the root mean square (rms) of the observation was obtained. 
This was done from regions in the image devoid of any emission. 
Furthermore, we visually inspected the individual SPWs to check for the presence of bright emission lines and found no evidence for line contamination in any of the sources.
To obtain cleaned images, we used the CASA task \texttt{tclean} in multi-frequency synthesis (mfs) mode with a cleaning threshold set to 1.5$\sigma$, which ensured that the residual images were free of any source emission. We used Briggs weighting and a robust\footnote{The \textsc{CASA} robust parameter controls the visibility weighting scheme used during imaging. It ranges between $-$2 and $+$2. At $-$2, uniform weighting prioritizes resolution, while at $+$2, natural weighting prioritizes sensitivity. A value of $+$0.5 is commonly used as the best trade-off between the two types of weighting.} parameter of $+$0.5 to obtain the best trade-off between sensitivity and resolution. 
To ensure the images captured enough detail, we chose pixel sizes that satisfy the Nyquist sampling criterion.
Finally, a mask was created around the source in the dirty image with the purpose of only cleaning the source emission. 
Images for each SPW were created together with images of all of the SPWs combined. We inspected the residuals to ensure sufficient cleaning with no residual flux. 
Then, the primary beam correction was applied. 
These primary beam corrected images and the respective beams and physical scales are displayed in Figure\,\ref{fig:ALMAimages} (Appendix\,\ref{AppendixALMAcont}). 
For each source, we obtained the peak flux density ($S_{\nu}$), as the sources are predominantly unresolved. 
Furthermore, we adopted an uncertainty of $\sigma_{\rm S_{100GHz}^{peak}}=\sqrt{(\rm rms)^2+(0.05\times S_{\rm 100GHz}^{\rm peak})^2}$, where a 5\% flux uncertainty agrees with the ALMA guidelines\footnote{The flux density uncertainties can be found in the ALMA Cycle 10 Technical Handbook (\url{https://almascience.eso.org/documents-and-tools/cycle10/alma-technical-handbook}).} for flux observations in Band\,3. 
The resulting images had synthesized beam sizes in the range of $0.11\arcsec$--$0.14\arcsec$ (corresponding to physical scales of $\sim128$--412\,pc). The 1$\rm \sigma$ rms noise was in the range 6--14 $\rm \mu Jy\,beam^{-1}$. 
The angular resolutions achieved and the 1$\sigma$ rms noise levels for each target are listed in Table\,\ref{tab:ObsALMA}. 

Although not all sources are completely unresolved, as shown in Figure\,\ref{fig:ALMAimages} (Appendix\,\ref{AppendixALMAcont}), our focus is on the nuclear millimeter emission. Therefore, we use the peak flux density rather than the integrated flux.
For eight sources, the peak-to-integrated flux ratio ranges between 0.80 and 1.14, with most values close to unity, indicating 
predominantly unresolved nuclear emission. 
The main outlier is LEDA\,12773 with a ratio of 0.65. Its ALMA image 
(Figure\,\ref{ALMAimages-LEDA127}) reveals a secondary component 
at $\sim$0.08$\arcsec$ ($\sim$101\,pc) from the nucleus, detected 
at a $7\sigma$ significance compared to the nuclear peak at $28\sigma$.
To further investigate this secondary component, we re-imaged LEDA\,12773 using natural weighting (\texttt{robust}$=2$) to maximize sensitivity to extended emission. The source appears unresolved at this weighting, suggesting the secondary component is compact and only marginally resolved at $\texttt{robust}=+0.5$. No archival high-resolution observations of this source are available for comparison. Future observations will be needed to determine the nature of this feature. 

In addition to determining the flux densities of our sample, we computed the intraband spectral indices ($\alpha_{\text{mm}}$), which are defined by the power--law relation $\log (S_{\nu})=\alpha_{\rm mm} \log(\nu)+b$.
We derive $\rm \alpha_{mm}$ for each of our nine sources by measuring the peak flux densities across all four available SPWs and fitting the aforementioned power-law relation. 
Uncertainties on the values of $\rm \alpha_{mm}$ were determined by centering the fit at $\nu=100$\,GHz to reduce the correlation between $\alpha_{\text{mm}}$ and $b$ to better reflect the actual dispersion of the data and the uncertainties.
The spectral indices for each source can be found in Table\,\ref{tab:ObsALMA}, while the resulting SEDs can be found in Figure\,\ref{fig:alpha} (Appendix\,\ref{ApSpectralindex}). 
We find a range of spectral indices between $\rm \alpha_{mm}
=0.15\pm0.69$ and $\rm \alpha_{mm}=1.41\pm0.35$, with an average of $ \rm \alpha_{mm}^{av}= 0.55\pm 0.15$.
This is consistent with the average spectral index of $\rm \alpha_{230GHz}^{av}=0.5\pm1.2$ at 230\,GHz reported by \cite{kawamuro_bass_2023}.
However, the derived spectral indices in this work are subject to large uncertainties, as the four SPWs span only a narrow frequency range and are prone to scatter. 
Broadband observations are needed to confidently constrain the spectral slopes in these sources.

\subsection{The X-ray data}\label{sect:XrayDataReduction}

Spectral analysis of the X-ray observations was performed with \texttt{XSPEC} version 12.14.0h \citep{arnaud_XSPEC_1996} and using \textsc{XRTPIPELINE} following the standard guidelines \citep{evans09}.
The XRT data cover energies between 0.3\,keV and 10\,keV. 
For this work, we used flux densities and luminosities in the 2--10\,keV range to mitigate the effects of absorption.
We binned the spectra at a minimum of one count per energy bin over the whole energy range and subtracted the background.
For the spectral fit, we fitted the XRT data to the power-law model \texttt{ZPOW} including \texttt{TBABS} to account for Galactic absorption in the direction of the sources \citep{wilms_absorption_2000}. 
The parameters for this model are $z$, the photon index ($\Gamma$) and the Galactic absorption\footnote{The values for the Galactic absorption were obtained from the HEASARC $ N_{\rm H, Gal}$ calculator (\url{https://heasarc.gsfc.nasa.gov/cgi-bin/Tools/w3nh/w3nh.pl})} ($ N_{\rm H, Gal}$), which we fixed to the values presented in Table\,\ref{tab:AdditionalSwiftData} (Appendix\,\ref{AppendixXray}). 
Cash statistics \citep[C-stat,][]{cash_parameter_1979} was applied to the XRT data because of the low number of counts in the data.
The values of $\Gamma$ resulting from the fit are listed in Table\,\ref{tab:AdditionalSwiftData} (Appendix\,\ref{AppendixXray}), along with their 90\% confidence uncertainties.

Eight out of nine sources were fitted with this model.
However, for the source 2MASX\,J02223523+2508143, we found evidence for the presence of an absorber. Therefore, we fitted an additional absorber component using \texttt{ZXIPCF} \citep{reeves08}. The presence of this absorber is further discussed in Section\,\ref{AppXrayWinds} in the Appendix. 

To determine the quality of the fits, we calculated the ratio between C-stat and the degrees of freedom.
For eight sources, the ratios range from 0.8--1, as shown in Table\,\ref{tab:AdditionalSwiftData}, indicating generally acceptable fits. One source, Q\,0119--286, stands out as an outlier with a lower ratio of 0.6, presumably due to the low number of counts (39) in the observation.

From these spectral fits, we obtain the intrinsic fluxes in the 2--10\,keV band, and we determined the luminosities including a $k$--correction which accounts for redshifting effects. This was done using the following relation:
\begin{equation}
     L_{\rm 2-10keV} = 4 \pi D_{\rm L}^2 \frac{F_{\rm 2-10keV}}{(1+z)^{2-\Gamma}},
\end{equation}
where $ D_{\rm L}$ is the luminosity distance.
Because the X-ray spectrum of the source 2MASX\,J02223523+2508143 is affected by an absorber
, the observed luminosity underestimates the intrinsic emission; we therefore compute the unabsorbed luminosity with \texttt{XSPEC} to recover the true AGN power.

The intrinsic X-ray flux densities and luminosities are presented in Table\,\ref{tab:ObsSwift}.
The broadband X-ray spectra and our best fits can be found in Figure\,\ref{fig:Xray_spectra} (Appendix\,\ref{AppendixXrayspectra}).

\subsection{The UV data}\label{sect:UVDATARED}

For the UV data, we followed the data reduction procedure recommended by the \textit{Swift}/UVOT Software Guide Version 2.2\footnote{\url{https://swift.gsfc.nasa.gov/analysis/UVOT_swguide_v2_2.pdf}} and the procedure described by \citetalias{gupta_bass_2024}. 
Each source was observed in a single filter, as listed in Table\,\ref{tab:ObsSwift}, 
except for RHS\,61, which was not observed by UVOT.

From the raw UVOT images, we produced calibrated sky images to derive magnitudes and, subsequently, the fluxes of our sources.
Using the latest UVOT CALDB calibration files, available at 2024--02--01\footnote{ \url{https://heasarc.gsfc.nasa.gov/docs/heasarc/caldb/swift/}}, we generated bad pixel maps with \texttt{uvotbadpix} and applied flat-field corrections using \texttt{uvotflatfield} along with the CALDB file.
These corrections were then applied to the raw images to produce the final sky images and obtain the UV flux densities.

Mrk\,813, LEDA\,126226, and LEDA\,12773 consisted of two, two, and three separate observations, respectively. These were stacked to obtain deeper images, after which the UV flux density was determined from the final sky image. However, before stacking the images, any misalignment between the separate observations had to be corrected for with \texttt{uvotimsum}. This was achieved by aligning known reference points within the images. This aspect correction, performed using \texttt{uvotskycorr}, was successful for LEDA\,12773, which enabled the stacking of the three exposures. The resulting image was visually inspected to determine whether the source was indeed aligned.
For Mrk\,813, one of the two exposures lacked sufficient detections, making aspect correction impossible. Consequently, we proceeded with the single exposure that had enough detections. Similarly, for LEDA\,126226, the image with the longest exposure time was used as the final sky image, since the aspect correction could not be performed.

To obtain the magnitudes ($m$) and background noise from the final sky images, we defined regions with a radius of 5$\arcsec$ around the source and 20$\arcsec$ for the background. The magnitudes were then determined using \texttt{uvotsource}, and fluxes were calculated using the relation: 
\begin{equation}
     m-m_0 = -2.5 \log(F[\rm counts]),
\end{equation} 
where the zero-point magnitude ($m_0$) differed for each filter. 
This formula produced a flux in count units, which was then converted to mJy using a conversion factor\footnote{For the zero-point magnitudes ($m_0$) and conversion factors for each UVOT filter see \url{https://heasarc.gsfc.nasa.gov/docs/heasarc/caldb/swift/docs/uvot/uvot_caldb_AB_10wa.pdf}}. All magnitudes are presented in Table\,\ref{tab:AdditionalSwiftData} (Appendix\,\ref{AppendixXray}).

The measured UV fluxes include contributions from both the AGN and the host galaxy, with the latter expected to remain constant over time. Therefore, the flux from the host galaxy was subtracted. We used host galaxy fluxes calculated by \citetalias{gupta_bass_2024}, who obtained these values for our sources using \texttt{GALFIT} \citep{peng_detailed_2002,peng_detailed_2010}. 
Finally, the intrinsic AGN flux was obtained by correcting the observed flux for dust extinction from both the host galaxy, characterized by $ E(B-V)_{\rm HG}$, and the Milky Way, characterized by a reddening constant of $ R_V=3.08$ \citep{pei_interstellar_1992}. Additionally, we accounted for the wavelength-dependent variation of extinction, as described by the extinction curve $ k(\lambda)$. The values of $ E(B-V)_{\rm HG}$ and $ k(\lambda)$ for each source and UVOT filter were determined by \citetalias{gupta_bass_2024} through broadband SED fitting. These values are presented in Table\,\ref{tab:AdditionalSwiftData} (Appendix\,\ref{AppendixXray}).
We used these corrections to obtain the intrinsic flux as follows: 
\begin{equation}
    \ {F_{\rm AGN,intr}}= {F_{\rm AGN, obs}}\times 10^{\ 0.4\ {R_V}\ E(B-V)\ k(\lambda)}.
\end{equation} 
The final intrinsic UV luminosities are listed in Table\,\ref{tab:ObsSwift}.

\section{Results}\label{Results}

Here, we present the observations of the nine AGN with higher bolometric luminosities, Eddington ratios, and bolometric corrections to investigate the millimeter/X-ray relation. 
In this work, we compare our sample to the lower-luminosity sample of \citetalias{ricci_tight_2023}. 
For an accurate comparison, we will express some of our results in terms of the ratio between the millimeter and X-ray luminosity ($\nu L_{\mathrm{100GHz}}/L_{\rm 2-10keV}$).

\subsection{The millimeter vs X-ray relation}\label{millimeter/Xray}

\begin{figure}[ht!]
\centering
\includegraphics[width=1\columnwidth]{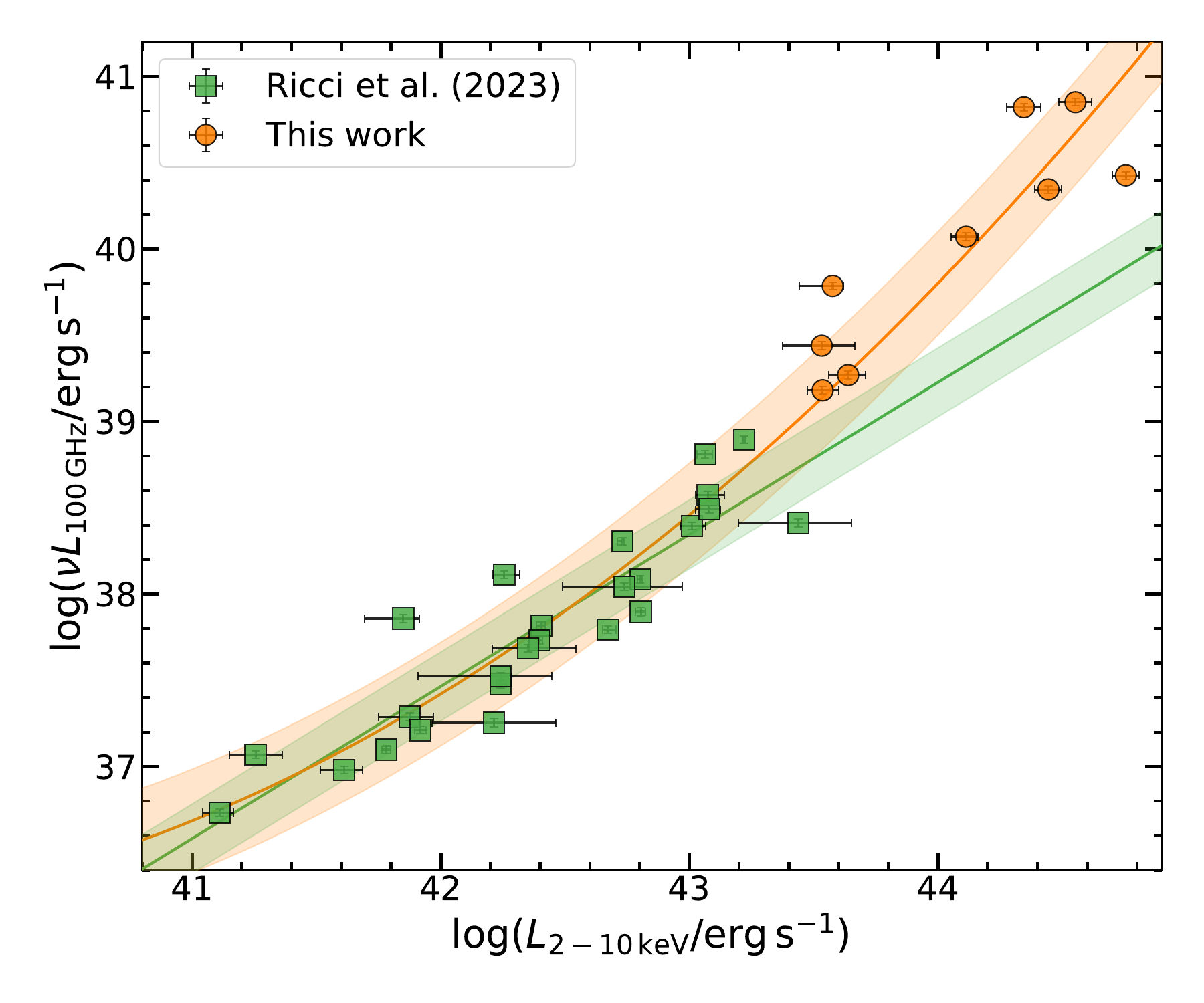}
\caption{The 100\,GHz vs. 2--10\,keV emission for the AGN sample in this work (orange circles) and in \citetalias{ricci_tight_2023} (green squares). 
We have fitted a second-degree polynomial relation between the millimeter and X-ray emission, as presented in Equation\,\ref{eq:mmXray}. This relation has a $p$-value of $2\times10^{-19}$ and an intrinsic scatter of 0.28\,dex.}
\label{fig:corrFL}
\end{figure}

Figure\,\ref{fig:corrFL} presents the relation between the 100\,GHz emission and the intrinsic 2--10\,keV emission for our nine sources and the sample by \citetalias{ricci_tight_2023}.  
Since the ALMA and \textit{Swift}/XRT observations were obtained quasi-simultaneously, while the \textit{Swift}/BAT fluxes are time-averaged over several years, we adopt the 2--10\,keV band for the X-ray luminosities rather than the 14--150\,keV band, although both are reported by \citetalias{ricci_tight_2023}. 
To compare our results with \citetalias{ricci_tight_2023}, we converted their 14--150\,keV fluxes into the 2--10\,keV band by assuming a photon index of $\Gamma=1.8$, which is the median value found for nearby AGN by \citet{ricci_bat_2017}.
This also minimizes the effects of absorption, which affect more strongly at $ E<\rm20\,keV$.

Figure\,\ref{fig:corrFL} shows that the high-luminosity AGN deviate from the linear correlation obtained by \citetalias{ricci_tight_2023}. Therefore, we computed the new relation between millimeter and X-ray emission over the full available range of X-ray luminosities $\log (L_{\rm 2-10keV}/\rm erg\,s^{-1})=41$--45.
We fitted a second-degree polynomial and obtained the relation:
\begin{align}
\log \left( \dfrac{L_\mathrm{100\,GHz}}{10^{38}\,\mathrm{erg\,s}^{-1}} \right) 
  &= (0.15\pm0.05) \, \log \left( \dfrac{L_\mathrm{2-10\,keV}}{10^{43}\,\mathrm{erg\,s}^{-1}} \right)^{2} \notag \\
  &\quad + (1.20\pm0.05) \, \log \left( \dfrac{L_\mathrm{2-10\,keV}}{10^{43}\,\mathrm{erg\,s}^{-1}} \right) \\&+ (0.54\pm0.07)
    \label{eq:mmXray}
\end{align}
with a $p$-value of $\rm 1.2\times 10^{-19}$ and an intrinsic 1$\sigma$ scatter of 0.28\,dex. 
When considering only the high-luminosity AGN, we find a slightly larger intrinsic scatter of 0.33\,dex.
We emphasize that the second-degree polynomial fit is intended as a phenomenological description of the combined dataset, rather than a physically motivated model. 
The key result is the systematic deviation of the high-luminosity sources from the linear relation defined for the low-luminosity AGN.

In addition to the 100\,GHz versus 2--10\,keV emission relation, we investigated the potential relation with the 0.3--2\,keV range. This was done only for the unobscured AGN in both samples, to avoid the strong effect of absorption in that energy range.
These results can be found in Figure\,\ref{fig:corrFL032} (Appendix\,\ref{soft}). 
The 100\,GHz emission appears to show a similar trend with the 0.3--2\,keV band compared to the 2--10\,keV emission, although with a larger scatter of 0.46\,dex, which might be explained by the smaller sample size. 
Although the 14--150\,keV luminosities listed in Table\,\ref{tab:Sources} are not simultaneous with our 100\,GHz observations, we nonetheless present their relation in Figure\,\ref{fig:corrFL032} in Appendix\,\ref{App14150}. 
We find a similar trend as seen for the 0.3--2\,keV and 2--10\,keV bands, with our high-luminosity sources generally deviating from the \citetalias{ricci_tight_2023} relation, though the observed scatter may be partly attributable to the nonsimultaneity of the observations.

\subsection{The millimeter vs X-ray ratio}\label{millimeter/Xrayratio}

\begin{figure*}[th!]
\centering
\includegraphics[width=1\textwidth]{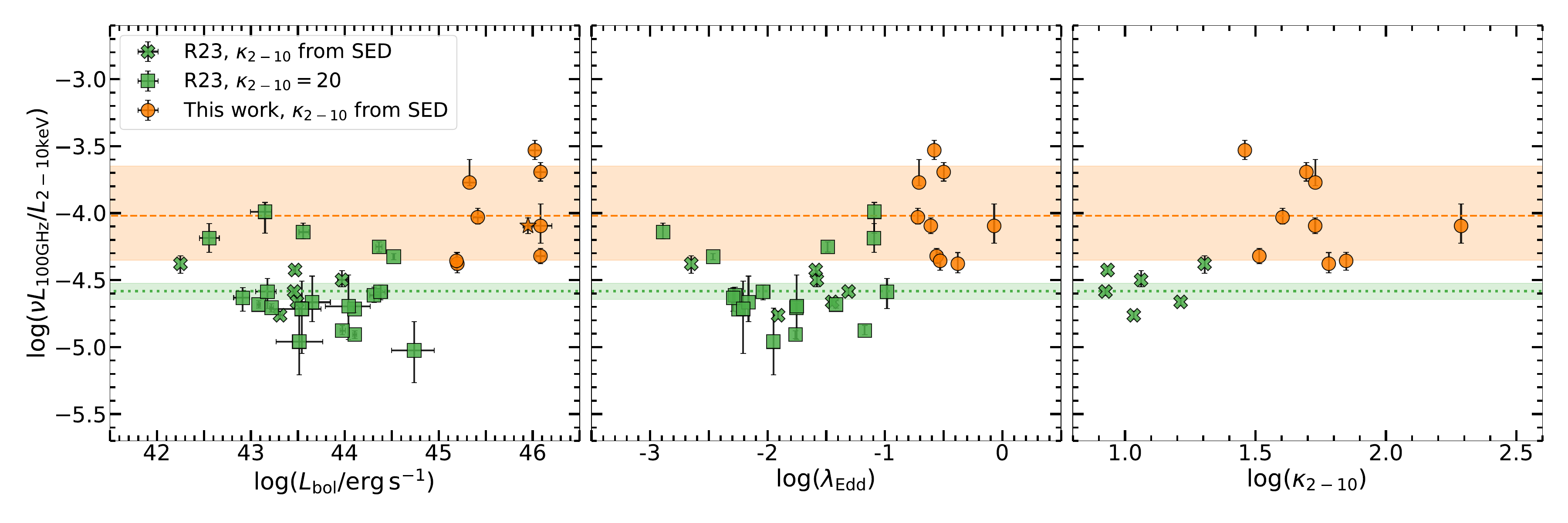}
\caption{Ratio of the 100\,GHz emission over the intrinsic 2--10\,keV emission as a function of AGN parameters for the sources in this work (orange) and in \citetalias{ricci_tight_2023} (green). Mean ratios are shown as dashed/dotted horizontal lines.
\textbf{Left:} Luminosity ratio vs. bolometric luminosity $L_{\rm bol}$.
The bolometric luminosities for our sample were determined by \citetalias{gupta_bass_2024} and normalized using UV observations from both this work and \citetalias{gupta_bass_2024} to account for variability; RHS\,61 was not normalized and is indicated with a star. 
For the \citetalias{ricci_tight_2023} sample, bolometric luminosities were estimated either through $\kappa_{2-10} \times L_{\rm 2-10keV}$ with $\kappa_{2-10}=20$ (squares) or via SED fitting by \citetalias{gupta_bass_2024} (crosses). 
\textbf{Middle:} Luminosity ratio vs. Eddington ratio $\lambda_{\rm Edd}$.
\textbf{Right:} Luminosity ratio vs. bolometric correction $\kappa_{2-10}$.}
\label{fig:ratio_vs_params}
\end{figure*}

Figure\,\ref{fig:ratio_vs_params} presents the millimeter/X-ray luminosity ratio ($\nu L_{\mathrm{100GHz}}/L_{\rm 2-10keV}$) as a function of $L_{\rm bol}$, $\lambda_{\rm Edd}$, and $\kappa_{2-10}$.
In the left panel, for the nine sources in this work, $L_{\rm bol}$ values were taken from \citetalias{gupta_bass_2024}. To account for possible variability between the epochs of \citetalias{gupta_bass_2024} and our new ALMA observations, these bolometric luminosities were re-normalized using the ratio of the UV fluxes measured in the two studies. 
One source, RHS\,61, was not observed with \textit{Swift}/UVOT and could therefore not be re-normalized; this source is marked with a star symbol in Figure\,\ref{fig:ratio_vs_params}. For the comparison sample of \citetalias{ricci_tight_2023}, bolometric luminosities were estimated as:
\begin{equation}
    L_{\rm bol} = \kappa_{2-10} \times L_{\rm 2-10keV},
\end{equation}
assuming a constant $\kappa_{2-10}=20$ \citep{vasudevan_simultaneous_2009}. However, for six sources, \citetalias{gupta_bass_2024} provided updated values of $L_{\rm bol}$ and $\lambda_{\rm Edd}$ based on source-specific $\kappa_{2-10}$ values derived from simultaneous optical/UV/X-ray SED fitting. The method used to estimate $L_{\rm bol}$ for each source in the sample by \citetalias{ricci_tight_2023} is indicated in the figure.



The middle panel shows the millimeter/X-ray luminosity ratio as a function of $\lambda_{\rm Edd}$. 
We adopt the $\lambda_{\rm Edd}$ values from \citetalias{gupta_bass_2024} for both samples. 
ESO\,138--1 is excluded from the \citetalias{ricci_tight_2023} sample due to the large uncertainty in its black hole mass, with published estimates spanning $10^{5}$--$10^{7}\,\rm M_{\odot}$ \citep[e.g.,][]{piconcelli_x-ray_2011, cerqueira-campos_coronal-line_2021, rodriguez-ardila_narrow-line_2024}.

The right panel presents the millimeter/X-ray luminosity ratio as a function of $\kappa_{2-10}$. Since \citetalias{ricci_tight_2023} assumed a constant $\kappa_{2-10}=20$ for most sources, only sources for which \citetalias{gupta_bass_2024} derived individual $\kappa_{2-10}$ values from broadband SED fitting are included.

As shown in all panels of Figure\,\ref{fig:ratio_vs_params}, our sample exhibits systematically higher millimeter/X-ray luminosity ratios, with a mean 
of $-4.02^{+0.37}_{-0.33}$, compared to $-4.58 \pm 0.06$ reported by \citetalias{ricci_tight_2023}. These averages are indicated by the orange dashed (this work) and green dotted \citepalias{ricci_tight_2023} lines, respectively.
The two-sample Anderson--Darling test\footnote{We used the 2-sample \textsc{scipy.stats.anderson} function from \url{https://docs.scipy.org/doc/scipy/reference/generated/scipy.stats.anderson_ksamp.html\#scipy.stats.anderson_ksamp}} \citep{scholz_anderson_1987} yields $\rm p < 0.001$, where $\rm p$ is the probability that the two samples are drawn from the same parent distribution. This further confirms that the two samples are indeed statistically distinct. 

\vspace{-0.5\baselineskip}

\subsection{The millimeter vs UV and bolometric emission}\label{mmvsUV}

In addition to millimeter and X-ray observations, we have obtained UV observations. 
However, as described in Section\,\ref{sect:UVDATARED}, each source from our sample was observed with \textit{Swift}/UVOT using a different filter. 
Since each filter traces a distinct part of the accretion disk emission, no observation represents the full disk emission. Therefore, it is not useful to directly compare the new UV observations among different sources and to search for a correlation with millimeter emission.
To enable a consistent comparison with the mm, we instead consider the total disk emission ($\nu F_{\rm UV,disk}$) integrated over the range $10^{-7}$\,keV to 0.1\,keV as determined by \citetalias{gupta_bass_2024}.
For our sources, we re-normalize this disk emission from \citetalias{gupta_bass_2024} using the fluxes from the new UV observations to take variability into account.

\begin{figure*}[ht!]
\centering
\includegraphics[width=0.49\textwidth]{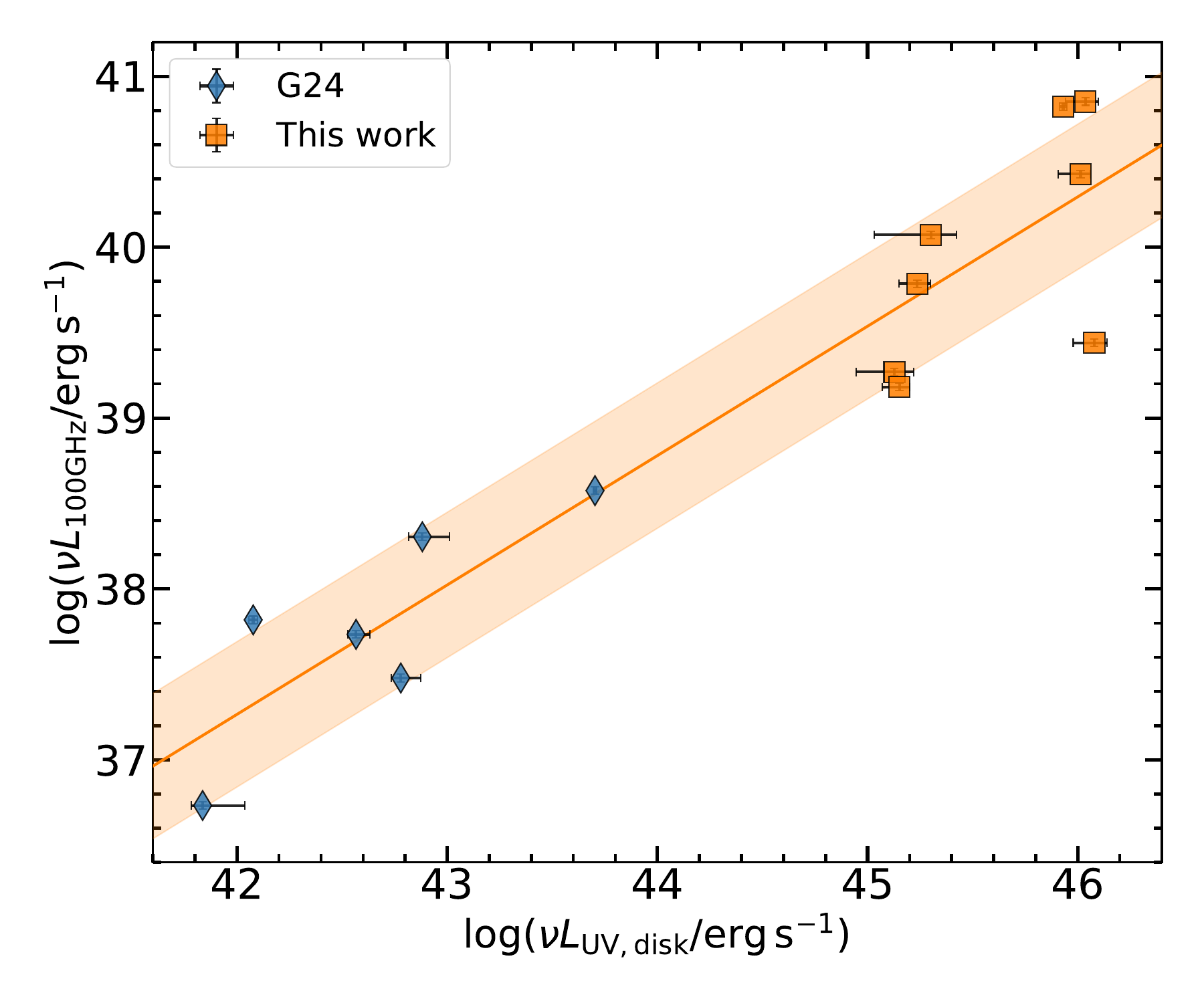}
\hfill
\includegraphics[width=0.49\textwidth]{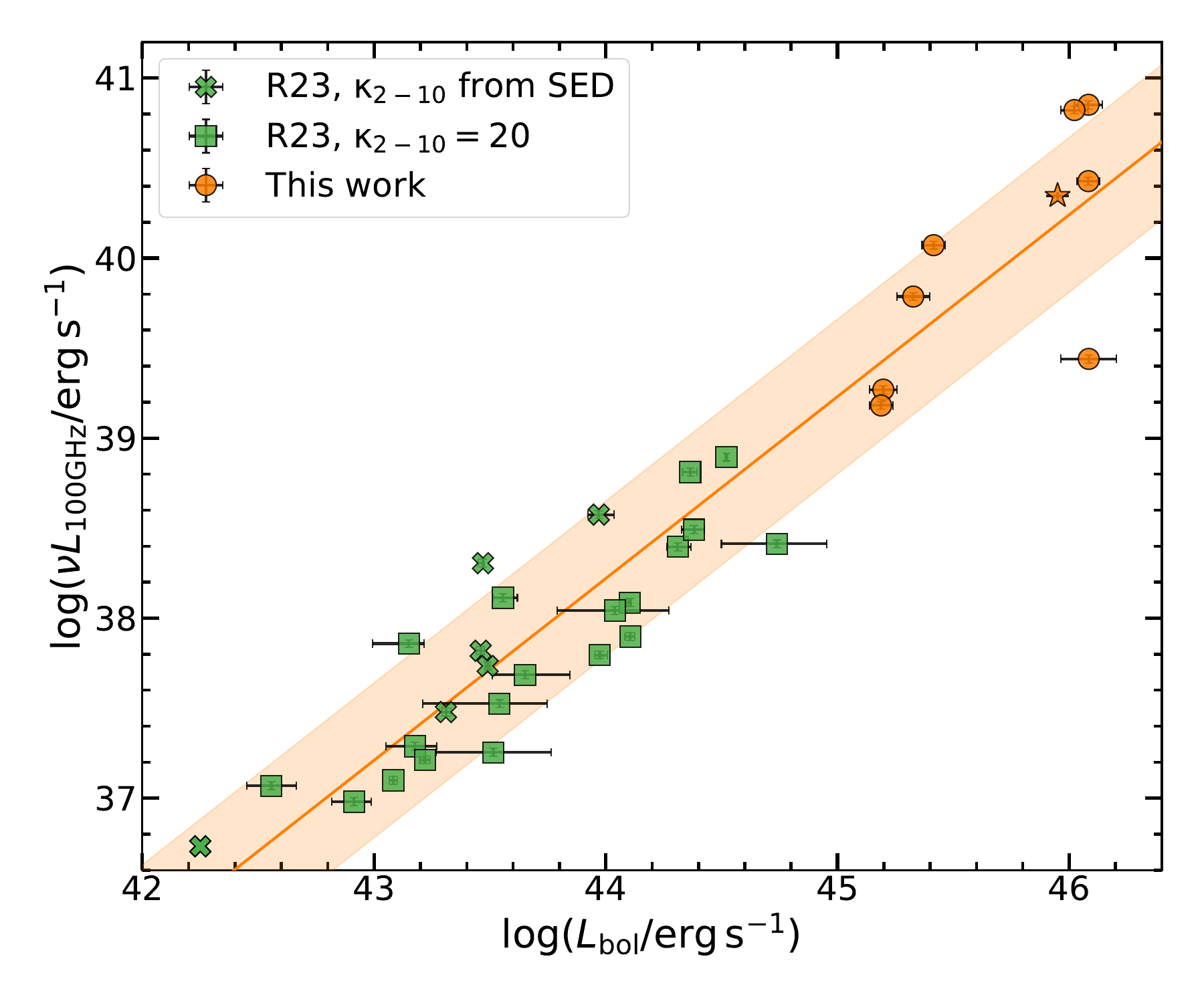}
\caption{
\textbf{Left:} The 100\,GHz luminosity ($\nu L_{\text{100GHz}}$) vs. the optical/UV disk luminosity integrated over $10^{-7}$--0.1\,keV ($\nu L_{\rm UV, disk}$) for the sources in this work (orange squares). 
The disk emission was re-normalized using the new UVOT observations. The source RHS\,61 is not shown, as it was not observed with UVOT. For comparison, six sources from \citetalias{ricci_tight_2023} that were observed in the optical/UV by \citetalias{gupta_bass_2024} are also included (blue diamond).
We find a linear relation between the millimeter and UV emission, with a $p$-value of $7\times10^{-6}$ and an intrinsic scatter of 0.45\,dex.
\textbf{Right:} The 100\,GHz luminosity ($\nu L_{\rm 100GHz}$) vs. bolometric luminosity ($L_{\rm bol}$) for our sample and the sources from \citetalias{ricci_tight_2023}. 
Bolometric emission was estimated either using $\kappa_{2-10} \times L_{2-10,\mathrm{keV}}$ with $\kappa_{2-10}=20$ (squares) or from SED fitting by \citetalias{gupta_bass_2024} (crosses).
A linear fit to the millimeter/bolometric relation (Equation\,\ref{eq:LmmBol}) yields a correlation with a scatter of 0.35\,dex.}
\label{fig:UV_Bol}
\end{figure*}

Figure\,\ref{fig:UV_Bol} presents the millimeter versus total disk emission. Here, we do not include RHS\,61, since this source was not observed with UVOT.
Furthermore, we only included the six sources from \citetalias{ricci_tight_2023} for which optical/UV observations were reported by \citetalias{gupta_bass_2024}.  
Between the millimeter and disk luminosities, we obtain a linear correlation of the form:
\begin{equation}
    \log (\nu L_{\mathrm{100GHz}}) = (0.76 \pm 0.08) \log (\nu L_{\rm UV,disk}) + (5.47 \pm 3.57),
\label{eq:LmmUV}
\end{equation}
with a $p$-value of $\rm 7\times 10^{-6}$, suggesting a significant correlation, and a large intrinsic scatter of 0.45\,dex. 

Figure\,\ref{fig:UV_Bol} also shows the relation between millimeter and bolometric emission, revealing a linear trend, though also accompanied by significant scatter. 
We obtain a linear relation of the form:
\begin{equation}
    \log (\nu L_{\mathrm{100GHz}}) = (1.01 \pm 0.06) \log (L_{\rm bol}) - (6.22 \pm 2.54),
\label{eq:LmmBol}
\end{equation}
with a $p$-value of $\rm 1.5\times 10^{-15}$ and an intrinsic scatter of 0.35\,dex.
Given the linear relation observed between the millimeter and UV emission, the corresponding correlation with bolometric luminosity is unsurprising, as the bulk of the bolometric output is expected to emerge in the UV.

\subsection{Robustness checks}

\subsubsection{Physical scale / beam size effects}

Since our sources lie at higher redshifts than those in \citetalias{ricci_tight_2023}, the same beam corresponds to larger physical scales in our observations. 
The physical scales explored in this research range from 128\,pc to 412\,pc, while \citetalias{ricci_tight_2023} focused on scales $<23$\,pc. 
This difference in scale could potentially lead to the observation of millimeter emission from outside of the nuclear region. Therefore, when stating that we observe higher millimeter/X-ray emission ratios in these highly luminous AGN, we must confirm that we are not detecting additional millimeter emission from regions further out.

The ALMA archive does not contain observations of the sources in the sample by \citetalias{ricci_tight_2023} at physical scales comparable to ours, which would be needed to assess whether millimeter emission increases significantly at larger scales. 
Only data with beam sizes up to three times larger are currently available (C.S. Chang et al. 2026, in preparation), probing scales up to $\sim50$\,pc. Furthermore, some of these sources show enhanced millimeter emission at larger scales while others do not, preventing a conclusive assessment. 
We will further discuss the potential contribution of larger-scale emission from star-formation, which we suspect to be low, in Section\,\ref{MbhSFR}.

In Figure\,\ref{fig:ratioLbeam} (Appendix\,\ref{physicalscale}), 
we illustrate the millimeter/X-ray luminosity ratio versus the physical beam size ($ \theta_{\rm pc}$) of our ALMA observations. We find that the ratio does not significantly increase with the beam size ($p$-value=0.93), arguing against a dominant contribution from larger-scale emission. 
We note, however, that even in the absence of a trend, a small host galaxy contribution could still bias the normalization of the relation, particularly at high $L_{\rm bol}$ (i.e., a small host contribution would systematically shift the ratio upward for all sources). This possibility can be robustly tested with higher-resolution ALMA observations. 
Similarly, Figure\,15 from \citet{kawamuro_bass_2022} showed that the millimeter/X-ray luminosity ratio remains nearly constant across a wide range of physical resolutions ($10\,\rm pc<\theta_{\rm beam}<220\,\rm pc$) in their ALMA data. In particular, they found that the ratio at 230\,GHz changes with physical resolution with a slope of only $7\times 10^{-4}$. 
Based on this result, we do not expect the ratio to vary significantly when probing larger physical scales in our work compared to \citetalias{ricci_tight_2023}.
Therefore, the extra millimeter emission expected at a larger beam of $\sim412$\,pc would be small. 
Additionally, \cite{kawamuro_bass_2022} determined that the nuclear millimeter emission dominated the observed emission in their observations at both high and low resolution. As a result, we anticipate that the contribution of diffuse emission, which does not originate from the nucleus, will be minimal. 

However, based on these arguments, we cannot firmly establish whether the larger physical scales affect the measured level of millimeter emission. Future, higher spatial resolution studies will be needed to confirm this.

\subsubsection{Rest-frame frequency effects}\label{restfreq}

The millimeter observations in this work and in \citetalias{ricci_tight_2023} were both obtained at an observed frequency of $\sim$100\,GHz. However, due to the higher redshifts of our sources ($z=0.058$--0.155) compared to those of \citetalias{ricci_tight_2023} ($z=0.001$--0.011), the corresponding rest-frame frequencies differ slightly.
Since the millimeter spectra are approximately flat at these wavelengths (Section\,\ref{Datamm}), this difference is expected to have minimal impact. 
But, to verify this, we compare the emission in overlapping SPWs from both studies: the second-highest frequency SPW at 102.5\,GHz in \citetalias{ricci_tight_2023} ($\nu_{\rm rest}=102.6$--103.7\,GHz, depending on the redshift of the source) and the second-lowest frequency SPW in our data, centered at 92.5\,GHz ($\nu_{\rm rest}=97.8$--106.8\,GHz).
This comparison shows that the flux ratios are consistent at similar rest-frame frequencies and are, on average, higher in our sample, confirming that the observed differences in millimeter/X-ray luminosity ratios are not due to rest-frequency effects.

\section{Discussion}\label{Discussion}

We have analyzed the millimeter/X-ray relation for nine high-luminosity AGN and compared them with the \citetalias{ricci_tight_2023} sample. 
Relative to that sample, our sources have higher $L_{\rm bol}$, $\lambda_{\rm Edd}$, and $\kappa_{2-10}$.
As shown in Figure\,\ref{fig:corrFL}, our nine AGN lie above the linear millimeter/X-ray correlation established by \citetalias{ricci_tight_2023}. 
Consequently, we derived a new second-degree polynomial relation, presented in Equation\,\ref{eq:mmXray}, spanning the full X-ray luminosity range explored ($\log(L_{2-10}/\rm erg\,s^{-1})=41.1$--44.8), with a scatter of 0.28\,dex. 
Consistently, in Figure\,\ref{fig:ratio_vs_params}, we find on average higher millimeter/X-ray luminosity ratios for our sample.
Turning to the comparison with disk emission (see Figure\,\ref{fig:UV_Bol}), we find a linear relation, although with substantial scatter (0.45\,dex), which may be partly driven by the smaller sample size.
Furthermore, when comparing millimeter and bolometric emission (see Figure\,\ref{fig:UV_Bol}), we observe a significant linear correlation as well, with a scatter of 0.35\,dex. 

In the following Sections\,\ref{spectralindex} and \ref{MbhSFR}, we will assess the millimeter origin by discussing the spectral index and examine whether the observed millimeter/X-ray luminosity ratios show any dependence on $M_{\rm BH}$ and SFR. This is particularly relevant for evaluating the reliability of millimeter emission as a tracer of AGN power, as discussed in Section \ref{sect:intro}.
In Section\,\ref{origin}, we will discuss in more detail the possible origins of the deviation of the high-luminosity sources from the relation established by \citetalias{ricci_tight_2023}, specifically addressing whether it might be driven by a relative decrease in X-ray emission (Section\,\ref{decrXray}), or by a relative increase of the millimeter emission (Section\,\ref{sec:outflows}). Finally, we will further study the origin of the millimeter emission through radio-to-submillimeter SEDs (Section\,\ref{SEDmod}).



\subsection{Spectral index $ \alpha_{\rm mm}$}\label{spectralindex}

To determine the origin of the millimeter emission, we can investigate the spectral index ($\alpha_{\text{mm}}$).
%
%
The average spectral index of $\alpha_{\rm mm}^{\rm av} =  0.55 \pm 
 0.15$ (Section\,\ref{Datamm}) is inconsistent with thermal dust 
emission, which follows a modified blackbody spectrum with 
$\alpha_{\rm dust} \approx -3.5$ \citep[e.g.,][]{condon_nrao_1998, 
mullaney_defining_2011}, confirming that the observed millimeter emission 
is not dust-dominated. 

The range of spectral indices observed across our sample 
($0.15 \leq \alpha_{\rm mm} \leq 1.41$) is broadly consistent with 
a mixture of synchrotron and free--free processes. 
Negative spectral indices would suggest a contribution from optically thick synchrotron emission, for which self-absorption can produce $\alpha_{\rm synchr}^{\rm opt.thick} \approx -2.5$.
Spectral indices in the range $0$--$0.5$ 
are consistent with a combination of optically thin synchrotron emission ($\alpha_{\rm synchr}^{\rm opt.thin} \approx 0.5$--$1.0$) and free--free emission ($\alpha_{\rm ff} \sim 0.1$; \citealt{panessa_origin_2019}).
Indices in the range $0.5$--$1.0$ are most naturally explained by optically thin synchrotron emission. 
The steepest index in our sample ($\alpha_{\rm mm} = 1.41$) exceeds what is typically expected for optically thin synchrotron emission, though we caution that all spectral indices carry large uncertainties (up to $\sigma_{\alpha_{\rm mm}}\sim1.15$) due to the narrow intraband frequency coverage of our ALMA observations (Section\,\ref{Datamm}), limiting firm conclusions about the physical origin of the millimeter emission. 
The reliability of intraband $\alpha_{\rm mm}$ as a diagnostic tool will be further discussed in \mbox{S.M.\,Venselaar et al. (2026, in preparation)}

\subsection{Dependence on $M_\mathrm{BH}$ and SFR}\label{MbhSFR}


As discussed in Section\,\ref{sect:intro}, \cite{kawamuro_bass_2022} and \citetalias{ricci_tight_2023} found that the millimeter/X-ray luminosity ratio is independent of $M_{\rm BH}$ and SFR, implying that millimeter emission can serve as a proxy for X-ray emission across a wide range of these AGN parameters. 
To test whether this holds for our sample as well, we examined possible correlations between $\log (\nu L_{\rm 100GHz}/L_{\rm 2-10keV})$ and $M_{\rm BH}$ and SFR. This is especially important for the comparison with SFR, since there is the possibility that the larger scales we probe in this work include extra millimeter emission from star-formation.

Figure\,\ref{fig:ratio_vs_MBH_SFR} (Appendix\,\ref{AppendixMSFR}) shows the luminosity ratio versus $M_{\rm BH}$ and SFR. 
No significant correlation is observed between $\log (\nu L_{\rm 100GHz}/L_{\rm 2-10keV})$ and $M_{\rm BH}$, obtaining a $p$-value of 0.03.
Although, it should be noted that the relation between $\nu L_{\rm 100GHz}/L_{\rm 2-10keV}$ and $M_{\rm BH}$ increases linearly if one does not consider data points at $\log(M_{\rm BH}/\rm M_{\odot})<6.3$. However, it is expected that more luminous AGN host more-massive SMBHs, therefore, in order to clearly assess any intrinsic dependence on $M_{\rm BH}$, we would need to compare sources with similar $L_{\rm bol}$ and $\lambda_{\rm Edd}$.

Furthermore, no significant correlation is observed between $\log (\nu L_{\rm 100GHz}/L_{\rm 2-10keV})$ and SFR, with a $p$-value of 0.22. 
Six of our nine AGN have only upper limits for the SFR, therefore, to determine any correlation, we used Kendall's tau test to account for these upper limits (\texttt{pymccorrelation}; \citealp{isobe_statistical_1986,curran_monte_2015,harris2020array,privon_hard_2020,2020SciPy-NMeth}).
Two of the three objects for which SFRs could be inferred have higher values than the R23 sample. This may be attributed to the relatively large beam size of our observations and to the fact that more-massive systems tend to have higher SFRs, as expected from the galaxy main sequence \citep[e.g.,][]{Brinchmann2004,Noeske2007}.

\subsection{Origin of the millimeter/X-ray deviation for luminous AGN}\label{origin}

In the following subsections, we discuss the origin of the higher millimeter/X-ray luminosity ratios in our sources compared to those reported by \citetalias{ricci_tight_2023} (Figures\,\ref{fig:corrFL} and \ref{fig:ratio_vs_params}). 
In particular, we address whether this offset might primarily be driven by a decrease in the X-ray emission (see Section\,\ref{decrXray}) or an increase in the millimeter emission (see Section\,\ref{sec:outflows}), and we employ SED modeling to further constrain the origin of the millimeter emission (see Section\,\ref{SEDmod}).

\subsubsection{Relative decrease in X-ray emission}\label{decrXray}

We know that as $L_{\rm bol}$ increases, the absolute X-ray luminosity still rises, but its relative contribution to the total radiative output decreases, demonstrated by larger $\kappa_{2-10}$ \citep[e.g.,][]{lusso_x-ray_2010,duras_universal_2020,gupta_bass_2024,gupta_bass_2025}. 
This behavior has been linked to changes in the physical state of the corona at high luminosities \citep[e.g.,][]{martocchia_wissh_2017, zappacosta_wissh_2020}, although the underlying mechanisms remain uncertain.
Consequently, the increased millimeter/X-ray ratios that we observe in this work might be due to this declining X-ray fraction at higher luminosities (Figures\,\ref{fig:corrFL} and \ref{fig:ratio_vs_params}).
At the same time, we find that the millimeter emission correlates linearly with $L_{\rm UV,disk}$ and $L_{\rm bol}$ (Figure\,\ref{fig:UV_Bol}), suggesting that the millimeter closely traces the total accretion power, while the X-rays start to deviate.

A physical interpretation for this behavior could be framed in terms of the coronal electron distribution. 
The hot corona likely consists of a dominant thermal population and a small nonthermal tail \citep[e.g.,][]{Fabian2017}. 
The thermal electrons are responsible for the X-ray emission, whereas the millimeter emission in RQ AGN is thought to arise from the nonthermal component \citep[e.g.,][]{inoue_unveiling_2014}. 
If the thermal electrons cool more efficiently at higher $L_{\rm bol}$, reducing the relative X-ray output as reflected in the rise of $\kappa_{2-10}$, while the nonthermal tail remains largely unaffected, the millimeter/X-ray ratio naturally increases. In this scenario, the nonthermal electrons continue to closely trace the bolometric output, explaining the observed linear $L_{\rm 100GHz}$--$L_{\rm bol}$ relation.
This might raise the question of why cooling affects the two electron populations differently. A possible explanation is that the nonthermal electrons originate from a spatially extended region outside the compact X-ray corona, as proposed by \cite{Hankla2026}.

Another commonly proposed explanation for the millimeter emission in RQ AGN is a compact, low-power jet \citep[e.g.,][]{panessa_origin_2019}. Based on our observations, it remains difficult to distinguish between a coronal and a jet origin for the millimeter emission. 
However, some studies have suggested that the compact jet and the corona might physically be the same structure (e.g., \citealp{Markoff2005}). Therefore, the nonthermal coronal electrons we propose as the source of the millimeter emission could also be associated with this compact jet.
Recently, \cite{Paul2026} tested this scenario for 69 RQ AGN ($z<0.2$), finding that compact jets could not account for the observed 5\,GHz emission, and instead favor a coronal origin, with possible contributions from outflows in $\lambda_{\rm Edd}$ sources.

\begin{figure*}[ht!]
\centering
\includegraphics[width=0.48\textwidth]{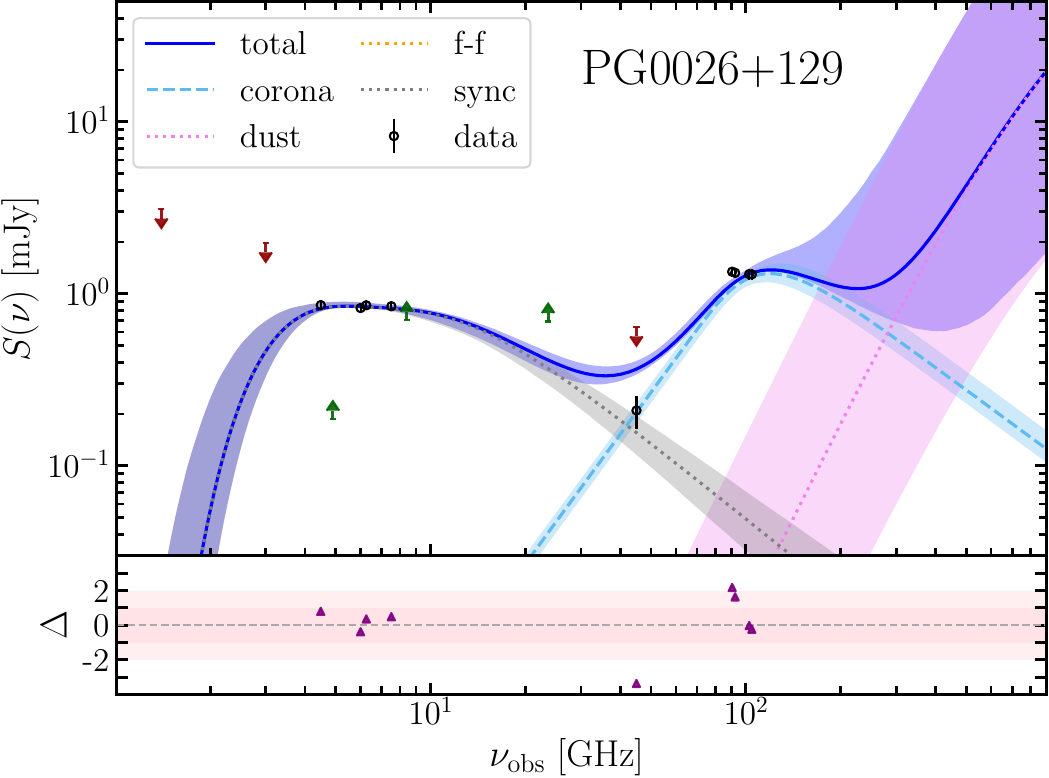}
\hfill
\includegraphics[width=0.48\textwidth]{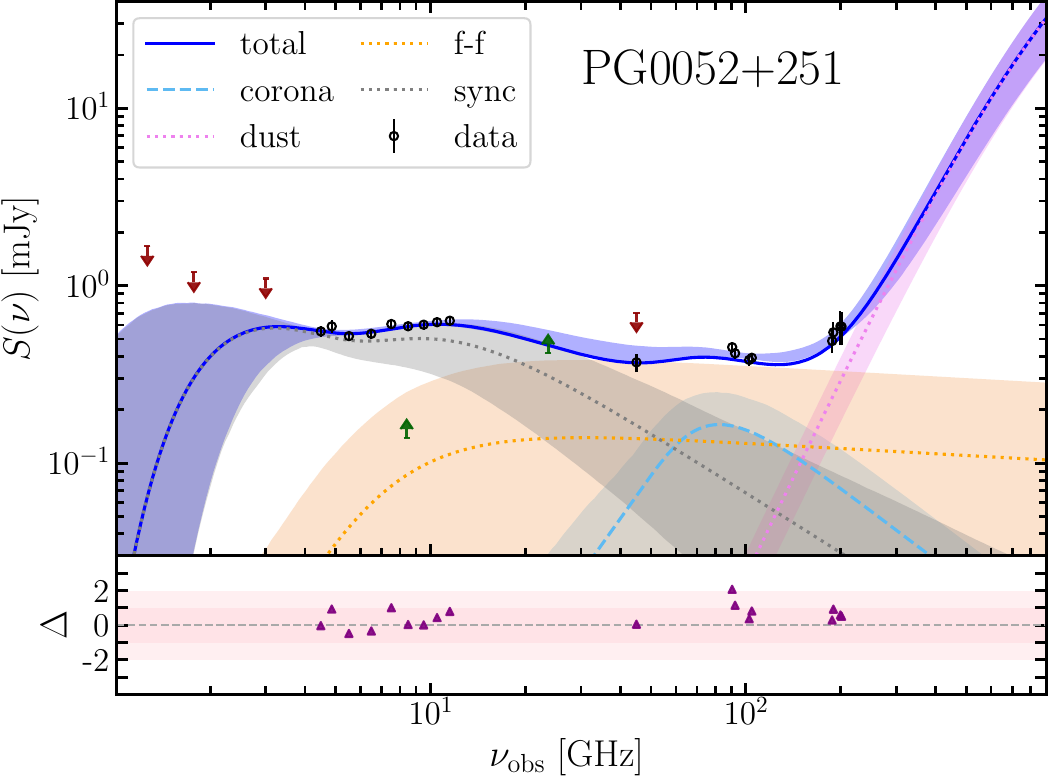}
\caption{The best-constrained SEDs, including the new 100\,GHz observations presented in this work and archival VLA, VLBA and ALMA Band\,5 ($\nu_{\rm center}\sim195$\,GHz) data. The separate components potentially contributing to the observed millimeter emission are indicated: millimeter emission from the corona, free--free emission, diffuse synchrotron emission, and dust.}
\label{fig:SEDs}
\end{figure*}

\subsubsection{Relative increase in millimeter emission}\label{sec:outflows}

The AGN in our sample have substantially higher Eddington ratios ($\lambda_{\rm Edd} = 0.19$--0.85) than those studied by \citetalias{ricci_tight_2023} ($\lambda_{\rm Edd} = 10^{-3}$--0.10). 
High accretion rates are known to power stronger outflows; for example, \citet{fiore_agn_2017} showed that mass outflow rates increase steeply with $L_{\rm bol}$. 
Such outflows can interact with the surrounding interstellar medium, producing shocks \citep[e.g.,][]{nims_observational_2015} that accelerate electrons in the ambient magnetic field and generate synchrotron emission \citep{jiang_synchrotron_2010, hwang_winds_2018, kawamuro_bass_2022}.
Consequently, additional millimeter emission could arise from outflow-induced shocks in our high-$\lambda_{\rm Edd}$ sources, potentially contributing to the enhanced millimeter/X-ray luminosity ratios we observe. 
Consistent with this picture, Figure \ref{fig:ratio_vs_params} shows that the millimeter emission increases with both $L_{\rm bol}$ and $\lambda_{\rm Edd}$, suggesting that shock-related synchrotron emission may play a role in the observed trend.

%

Observational support for this scenario comes from the work of \cite{Shablovinskaia25}, who detected an unresolved, polarized source located $\sim20$\,pc from the nucleus of the RQ AGN NGC\,3783 ($z \sim 0.009$). 
This source was part of the sample studied by \citetalias{ricci_tight_2023} and exhibited the highest $L_{\rm 100GHz}$ among all sources.
The polarized structure coincides with extended millimeter emission, contributing approximately 10\% to the total millimeter flux, and aligns with a known narrow-line outflow previously detected with MUSE \citep{den_brok_muse_2020} and GRAVITY \citep{gravity_collaboration_central_2021}. 
They concluded that the observed extended millimeter emission likely originates from an AGN-driven outflow, with the polarization resulting from synchrotron emission produced by shocks.  
Future work with ALMA polarimetry will investigate whether, for example, the extended millimeter emission observed in LEDA\,12773 has a similar origin.

To further explore the potential link between millimeter emission and outflows in our sample, we analyzed archival high-quality X-ray spectra of our AGN for signatures of ionized absorption (Appendix\,\ref{AppXrayWinds}).
We find that six of the nine AGN have sufficiently high signal-to-noise X-ray observations for meaningful spectroscopic analysis, 
and in five of these six cases, we identified X-ray absorption features consistent with warm ionized outflows. 
Although this analysis is limited by small number statistics, the high incidence of X-ray outflows among sources with elevated millimeter/X-ray luminosity ratios might hint at a possible connection.

We also investigated potential correlations between $\alpha_{\rm mm}$ and $\lambda_{\rm Edd}$ or $L_{\rm bol}$, as presented in Figure\,\ref{fig:alpha_vs_x} (Appendix\,\ref{Appendix_alpha_dep}). 
Since more luminous AGN are capable of driving stronger outflows, a correlation between $\alpha_{\rm mm}$ and $\lambda_{\rm Edd}$ or $L_{\rm bol}$ could indicate a contribution from outflow-related synchrotron emission. 
However, as shown in Figure\,\ref{fig:alpha_vs_x} (Appendix\,\ref{Appendix_alpha_dep}), no significant correlation is observed, with a $p$-value of 0.17 for $\alpha_{\rm mm}$ versus $\lambda_{\rm Edd}$ and a $p$-value of 0.37 for $\alpha_{\rm mm}$ versus $L_{\rm bol}$.
This figure also includes the spectral indices for the sources from \citetalias{ricci_tight_2023}, which have an average value of $\rm \alpha_{mm}^{av,R23}=0.63 \pm 0.14$ (C.S. Chang et al. 2026, in preparation). 
The absence of a correlation between $\alpha_{\rm mm}$ and $\lambda_{\rm Edd}$ at $\sim$230\,GHz was previously reported by \citet{kawamuro_bass_2022}. 
These results suggest that, although outflows may contribute to the millimeter flux, they are unlikely to dominate the emission, which is instead primarily associated with the corona.

Supporting this conclusion, most of our sources exhibit relatively flat $\alpha_{\rm mm}$ (Section\,\ref{spectralindex}), consistent with SSA in the corona. 
Such flat slopes are inconsistent with emission from outflow-driven shocks, which would produce steeper spectra \citep[e.g.,][]{jiang_synchrotron_2010}.
Moreover, emission associated with AGN-driven outflows is generally expected to be most prominent at centimeter (cm) wavelengths rather than in the millimeter regime \citep[e.g.,][]{Yamada2024}. 
In line with this, \cite{Paul2026} reported a radio excess at 5\,GHz in RQ AGN observed with the Karl G. Jansky Very Large Array (VLA) that appears to increase with $\lambda_{\rm Edd}$, and which they suggested may originate from outflow-related synchrotron emission.
\cite{Hankla2026} recently proposed a model in which millimeter emission could arise from an extended, outflowing region physically connected to the corona via magnetic fields, providing a mechanism for coronal-related millimeter emission beyond the compact corona. Notably, this coronal outflow was found to produce flat spectra, which agree with the values for $\alpha_{\rm mm}$ we observe.


Finally, Figure\,\ref{fig:alpha_vs_x} also presents $\alpha_{\rm mm}$ versus SFR, where we do not observe any correlation either ($p$-value=0.85). 
This supports that the millimeter emission is presumably not dominated by star-formation from the larger physical scales probed in this work.

\subsubsection{Radio to submillimeter SED modeling}\label{SEDmod}


To further investigate the origin of the millimeter emission we observe in our sample, we performed SED modeling. The radio to submillimeter SED of an RQ AGN could consist of several components, namely: (1) optically thin synchrotron emission from a diffuse population of relativistic electrons, (2) diffuse free--free emission from ionized gas, (3) synchrotron emission from a compact corona, and (4) thermal emission from dust. 
Of these, the most relevant at frequencies of $\sim$100\,GHz is expected to be the corona, provided that the resolution of the observations is high enough to filter out most of the diffuse emission. 
However, in a more general case, one needs to account for all emission components in the SED. To achieve this, we use the SED fitting procedure presented by \cite{del_palacio_millimeter_2025}. 
This method was originally tested in seven AGN with $\rm \lambda_{Edd}\sim 0.003$--0.035, and it can allow us to infer coronal parameters for our sample. 
In particular, the corona synchrotron SED model depends mainly on two parameters: the radius of the corona ($r_\mathrm{c}=R_{\rm c}/R_{\rm g}$, where $R_{\rm g} = GM_{\rm BH}/c^2$ is the gravitational radius) and the relativistic electron content, defined via the ratio ($\delta$) between the energy density in nonthermal and thermal electrons. 
However, a well-sampled SED is required to disentangle the corona component in the SED and constrain these two main parameters. 

For all of the objects in our sample, we looked for archival radio data. The archival data available for each source are described in Appendix\,\ref{AppSED}.
The objects with the most complete radio to submillimeter SEDs are \pgtw and \pgfi, for which extensive VLA \citep{baldi_pg-rqs_2022} and Very Long Baseline Array \citep[VLBA;][]{Chen2025} data exists. Additionally, ALMA Band\,5 ($\nu_{\rm center}\sim195$\,GHz) observations are available for \pgfi (Proposal ID 2023.1.01062.S; PI: F. Bauer). 
We show their SEDs and the fits in Figure\,\ref{fig:SEDs}. 
For the case of \pgtw, the fit yields well--constrained values of $r_\mathrm{c} = 217\pm13$ and $\log{\delta}=-1.06\pm0.11$, with a peak of the corona component at $\nu_\mathrm{p} = 120\pm10$\,GHz and a flux $S_\mathrm{p} = 1.34\pm0.17$\,mJy. Furthermore, a magnetic field strength ($B$) of $B\sim 19$\,G was derived. 
The observed millimeter peak fluxes across the four SPWs range from $1.30 \pm 0.06$ to $1.34 \pm 0.07$\,mJy, as shown in Figure\,\ref{fig:SEDs}. Therefore, this strongly supports the interpretation of the millimeter flux at $\sim$100\,GHz coming almost exclusively from the corona. 
For \pgfi, 
the fit is more ambiguous, with possible contributions from free--free emission or diffuse synchrotron radiation. This demonstrates the need for multiband observations with similar resolutions.
Nevertheless, we infer $\log \delta=-1.68^{+0.31}_{-1.17}$, a coronal size of $r_\mathrm{c} = 131 \pm 66$, a peak flux and frequency of $S_\mathrm{p}=0.17\pm0.10$\,mJy and $\nu_\mathrm{p} = 82\pm31$\,GHz, respectively, and a magnetic field strength of $B\sim12$\,G.
These coronal size values are consistent with those found for RQ AGN by \citet{del_palacio_millimeter_2025}, where the reported sizes ranged from $r_\mathrm{c} = 60$--250.
For sources in our sample with sparse high-resolution archival data, it remains uncertain whether the 100\,GHz emission originates entirely from the corona.

\section{Conclusions}\label{Conclusion}

In this work, we studied a sample of nine RQ AGN, which were selected for their high bolometric luminosities ($L_{\rm bol}$), bolometric corrections ($ \kappa_{2-10}=L_{\rm bol}/L_{\rm 2-10keV}$), and Eddington ratios ($ \lambda_{\rm Edd}=L_{\rm bol}/L_{\rm Edd}$) to investigate their millimeter/X-ray relation. 
These sources were chosen from the BAT hard-X-ray survey as the brightest AGN that are observable with ALMA. 
For this study, we obtained new quasi-simultaneous observations with ALMA and \textit{Swift}/XRT at 100\,GHz and 2--10\,keV, respectively. 

\begin{enumerate}

    \item We find that the high-luminosity AGN with $ \log(L_{\rm bol}/\rm erg\,s^{-1})=45.3$--46.3 are located above the millimeter/X-ray correlation that was obtained for lower-luminosity sources with $ \log(L_{\rm bol}/\rm erg\,s^{-1})<45$ by \citetalias{ricci_tight_2023} (Figures\,\ref{fig:corrFL} and \ref{fig:ratio_vs_params}). We fit a second-degree polynomial to both samples (Equation\,\ref{eq:mmXray}) and determine an intrinsic scatter of 0.28\,dex.
    We obtain an average millimeter/X-ray luminosity ratio of $-4.02^{+0.37}_{-0.33}$, compared to a lower average ratio of $-4.58\pm0.06$ for the less luminous AGN (Sections\,\ref{millimeter/Xray} and \ref{millimeter/Xrayratio}).

    

    \item 
    We observe a linear relation between the millimeter and disk luminosities (Figure\,\ref{fig:UV_Bol} and Equation\,\ref{eq:LmmUV} in Section\,\ref{mmvsUV}), accompanied by a large intrinsic scatter of 0.45\,dex. 
    We find a significant linear relation between the millimeter and bolometric emission with a scatter of 0.35\,dex (Figure\,\ref{fig:UV_Bol} and Equation\,\ref{eq:LmmBol} in Section\,\ref{mmvsUV}). 

    \item Since no clear correlation is observed between the millimeter/X-ray luminosity ratio and either black hole mass ($M_{\rm BH}$) or star-formation rate (SFR), we conclude that the millimeter/X-ray relation derived in this work might hold across a broad range of AGN properties (Section\,\ref{MbhSFR}). 

    \item As $L_{\rm bol}$ rises, $\kappa_{2-10}$ is known to increase. This declining X-ray fraction may drive the observed rise in the millimeter/X-ray ratio, while the millimeter emission continues to scale linearly with $L_{\rm bol}$. 
    The latter suggests that the millimeter emission traces the total accretion power and is largely insensitive to coronal changes. 
    We propose that this behavior can be explained if the corona hosts two electron populations: a thermal component that becomes less efficient at producing X-rays as $L_{\rm bol}$ rises, and a nonthermal population that continues to scale with the total accretion power (Section\,\ref{decrXray}).

    \item Another possibility is that the higher millimeter/X-ray ratios observed in our high–$\lambda_{\rm Edd}$ AGN may be partly driven by an increase in millimeter emission associated with stronger outflows. Here, shock-induced synchrotron radiation can contribute to the millimeter flux, although the lack of a correlation between $\alpha_{\rm mm}$ and $\lambda_{\rm Edd}$ indicates that such outflow-related emission is likely a secondary component, with the corona remaining the dominant millimeter emitter (Section\,\ref{sec:outflows}).
    
    \item Furthermore, SED modeling for \pgtw indicates that the millimeter emission is dominated by the corona. For \pgfi, a contribution of free--free or diffuse synchrotron emission is present as well.
    The size of the compact component is $R_{\rm c}\sim130$--220$\,R_{\rm g}$, consistent with findings for other RQ AGN.
    For the remaining sources, sparse high-resolution archival data or poorly constrained SEDs prevent robust determination of the coronal contribution. 

\end{enumerate}

\vspace{0.2\baselineskip}

In this work, we probe larger physical scales than the low-luminosity AGN studied by \citetalias{ricci_tight_2023}. 
Despite this, the millimeter/X-ray luminosity ratio remains roughly constant with increasing scale. However, we cannot entirely rule out that the larger physical scales may influence the observed millimeter emission.
Future observations will be essential for further constraining the exact contributions of the millimeter emission in high-luminosity RQ AGN. 
The ALMA Wideband Sensitivity Upgrade (WSU), which will be implemented in the early 2030s, will allow for higher sensitivity observations and could decrease the significant uncertainties on, e.g., the millimeter spectral index $\rm \alpha_{mm}$. 
Additionally, higher-resolution observations will allow us to trace smaller physical scales in these sources, allowing for a more accurate comparison to the lower-luminosity sources from \citetalias{ricci_tight_2023}.
Furthermore, millimeter polarimetry could shed light on the potential contribution of outflows to the observed millimeter emission, as was previously done by \cite{Shablovinskaia25}.
This will be investigated in an accepted Cycle\,12 ALMA proposal (Proposal\,ID\,2025.1.00150.S; PI:\, S.\,Venselaar). 


\section*{Acknowledgments}

We thank the referee, Ehud Behar, and the AAS Statistics Editor for their helpful comments, which improved the quality of this work.

We thank Brad Cenko and the \textit{Swift} team for carrying out the observations of the sources in our sample.

S.V. acknowledges support from SNSF Consolidator grant F01$-$13252 and the China-Chile joint research fund. 
C.R. acknowledges support from SNSF Consolidator grant F01$-$13252, Fondecyt Regular grant 1230345, ANID BASAL project FB210003 and the China-Chile joint research fund.
K.K.G. acknowledges financial support from the Belgian Federal Science Policy Office (BELSPO) in the framework of the PRODEX Programme of the European Space Agency.
R.S. acknowledges funding from the CAS-ANID grant No. CAS220016.
E.S. acknowledges a Humboldt Research Fellowship by the Alexander von Humboldt Foundation. 
E.T. acknowledges support from FONDECYT Regular 1250821.

This paper makes use of the following ALMA data: ADS/JAO.ALMA\#2023.1.01046. 
ALMA is a partnership of ESO (representing its member states), NSF (USA), and NINS (Japan), together with NRC (Canada), MOST and ASIAA (Taiwan), and KASI (Republic of Korea), in cooperation with the Republic of Chile. 
The Joint ALMA Observatory is operated by ESO, AUI/NRAO, and NAOJ.
The National Radio Astronomy Observatory and Green Bank Observatory are facilities of the U.S. National Science Foundation operated under cooperative agreement by Associated Universities, Inc.

Facilities: \textit{Swift}, ALMA.

\vspace{-1\baselineskip}

\bibliography{ref.bib}{}

@INPROCEEDINGS{Ricci2026,
       author = {{Ricci}, Claudio},
        title = "{Unification models of active galactic nuclei}",
    booktitle = {Encyclopedia of Astrophysics},
         year = 2026,
       volume = {4},
        month = jan,
        pages = {210-235},
          doi = {10.1016/B978-0-443-21439-4.00080-8},
       adsurl = {https://ui.adsabs.harvard.edu/abs/2026enap....4..210R}
}

@article{ricci_bat_2017,
	title = {{BAT} {AGN} Spectroscopic Survey. V. X-Ray Properties of the Swift/{BAT} 70-month {AGN} Catalog},
	volume = {233},
	issn = {0067-0049},
	url = {https://ui.adsabs.harvard.edu/abs/2017ApJS..233...17R},
	doi = {10.3847/1538-4365/aa96ad},
	pages = {17},
	journal = {\apj\,Supplement Series},
	author = {Ricci, C. and Trakhtenbrot, B. and Koss, M. J. and Ueda, Y. and Del Vecchio, I. and Treister, E. and Schawinski, K. and Paltani, S. and Oh, K. and Lamperti, I. and Berney, S. and Gandhi, P. and Ichikawa, K. and Bauer, F. E. and Ho, L. C. and Asmus, D. and Beckmann, V. and Soldi, S. and Baloković, M. and Gehrels, N. and Markwardt, C. B.},
	urlyear = {2024-08-08},
	year = {2017},
	note = {Publisher: {IOP}
{ADS} Bibcode: 2017ApJS..233...17R},
}

@article{kawamuro_bass_2023,
	title = {{BASS}. {XXXIV}. A Catalog of the Nuclear Millimeter-wave Continuum Emission Properties of {AGNs} Constrained on Scales ≤ 100–200 pc},
	volume = {269},
	issn = {0067-0049, 1538-4365},
	url = {https://iopscience.iop.org/article/10.3847/1538-4365/acf467},
	doi = {10.3847/1538-4365/acf467},
	pages = {24},
	number = {1},
	journal = {\apj\,Supplement Series},
	shortjournal = {{ApJS}},
	author = {Kawamuro, Taiki and Ricci, Claudio and Mushotzky, Richard F. and Imanishi, Masatoshi and Bauer, Franz E. and Ricci, Federica and Koss, Michael J. and Privon, George C. and Trakhtenbrot, Benny and Izumi, Takuma and Ichikawa, Kohei and Rojas, Alejandra F. and Smith, Krista Lynne and Shimizu, Taro and Oh, Kyuseok and Den Brok, Jakob S. and Baba, Shunsuke and Baloković, Mislav and Chang, Chin-Shin and Kakkad, Darshan and Pfeifle, Ryan W. and Temple, Matthew J. and Ueda, Yoshihiro and Harrison, Fiona and Powell, Meredith C. and Stern, Daniel and Urry, Meg and Sanders, David B.},
	urlyear = {2024-07-15},
	year = {2023},
}

@article{inoue_unveiling_2014,
	title = {Unveiling the nature of coronae in active galactic nuclei through submillimeter observations},
	volume = {66},
	issn = {0004-6264},
	url = {https://ui.adsabs.harvard.edu/abs/2014PASJ...66L...8I},
	doi = {10.1093/pasj/psu079},
	pages = {L8},
	journal = {\pasj},
	author = {Inoue, Yoshiyuki and Doi, Akihiro},
	urlyear = {2024-07-15},
	year = {2014},
	note = {Publisher: {OUP}
{ADS} Bibcode: 2014PASJ...66L...8I},
}

@article{inoue_detection_2018,
	title = {Detection of Coronal Magnetic Activity in nearby Active Supermassive Black Holes},
	volume = {869},
	issn = {0004-637X},
	url = {https://ui.adsabs.harvard.edu/abs/2018ApJ...869..114I},
	doi = {10.3847/1538-4357/aaeb95},
	pages = {114},
	journal = {\apj},
	author = {Inoue, Yoshiyuki and Doi, Akihiro},
	urlyear = {2024-07-15},
	year = {2018},
	note = {Publisher: {IOP}
{ADS} Bibcode: 2018ApJ...869..114I},
}

@article{ricci_tight_2023,
	title = {A Tight Correlation between Millimeter and X-Ray Emission in Accreting Massive Black Holes from {\textless}100 mas Resolution {ALMA} Observations},
	volume = {952},
	issn = {0004-637X},
	url = {https://ui.adsabs.harvard.edu/abs/2023ApJ...952L..28R},
	doi = {10.3847/2041-8213/acda27},
	pages = {L28},
	journal = {\apj},
	author = {Ricci, Claudio and Chang, Chin-Shin and Kawamuro, Taiki and Privon, George C. and Mushotzky, Richard and Trakhtenbrot, Benny and Laor, Ari and Koss, Michael J. and Smith, Krista L. and Gupta, Kriti K. and Dimopoulos, Georgios and Aalto, Susanne and Ros, Eduardo},
	urlyear = {2024-07-15},
	year = {2023},
	note = {Publisher: {IOP}
{ADS} Bibcode: 2023ApJ...952L..28R},
}

@ARTICLE{shablovinskaya_joint_2024,
       author = {{Shablovinskaya}, E. and {Ricci}, C. and {Chang}, C.-S. and {Tortosa}, A. and {del Palacio}, S. and {Kawamuro}, T. and {Aalto}, S. and {Arzoumanian}, Z. and {Balokovic}, M. and {Bauer}, F.~E. and {Gendreau}, K.~C. and {Ho}, L.~C. and {Kakkad}, D. and {Kara}, E. and {Koss}, M.~J. and {Liu}, T. and {Loewenstein}, M. and {Mushotzky}, R. and {Paltani}, S. and {Privon}, G.~C. and {Smith}, K. and {Tombesi}, F. and {Trakhtenbrot}, B.},
        title = "{Joint ALMA/X-ray monitoring of the radio-quiet type 1 active galactic nucleus IC 4329A}",
      journal = {\aap},
         year = 2024,
        month = oct,
       volume = {690},
          eid = {A232},
        pages = {A232},
          doi = {10.1051/0004-6361/202450133},
archivePrefix = {arXiv},
       eprint = {2403.19524},
 primaryClass = {astro-ph.HE},
       adsurl = {https://ui.adsabs.harvard.edu/abs/2024A&A...690A.232S}
}

@article{laor_origin_2008,
	title = {On the origin of radio emission in radio-quiet quasars},
	volume = {390},
	issn = {0035-8711},
	url = {https://ui.adsabs.harvard.edu/abs/2008MNRAS.390..847L},
	doi = {10.1111/j.1365-2966.2008.13806.x},
	pages = {847--862},
	journal = {\mnras},
	author = {Laor, Ari and Behar, Ehud},
	urlyear = {2024-07-15},
	year = {2008},
	note = {Publisher: {OUP}
{ADS} Bibcode: 2008MNRAS.390..847L},
}

@article{casa_team_casa_2022,
	title = {{CASA}, the Common Astronomy Software Applications for Radio Astronomy},
	volume = {134},
	issn = {0004-6280},
	url = {https://ui.adsabs.harvard.edu/abs/2022PASP..134k4501C},
	doi = {10.1088/1538-3873/ac9642},
	pages = {114501},
	journal = {Publications of the Astronomical Society of the Pacific},
	author = {{CASA Team} and Bean, Ben and Bhatnagar, Sanjay and Castro, Sandra and Donovan Meyer, Jennifer and Emonts, Bjorn and Garcia, Enrique and Garwood, Robert and Golap, Kumar and Gonzalez Villalba, Justo and Harris, Pamela and Hayashi, Yohei and Hoskins, Josh and Hsieh, Mingyu and Jagannathan, Preshanth and Kawasaki, Wataru and Keimpema, Aard and Kettenis, Mark and Lopez, Jorge and Marvil, Joshua and Masters, Joseph and McNichols, Andrew and Mehringer, David and Miel, Renaud and Moellenbrock, George and Montesino, Federico and Nakazato, Takeshi and Ott, Juergen and Petry, Dirk and Pokorny, Martin and Raba, Ryan and Rau, Urvashi and Schiebel, Darrell and Schweighart, Neal and Sekhar, Srikrishna and Shimada, Kazuhiko and Small, Des and Steeb, Jan-Willem and Sugimoto, Kanako and Suoranta, Ville and Tsutsumi, Takahiro and van Bemmel, Ilse M. and Verkouter, Marjolein and Wells, Akeem and Xiong, Wei and Szomoru, Arpad and Griffith, Morgan and Glendenning, Brian and Kern, Jeff},
	urlyear = {2023-06-28},
	year = {2022},
	note = {{ADS} Bibcode: 2022PASP..134k4501C},
}

@article{mcmullin_casa_2007,
	title = {{CASA} Architecture and Applications},
	volume = {376},
	url = {https://ui.adsabs.harvard.edu/abs/2007ASPC..376..127M},
	pages = {127},
	author = {{McMullin}, J. P. and Waters, B. and Schiebel, D. and Young, W. and Golap, K.},
	urlyear = {2023-06-28},
	year = {2007},
	note = {Conference Name: Astronomical Data Analysis Software and Systems {XVI}
{ADS} Bibcode: 2007ASPC..376..127M},
}

@article{ramos_almeida_nuclear_2017,
	title = {Nuclear obscuration in active galactic nuclei},
	volume = {1},
	issn = {2397-3366},
	url = {https://ui.adsabs.harvard.edu/abs/2017NatAs...1..679R},
	doi = {10.1038/s41550-017-0232-z},
	pages = {679--689},
	journal = {Nature Astronomy},
	author = {Ramos Almeida, Cristina and Ricci, Claudio},
	urlyear = {2024-09-30},
	year = {2017},
	note = {{ADS} Bibcode: 2017NatAs...1..679R},
}

@article{panessa_origin_2019,
	title = {The origin of radio emission from radio-quiet active galactic nuclei},
	volume = {3},
	issn = {2397-3366},
	url = {https://ui.adsabs.harvard.edu/abs/2019NatAs...3..387P},
	doi = {10.1038/s41550-019-0765-4},
	pages = {387--396},
	journal = {Nature Astronomy},
	author = {Panessa, Francesca and Baldi, Ranieri Diego and Laor, Ari and Padovani, Paolo and Behar, Ehud and {McHardy}, Ian},
	urlyear = {2024-10-04},
	year = {2019},
	note = {{ADS} Bibcode: 2019NatAs...3..387P},
}

@article{kawamuro_bass_2022,
	title = {{BASS} {XXXII}: Studying the Nuclear Millimeter-wave Continuum Emission of {AGNs} with {ALMA} at Scales ≲100–200 pc},
	volume = {938},
	issn = {0004-637X, 1538-4357},
	url = {https://iopscience.iop.org/article/10.3847/1538-4357/ac8794},
	doi = {10.3847/1538-4357/ac8794},
	shorttitle = {{BASS} {XXXII}},
	pages = {87},
	number = {1},
	journal = {\apj},
	shortjournal = {{ApJ}},
	author = {Kawamuro, Taiki and Ricci, Claudio and Imanishi, Masatoshi and Mushotzky, Richard F. and Izumi, Takuma and Ricci, Federica and Bauer, Franz E. and Koss, Michael J. and Trakhtenbrot, Benny and Ichikawa, Kohei and Rojas, Alejandra F. and Smith, Krista Lynne and Shimizu, Taro and Oh, Kyuseok and Den Brok, Jakob S. and Baba, Shunsuke and Baloković, Mislav and Chang, Chin-Shin and Kakkad, Darshan and Pfeifle, Ryan W. and Privon, George C. and Temple, Matthew J. and Ueda, Yoshihiro and Harrison, Fiona and Powell, Meredith C. and Stern, Daniel and Urry, Meg and Sanders, David B.},
	urlyear = {2024-11-08},
	year = {2022},
}

@article{vasudevan_piecing_2007,
	title = {Piecing together the X-ray background: bolometric corrections for active galactic nuclei},
	volume = {381},
	issn = {0035-8711},
	url = {https://ui.adsabs.harvard.edu/abs/2007MNRAS.381.1235V},
	doi = {10.1111/j.1365-2966.2007.12328.x},
	shorttitle = {Piecing together the X-ray background},
	pages = {1235--1251},
	journal = {\mnras},
	author = {Vasudevan, R. V. and Fabian, A. C.},
	urlyear = {2024-11-08},
	year = {2007},
	note = {Publisher: {OUP}
{ADS} Bibcode: 2007MNRAS.381.1235V},
}

@article{baumgartner_70_2013,
	title = {The 70 Month Swift-{BAT} All-sky Hard X-Ray Survey},
	volume = {207},
	issn = {0067-0049},
	url = {https://ui.adsabs.harvard.edu/abs/2013ApJS..207...19B},
	doi = {10.1088/0067-0049/207/2/19},
	pages = {19},
	journal = {\apj\,Supplement Series},
	author = {Baumgartner, W. H. and Tueller, J. and Markwardt, C. B. and Skinner, G. K. and Barthelmy, S. and Mushotzky, R. F. and Evans, P. A. and Gehrels, N.},
	urlyear = {2024-11-12},
	year = {2013},
	note = {Publisher: {IOP}
{ADS} Bibcode: 2013ApJS..207...19B},
}

@article{behar_mm-wave_2018,
	title = {The mm-wave compact component of an {AGN}},
	volume = {478},
	issn = {0035-8711},
	url = {https://ui.adsabs.harvard.edu/abs/2018MNRAS.478..399B},
	doi = {10.1093/mnras/sty850},
	pages = {399--406},
	journal = {\mnras},
	author = {Behar, Ehud and Vogel, Stuart and Baldi, Ranieri D. and Smith, Krista L. and Mushotzky, Richard F.},
	urlyear = {2024},
	year = {2018},
	note = {Publisher: {OUP}
{ADS} Bibcode: 2018MNRAS.478..399B},
}

@article{gupta_bass_2024,
	title = {{BASS}: {XLIII}. Optical, {UV}, and X-ray emission properties of unobscured Swift/{BAT} active galactic nuclei},
	volume = {691},
	issn = {0004-6361},
	url = {https://ui.adsabs.harvard.edu/abs/2024A&A...691A.203G},
	doi = {10.1051/0004-6361/202450567},
	shorttitle = {{BASS}},
	pages = {A203},
	journal = {A\&A},
	author = {Gupta, Kriti K. and Ricci, Claudio and Temple, Matthew J. and Tortosa, Alessia and Koss, Michael J. and Assef, Roberto J. and Bauer, Franz E. and Mushotzy, Richard and Ricci, Federica and Ueda, Yoshihiro and Rojas, Alejandra F. and Trakhtenbrot, Benny and Chang, Chin-Shin and Oh, Kyuseok and Li, Ruancun and Kawamuro, Taiki and Diaz, Yaherlyn and Powell, Meredith C. and Stern, Daniel and Megan Urry, C. and Harrison, Fiona and Cenko, Brad},
	urlyear = {2024-12-16},
	year = {2024},
	note = {Publisher: {EDP}
{ADS} Bibcode: 2024A\&A...691A.203G},
}

@article{hildebrand_determination_1983,
	title = {The determination of cloud masses and dust characteristics from submillimetre thermal emission.},
	volume = {24},
	issn = {0035-8738},
	url = {https://ui.adsabs.harvard.edu/abs/1983QJRAS..24..267H},
	pages = {267--282},
	journal = {Quarterly Journal of the Royal Astronomical Society},
	author = {Hildebrand, R. H.},
	urlyear = {2024-12-18},
	year = {1983},
	note = {{ADS} Bibcode: 1983QJRAS..24..267H},
}

@article{poole_photometric_2008,
	title = {Photometric calibration of the Swift ultraviolet/optical telescope},
	volume = {383},
	issn = {0035-8711},
	url = {https://ui.adsabs.harvard.edu/abs/2008MNRAS.383..627P},
	doi = {10.1111/j.1365-2966.2007.12563.x},
	pages = {627--645},
	journal = {\mnras},
	author = {Poole, T. S. and Breeveld, A. A. and Page, M. J. and Landsman, W. and Holland, S. T. and Roming, P. and Kuin, N. P. M. and Brown, P. J. and Gronwall, C. and Hunsberger, S. and Koch, S. and Mason, K. O. and Schady, P. and vanden Berk, D. and Blustin, A. J. and Boyd, P. and Broos, P. and Carter, M. and Chester, M. M. and Cucchiara, A. and Hancock, B. and Huckle, H. and Immler, S. and Ivanushkina, M. and Kennedy, T. and Marshall, F. and Morgan, A. and Pandey, S. B. and de Pasquale, M. and Smith, P. J. and Still, M.},
	urlyear = {2025-01-17},
	year = {2008},
	note = {Publisher: {OUP}
{ADS} Bibcode: 2008MNRAS.383..627P},
}

@article{breeveld_further_2010,
	title = {Further calibration of the Swift ultraviolet/optical telescope},
	volume = {406},
	issn = {0035-8711},
	url = {https://ui.adsabs.harvard.edu/abs/2010MNRAS.406.1687B},
	doi = {10.1111/j.1365-2966.2010.16832.x},
	pages = {1687--1700},
	journal = {\mnras},
	author = {Breeveld, A. A. and Curran, P. A. and Hoversten, E. A. and Koch, S. and Landsman, W. and Marshall, F. E. and Page, M. J. and Poole, T. S. and Roming, P. and Smith, P. J. and Still, M. and Yershov, V. and Blustin, A. J. and Brown, P. J. and Gronwall, C. and Holland, S. T. and Kuin, N. P. M. and {McGowan}, K. and Rosen, S. and Boyd, P. and Broos, P. and Carter, M. and Chester, M. M. and Hancock, B. and Huckle, H. and Immler, S. and Ivanushkina, M. and Kennedy, T. and Mason, K. O. and Morgan, A. N. and Oates, S. and de Pasquale, M. and Schady, P. and Siegel, M. and vanden Berk, D.},
	urlyear = {2025-01-17},
	year = {2010},
	note = {Publisher: {OUP}
{ADS} Bibcode: 2010MNRAS.406.1687B},
}

@article{wilms_absorption_2000,
	title = {On the Absorption of X-Rays in the Interstellar Medium},
	volume = {542},
	issn = {0004-637X},
	url = {https://ui.adsabs.harvard.edu/abs/2000ApJ...542..914W},
	doi = {10.1086/317016},
	pages = {914--924},
	journal = {\apj},
	author = {Wilms, J. and Allen, A. and {McCray}, R.},
	urlyear = {2025-01-17},
	year = {2000},
	note = {Publisher: {IOP}
{ADS} Bibcode: 2000ApJ...542..914W},
}

@article{cash_parameter_1979,
	title = {Parameter estimation in astronomy through application of the likelihood ratio.},
	volume = {228},
	issn = {0004-637X},
	url = {https://ui.adsabs.harvard.edu/abs/1979ApJ...228..939C},
	doi = {10.1086/156922},
	pages = {939--947},
	journal = {\apj},
	author = {Cash, W.},
	urlyear = {2025-01-17},
	year = {1979},
	note = {Publisher: {IOP}
{ADS} Bibcode: 1979ApJ...228..939C},
}

@article{burrows_swift_2005,
	title = {The Swift X-Ray Telescope},
	volume = {120},
	issn = {0038-6308},
	url = {https://ui.adsabs.harvard.edu/abs/2005SSRv..120..165B},
	doi = {10.1007/s11214-005-5097-2},
	pages = {165--195},
	journal = {Space Science Reviews},
	author = {Burrows, David N. and Hill, J. E. and Nousek, J. A. and Kennea, J. A. and Wells, A. and Osborne, J. P. and Abbey, A. F. and Beardmore, A. and Mukerjee, K. and Short, A. D. T. and Chincarini, G. and Campana, S. and Citterio, O. and Moretti, A. and Pagani, C. and Tagliaferri, G. and Giommi, P. and Capalbi, M. and Tamburelli, F. and Angelini, L. and Cusumano, G. and Bräuninger, H. W. and Burkert, W. and Hartner, G. D.},
	urlyear = {2025-01-17},
	year = {2005},
	note = {{ADS} Bibcode: 2005SSRv..120..165B},
}

@article{ricci_compton-thick_2015,
	title = {Compton-thick Accretion in the Local Universe},
	volume = {815},
	issn = {0004-637X},
	url = {https://ui.adsabs.harvard.edu/abs/2015ApJ...815L..13R},
	doi = {10.1088/2041-8205/815/1/L13},
	pages = {L13},
	journal = {\apj},
	author = {Ricci, C. and Ueda, Y. and Koss, M. J. and Trakhtenbrot, B. and Bauer, F. E. and Gandhi, P.},
	urldate = {2025-01-27},
	year = {2015},
	note = {Publisher: {IOP}
{ADS} Bibcode: 2015ApJ...815L..13R},
}

@article{behar_discovery_2015,
	title = {Discovery of millimetre-wave excess emission in radio-quiet active galactic nuclei},
	volume = {451},
	issn = {0035-8711},
	url = {https://ui.adsabs.harvard.edu/abs/2015MNRAS.451..517B},
	doi = {10.1093/mnras/stv988},
	pages = {517--526},
	journal = {\mnras},
	author = {Behar, Ehud and Baldi, Ranieri D. and Laor, Ari and Horesh, Assaf and Stevens, Jamie and Tzioumis, Tasso},
	urldate = {2025-01-27},
	year = {2015},
	note = {Publisher: {OUP}
{ADS} Bibcode: 2015MNRAS.451..517B},
}

@article{arnaud_xspec_1996,
	title = {{XSPEC}: The First Ten Years},
	volume = {101},
	url = {https://ui.adsabs.harvard.edu/abs/1996ASPC..101...17A},
	shorttitle = {{XSPEC}},
	pages = {17},
	author = {Arnaud, K. A.},
	urldate = {2025-01-27},
	year = {1996},
	note = {Conference Name: Astronomical Data Analysis Software and Systems V
{ADS} Bibcode: 1996ASPC..101...17A},
}

@article{vasudevan_simultaneous_2009,
	title = {Simultaneous X-ray/optical/{UV} snapshots of active galactic nuclei from {XMM}-Newton: spectral energy distributions for the reverberation mapped sample},
	volume = {392},
	issn = {0035-8711},
	url = {https://ui.adsabs.harvard.edu/abs/2009MNRAS.392.1124V},
	doi = {10.1111/j.1365-2966.2008.14108.x},
	shorttitle = {Simultaneous X-ray/optical/{UV} snapshots of active galactic nuclei from {XMM}-Newton},
	pages = {1124--1140},
	journal = {\mnras},
	author = {Vasudevan, R. V. and Fabian, A. C.},
	urldate = {2025-01-28},
	year = {2009},
	note = {Publisher: {OUP}
{ADS} Bibcode: 2009MNRAS.392.1124V},
}

@article{di_matteo_magnetic_1998,
	title = {Magnetic reconnection: flares and coronal heating in active galactic nuclei},
	volume = {299},
	issn = {0035-8711},
	url = {https://ui.adsabs.harvard.edu/abs/1998MNRAS.299L..15D},
	doi = {10.1046/j.1365-8711.1998.01950.x},
	shorttitle = {Magnetic reconnection},
	pages = {L15--l20},
	journal = {\mnras},
	author = {Di Matteo, T.},
	urldate = {2025-01-29},
	year = {1998},
	note = {Publisher: {OUP}
{ADS} Bibcode: 1998MNRAS.299L..15D},
}

@article{duras_universal_2020,
	title = {Universal bolometric corrections for active galactic nuclei over seven luminosity decades},
	volume = {636},
	issn = {0004-6361},
	url = {https://ui.adsabs.harvard.edu/abs/2020A&A...636A..73D},
	doi = {10.1051/0004-6361/201936817},
	pages = {A73},
	journal = {A\&A},
	author = {Duras, F. and Bongiorno, A. and Ricci, F. and Piconcelli, E. and Shankar, F. and Lusso, E. and Bianchi, S. and Fiore, F. and Maiolino, R. and Marconi, A. and Onori, F. and Sani, E. and Schneider, R. and Vignali, C. and La Franca, F.},
	urldate = {2025-02-03},
	year = {2020},
	note = {Publisher: {EDP}
{ADS} Bibcode: 2020A\&A...636A..73D},
}

@article{martocchia_wissh_2017,
	title = {The {WISSH} quasars project. {III}. X-ray properties of hyper-luminous quasars},
	volume = {608},
	issn = {0004-6361},
	url = {https://ui.adsabs.harvard.edu/abs/2017A&A...608A..51M},
	doi = {10.1051/0004-6361/201731314},
	pages = {A51},
	journal = {A\&A},
	author = {Martocchia, S. and Piconcelli, E. and Zappacosta, L. and Duras, F. and Vietri, G. and Vignali, C. and Bianchi, S. and Bischetti, M. and Bongiorno, A. and Brusa, M. and Lanzuisi, G. and Marconi, A. and Mathur, S. and Miniutti, G. and Nicastro, F. and Bruni, G. and Fiore, F.},
	urldate = {2025-02-03},
	year = {2017},
	note = {Publisher: {EDP}
{ADS} Bibcode: 2017A\&A...608A..51M},
}

@article{padovani_two_2017,
	title = {On the two main classes of active galactic nuclei},
	volume = {1},
	issn = {2397-3366},
	url = {https://ui.adsabs.harvard.edu/abs/2017NatAs...1E.194P},
	doi = {10.1038/s41550-017-0194},
	pages = {0194},
	journal = {Nature Astronomy},
	author = {Padovani, Paolo},
	urldate = {2025-01-29},
	year = {2017},
	note = {{ADS} Bibcode: 2017NatAs...1E.194P},
}

@article{katz_nonrelativistic_1976,
	title = {Nonrelativistic {Compton} scattering and models of quasars.},
	volume = {206},
	issn = {0004-637X},
	url = {https://ui.adsabs.harvard.edu/abs/1976ApJ...206..910K},
	doi = {10.1086/154455},
	urldate = {2025-02-04},
	journal = {\apj},
	author = {Katz, J. I.},
	month = jun,
	year = {1976},
	note = {Publisher: IOP
ADS Bibcode: 1976ApJ...206..910K},
	pages = {910--916},
}

@article{pei_interstellar_1992,
	title = {Interstellar {Dust} from the {Milky} {Way} to the {Magellanic} {Clouds}},
	volume = {395},
	issn = {0004-637X},
	url = {https://ui.adsabs.harvard.edu/abs/1992ApJ...395..130P},
	doi = {10.1086/171637},
	urldate = {2025-02-05},
	journal = {\apj},
	author = {Pei, Yichuan C.},
	month = aug,
	year = {1992},
	note = {Publisher: IOP
ADS Bibcode: 1992ApJ...395..130P},
	pages = {130},
}

@article{koss_bass_2022,
	title = {{BASS}. {XXI}. {The} {Data} {Release} 2 {Overview}},
	volume = {261},
	issn = {0067-0049, 1538-4365},
	url = {https://iopscience.iop.org/article/10.3847/1538-4365/ac6c8f},
	doi = {10.3847/1538-4365/ac6c8f},
	number = {1},
	urldate = {2025-02-10},
	journal = {\apj\,Supplement Series},
	author = {Koss, Michael J. and Trakhtenbrot, Benny and Ricci, Claudio and Bauer, Franz E. and Treister, Ezequiel and Mushotzky, Richard and Urry, C. Megan and Ananna, Tonima T. and Baloković, Mislav and Den Brok, Jakob S. and Cenko, S. Bradley and Harrison, Fiona and Ichikawa, Kohei and Lamperti, Isabella and Lein, Amy and Mejía-Restrepo, Julian E. and Oh, Kyuseok and Pacucci, Fabio and Pfeifle, Ryan W. and Powell, Meredith C. and Privon, George C. and Ricci, Federica and Salvato, Mara and Schawinski, Kevin and Shimizu, Taro and Smith, Krista L. and Stern, Daniel},
	month = jul,
	year = {2022},
	pages = {1},
}

@article{koss_bat_2017,
	title = {{BAT} {AGN} {Spectroscopic} {Survey}. {I}. {Spectral} {Measurements}, {Derived} {Quantities}, and {AGN} {Demographics}},
	volume = {850},
	issn = {0004-637X, 1538-4357},
	url = {https://iopscience.iop.org/article/10.3847/1538-4357/aa8ec9},
	doi = {10.3847/1538-4357/aa8ec9},
	number = {1},
	urldate = {2025-02-10},
	journal = {\apj},
	author = {Koss, Michael and Trakhtenbrot, Benny and Ricci, Claudio and Lamperti, Isabella and Oh, Kyuseok and Berney, Simon and Schawinski, Kevin and Baloković, Mislav and Baronchelli, Linda and Crenshaw, D. Michael and Fischer, Travis and Gehrels, Neil and Harrison, Fiona and Hashimoto, Yasuhiro and Hogg, Drew and Ichikawa, Kohei and Masetti, Nicola and Mushotzky, Richard and Sartori, Lia and Stern, Daniel and Treister, Ezequiel and Ueda, Yoshihiro and Veilleux, Sylvain and Winter, Lisa},
	month = nov,
	year = {2017},
	pages = {74},
}

@article{hwang_winds_2018,
	title = {Winds as the origin of radio emission in z = 2.5 radio-quiet extremely red quasars},
	volume = {477},
	issn = {0035-8711, 1365-2966},
	url = {https://academic.oup.com/mnras/article/477/1/830/4952005},
	doi = {10.1093/mnras/sty742},
	language = {en},
	number = {1},
	urldate = {2025-02-11},
	journal = {\mnras},
	author = {Hwang, Hsiang-Chih and Zakamska, Nadia L and Alexandroff, Rachael M and Hamann, Fred and Greene, Jenny E and Perrotta, Serena and Richards, Gordon T},
	month = jun,
	year = {2018},
	pages = {830--844},
}

@article{jiang_synchrotron_2010,
	title = {{SYNCHROTRON} {EMISSION} {FROM} {ELLIPTICAL} {GALAXIES} {CONSEQUENT} {TO} {ACTIVE} {GALACTIC} {NUCLEUS} {OUTBURSTS}},
	volume = {711},
	issn = {0004-637X, 1538-4357},
	url = {https://iopscience.iop.org/article/10.1088/0004-637X/711/1/125},
	doi = {10.1088/0004-637X/711/1/125},
	number = {1},
	urldate = {2025-02-11},
	journal = {\apj},
	author = {Jiang, Yan-Fei and Ciotti, Luca and Ostriker, Jeremiah P. and Spitkovsky, Anatoly},
	month = mar,
	year = {2010},
	pages = {125--137},
}

@article{lusso_x-ray_2010,
	title = {The {X}-ray to optical-{UV} luminosity ratio of {X}-ray selected type 1 {AGN} in {XMM}-{COSMOS}},
	volume = {512},
	issn = {0004-6361, 1432-0746},
	url = {http://www.aanda.org/10.1051/0004-6361/200913298},
	doi = {10.1051/0004-6361/200913298},
	urldate = {2025-02-11},
	journal = {A\&A},
	author = {Lusso, E. and Comastri, A. and Vignali, C. and Zamorani, G. and Brusa, M. and Gilli, R. and Iwasawa, K. and Salvato, M. and Civano, F. and Elvis, M. and Merloni, A. and Bongiorno, A. and Trump, J. R. and Koekemoer, A. M. and Schinnerer, E. and Le Floc'h, E. and Cappelluti, N. and Jahnke, K. and Sargent, M. and Silverman, J. and Mainieri, V. and Fiore, F. and Bolzonella, M. and Le Fèvre, O. and Garilli, B. and Iovino, A. and Kneib, J. P. and Lamareille, F. and Lilly, S. and Mignoli, M. and Scodeggio, M. and Vergani, D.},
	month = mar,
	year = {2010},
	pages = {A34},
}

@article{baldi_milimetre-band_2015,
	title = {Milimetre-band variability of the radio-quiet nucleus of {NGC} 7469},
	volume = {454},
	issn = {0035-8711},
	url = {https://ui.adsabs.harvard.edu/abs/2015MNRAS.454.4277B},
	doi = {10.1093/mnras/stv2284},
	urldate = {2025-02-19},
	journal = {\mnras},
	author = {Baldi, Ranieri D. and Behar, Ehud and Laor, Ari and Horesh, Assaf},
	month = dec,
	year = {2015},
	note = {Publisher: OUP
ADS Bibcode: 2015MNRAS.454.4277B},
	pages = {4277--4281},
}

@article{behar_simultaneous_2020,
	title = {Simultaneous {Millimetre}-wave and {X}-ray monitoring of the {Seyfert} galaxy {NGC} 7469},
	volume = {491},
	issn = {0035-8711},
	url = {https://ui.adsabs.harvard.edu/abs/2020MNRAS.491.3523B},
	doi = {10.1093/mnras/stz3273},
	urldate = {2025-02-19},
	journal = {\mnras},
	author = {Behar, Ehud and Kaspi, Shai and Paubert, Gabriel and Billot, Nicolas and Peretz, Uria and Baldi, Ranieri D. and Laor, Ari and Kaastra, Jelle and Mehdipour, Missagh},
	month = jan,
	year = {2020},
	note = {Publisher: OUP
ADS Bibcode: 2020MNRAS.491.3523B},
	pages = {3523--3534},
}

@article{petrucci_simultaneous_2023,
	title = {Simultaneous millimetric and {X}-ray intraday variability in the radio-quiet {AGN} {MCG}+08-11-11},
	volume = {678},
	issn = {0004-6361},
	url = {https://ui.adsabs.harvard.edu/abs/2023A&A...678L...4P},
	doi = {10.1051/0004-6361/202347495},
	urldate = {2025-02-19},
	journal = {A\&A},
	author = {Petrucci, P. -O. and Piétu, V. and Behar, E. and Clavel, M. and Bianchi, S. and Henri, G. and Barnier, S. and Chen, S. and Ferreira, J. and Malzac, J. and Belmont, R. and Corbel, S. and Coriat, M.},
	month = oct,
	year = {2023},
	note = {ADS Bibcode: 2023A\&A...678L...4P},
	pages = {L4},
}

@article{ackermann_search_2012,
	title = {{SEARCH} {FOR} {GAMMA}-{RAY} {EMISSION} {FROM} {X}-{RAY}-{SELECTED} {SEYFERT} {GALAXIES} {WITH} \textit{{FERMI}} -{LAT}},
	volume = {747},
	issn = {0004-637X, 1538-4357},
	url = {https://iopscience.iop.org/article/10.1088/0004-637X/747/2/104},
	doi = {10.1088/0004-637X/747/2/104},
	number = {2},
	urldate = {2025-02-20},
	journal = {\apj},
	author = {Ackermann, M. and Ajello, M. and Allafort, A. and Baldini, L. and Ballet, J. and Barbiellini, G. and Bastieri, D. and Bechtol, K. and Bellazzini, R. and Berenji, B. and Bloom, E. D. and Bonamente, E. and Borgland, A. W. and Bregeon, J. and Brigida, M. and Bruel, P. and Buehler, R. and Buson, S. and Caliandro, G. A. and Cameron, R. A. and Caraveo, P. A. and Casandjian, J. M. and Cavazzuti, E. and Cecchi, C. and Charles, E. and Chekhtman, A. and Cheung, C. C. and Chiang, J. and Ciprini, S. and Claus, R. and Cohen-Tanugi, J. and Conrad, J. and Cutini, S. and D'Ammando, F. and De Angelis, A. and De Palma, F. and Dermer, C. D. and Do Couto E Silva, E. and Drell, P. S. and Drlica-Wagner, A. and Enoto, T. and Favuzzi, C. and Fegan, S. J. and Ferrara, E. C. and Fortin, P. and Fukazawa, Y. and Fusco, P. and Gargano, F. and Gasparrini, D. and Gehrels, N. and Germani, S. and Giglietto, N. and Giommi, P. and Giordano, F. and Giroletti, M. and Godfrey, G. and Grove, J. E. and Guiriec, S. and Hadasch, D. and Hayashida, M. and Hays, E. and Hughes, R. E. and Jóhannesson, G. and Johnson, A. S. and Kamae, T. and Katagiri, H. and Kataoka, J. and Knödlseder, J. and Kuss, M. and Lande, J. and Garde, M. Llena and Longo, F. and Loparco, F. and Lott, B. and Lovellette, M. N. and Lubrano, P. and Madejski, G. M. and Mazziotta, M. N. and Michelson, P. F. and Mizuno, T. and Monte, C. and Monzani, M. E. and Morselli, A. and Moskalenko, I. V. and Murgia, S. and Nishino, S. and Norris, J. P. and Nuss, E. and Ohno, M. and Ohsugi, T. and Okumura, A. and Orlando, E. and Ozaki, M. and Paneque, D. and Pesce-Rollins, M. and Pierbattista, M. and Piron, F. and Pivato, G. and Porter, T. A. and Rainò, S. and Rando, R. and Razzano, M. and Reimer, A. and Reimer, O. and Ritz, S. and Roth, M. and Sanchez, D. A. and Sbarra, C. and Sgrò, C. and Siskind, E. J. and Spandre, G. and Spinelli, P. and Stawarz, Ł. and Strong, A. W. and Takahashi, H. and Takahashi, T. and Tanaka, T. and Thayer, J. B. and Thompson, D. J. and Tibaldo, L. and Tinivella, M. and Torres, D. F. and Tosti, G. and Troja, E. and Uchiyama, Y. and Usher, T. L. and Vandenbroucke, J. and Vasileiou, V. and Vianello, G. and Vitale, V. and Waite, A. P. and Winer, B. L. and Wood, K. S. and Wood, M. and Yang, Z. and Zimmer, S.},
	month = mar,
	year = {2012},
	pages = {104},
}

@article{malizia_integral_2014,
	title = {{THE} \textit{{INTEGRAL}} {HIGH}-{ENERGY} {CUT}-{OFF} {DISTRIBUTION} {OF} {TYPE} 1 {ACTIVE} {GALACTIC} {NUCLEI}},
	volume = {782},
	copyright = {http://iopscience.iop.org/info/page/text-and-data-mining},
	issn = {2041-8205, 2041-8213},
	url = {https://iopscience.iop.org/article/10.1088/2041-8205/782/2/L25},
	doi = {10.1088/2041-8205/782/2/L25},
	number = {2},
	urldate = {2025-02-20},
	journal = {\apj},
	author = {Malizia, A. and Molina, M. and Bassani, L. and Stephen, J. B. and Bazzano, A. and Ubertini, P. and Bird, A. J.},
	month = feb,
	year = {2014},
	pages = {L25},
}

@article{padovani_vla_2011,
	title = {The {VLA} {Survey} of {Chandra} {Deep} {Field} {South}. {V}. {Evolution} and {Luminosity} {Functions} of {Sub}-millijansky {Radio} {Sources} and the {Issue} of {Radio} {Emission} in {Radio}-quiet {Active} {Galactic} {Nuclei}},
	volume = {740},
	issn = {0004-637X},
	url = {https://ui.adsabs.harvard.edu/abs/2011ApJ...740...20P},
	doi = {10.1088/0004-637X/740/1/20},
	urldate = {2025-02-04},
	journal = {\apj},
	author = {Padovani, P. and Miller, N. and Kellermann, K. I. and Mainieri, V. and Rosati, P. and Tozzi, P.},
	month = oct,
	year = {2011},
	note = {Publisher: IOP
ADS Bibcode: 2011ApJ...740...20P},
	pages = {20},
}

@article{ichikawa_complete_2017,
	title = {The {Complete} {Infrared} {View} of {Active} {Galactic} {Nuclei} from the 70 {Month} {Swift}/{BAT} {Catalog}},
	volume = {835},
	issn = {0004-637X},
	url = {https://ui.adsabs.harvard.edu/abs/2017ApJ...835...74I},
	doi = {10.3847/1538-4357/835/1/74},
	urldate = {2025-02-20},
	journal = {\apj},
	author = {Ichikawa, Kohei and Ricci, Claudio and Ueda, Yoshihiro and Matsuoka, Kenta and Toba, Yoshiki and Kawamuro, Taiki and Trakhtenbrot, Benny and Koss, Michael J.},
	month = jan,
	year = {2017},
	note = {Publisher: IOP
ADS Bibcode: 2017ApJ...835...74I},
	pages = {74},
}

@article{zappacosta_wissh_2020,
	title = {The {WISSH} quasars project. {VII}. {The} impact of extreme radiative field in the accretion disc and {X}-ray corona interplay},
	volume = {635},
	issn = {0004-6361},
	url = {https://ui.adsabs.harvard.edu/abs/2020A&A...635L...5Z},
	doi = {10.1051/0004-6361/201937292},
	urldate = {2025-02-26},
	journal = {A\&A},
	author = {Zappacosta, L. and Piconcelli, E. and Giustini, M. and Vietri, G. and Duras, F. and Miniutti, G. and Bischetti, M. and Bongiorno, A. and Brusa, M. and Chiaberge, M. and Comastri, A. and Feruglio, C. and Luminari, A. and Marconi, A. and Ricci, C. and Vignali, C. and Fiore, F.},
	month = mar,
	year = {2020},
	note = {Publisher: EDP
ADS Bibcode: 2020A\&A...635L...5Z},
	pages = {L5},
}

@article{fiore_agn_2017,
	title = {{AGN} wind scaling relations and the co-evolution of black holes and galaxies},
	volume = {601},
	issn = {0004-6361},
	url = {https://ui.adsabs.harvard.edu/abs/2017A&A...601A.143F},
	doi = {10.1051/0004-6361/201629478},
	urldate = {2025-04-15},
	journal = {A\&A},
	author = {Fiore, F. and Feruglio, C. and Shankar, F. and Bischetti, M. and Bongiorno, A. and Brusa, M. and Carniani, S. and Cicone, C. and Duras, F. and Lamastra, A. and Mainieri, V. and Marconi, A. and Menci, N. and Maiolino, R. and Piconcelli, E. and Vietri, G. and Zappacosta, L.},
	month = may,
	year = {2017},
	note = {ADS Bibcode: 2017A\&A...601A.143F},
	pages = {A143},
}

@article{nims_observational_2015,
	title = {Observational signatures of galactic winds powered by active galactic nuclei},
	volume = {447},
	issn = {0035-8711},
	url = {https://ui.adsabs.harvard.edu/abs/2015MNRAS.447.3612N},
	doi = {10.1093/mnras/stu2648},
	urldate = {2025-06-26},
	journal = {\mnras},
	author = {Nims, Jesse and Quataert, Eliot and Faucher-Giguère, Claude-André},
	month = mar,
	year = {2015},
	note = {Publisher: OUP
ADS Bibcode: 2015MNRAS.447.3612N},
	pages = {3612--3622},
}

@article{ricci_bat_2018,
	title = {{BAT} {AGN} {Spectroscopic} {Survey} - {XII}. {The} relation between coronal properties of active galactic nuclei and the {Eddington} ratio},
	volume = {480},
	issn = {0035-8711},
	url = {https://ui.adsabs.harvard.edu/abs/2018MNRAS.480.1819R},
	doi = {10.1093/mnras/sty1879},
	urldate = {2025-07-07},
	journal = {\mnras},
	author = {Ricci, C. and Ho, L. C. and Fabian, A. C. and Trakhtenbrot, B. and Koss, M. J. and Ueda, Y. and Lohfink, A. and Shimizu, T. and Bauer, F. E. and Mushotzky, R. and Schawinski, K. and Paltani, S. and Lamperti, I. and Treister, E. and Oh, K.},
	month = oct,
	year = {2018},
	note = {Publisher: OUP
ADS Bibcode: 2018MNRAS.480.1819R},
	pages = {1819--1830},
}

@ARTICLE{del_palacio_millimeter_2025,
       author = {{del Palacio}, S. and {Yang}, C. and {Aalto}, S. and {Ricci}, C. and {Lankhaar}, B. and {K{\"o}nig}, S. and {Becker Tjus}, J. and {Magno}, M. and {Smith}, K.~L. and {Yang}, J. and {Barcos-Mu{\~n}oz}, L. and {Combes}, F. and {Linden}, S. and {Henkel}, C. and {Mangum}, J.~G. and {Mart{\'\i}n}, S. and {Olander}, G. and {Privon}, G. and {Wethers}, C. and {Baczko}, A.-K. and {Beswick}, R.~J. and {Garc{\'\i}a-Bernete}, I. and {Garc{\'\i}a-Burillo}, S. and {Gonz{\'a}lez-Alfonso}, E. and {Gorski}, M. and {Imanishi}, M. and {Izumi}, T. and {Muller}, S. and {Nishimura}, Y. and {Pereira-Santaella}, M. and {van der Werf}, P.~P.},
        title = "{Millimeter emission from supermassive black hole coronae}",
      journal = {\aap},
         year = 2025,
        month = sep,
       volume = {701},
          eid = {A41},
        pages = {A41},
          doi = {10.1051/0004-6361/202554936},
archivePrefix = {arXiv},
       eprint = {2504.07762},
 primaryClass = {astro-ph.HE},
       adsurl = {https://ui.adsabs.harvard.edu/abs/2025A&A...701A..41D}
}

@article{ichikawa_bat_2019,
	title = {{BAT} {AGN} {Spectroscopic} {Survey}. {XI}. {The} {Covering} {Factor} of {Dust} and {Gas} in {Swift}/{BAT} {Active} {Galactic} {Nuclei}},
	volume = {870},
	issn = {0004-637X},
	url = {https://ui.adsabs.harvard.edu/abs/2019ApJ...870...31I},
	doi = {10.3847/1538-4357/aaef8f},
	urldate = {2025-07-09},
	journal = {\apj},
	author = {Ichikawa, Kohei and Ricci, Claudio and Ueda, Yoshihiro and Bauer, Franz E. and Kawamuro, Taiki and Koss, Michael J. and Oh, Kyuseok and Rosario, David J. and Shimizu, T. Taro and Stalevski, Marko and Fuller, Lindsay and Packham, Christopher and Trakhtenbrot, Benny},
	month = jan,
	year = {2019},
	note = {Publisher: IOP
ADS Bibcode: 2019ApJ...870...31I},
	pages = {31},
}

@ARTICLE{curran_monte_2015,
       author = {{Curran}, P.~A.},
        title = "{Monte Carlo error analyses of Spearman's rank test}",
      journal = {arXiv e-prints},
         year = 2014,
        month = nov,
          eid = {arXiv:1411.3816},
        pages = {arXiv:1411.3816},
          doi = {10.48550/arXiv.1411.3816},
archivePrefix = {arXiv},
       eprint = {1411.3816},
 primaryClass = {astro-ph.IM},
       adsurl = {https://ui.adsabs.harvard.edu/abs/2014arXiv1411.3816C}
}

@article{privon_hard_2020,
	title = {A {Hard} {X}-{Ray} {Test} of {HCN} {Enhancements} {As} a {Tracer} of {Embedded} {Black} {Hole} {Growth}},
	volume = {893},
	issn = {0004-637X},
	url = {https://ui.adsabs.harvard.edu/abs/2020ApJ...893..149P},
	doi = {10.3847/1538-4357/ab8015},
	urldate = {2025-07-09},
	journal = {\apj},
	author = {Privon, G. C. and Ricci, C. and Aalto, S. and Viti, S. and Armus, L. and Díaz-Santos, T. and González-Alfonso, E. and Iwasawa, K. and Jeff, D. L. and Treister, E. and Bauer, F. and Evans, A. S. and Garg, P. and Herrero-Illana, R. and Mazzarella, J. M. and Larson, K. and Blecha, L. and Barcos-Muñoz, L. and Charmandaris, V. and Stierwalt, S. and Pérez-Torres, M. A.},
	month = apr,
	year = {2020},
	note = {Publisher: IOP
ADS Bibcode: 2020ApJ...893..149P},
	pages = {149},
}

@article{isobe_statistical_1986,
	title = {Statistical {Methods} for {Astronomical} {Data} with {Upper} {Limits}. {II}. {Correlation} and {Regression}},
	volume = {306},
	issn = {0004-637X},
	url = {https://ui.adsabs.harvard.edu/abs/1986ApJ...306..490I},
	doi = {10.1086/164359},
	urldate = {2025-07-09},
	journal = {\apj},
	author = {Isobe, T. and Feigelson, E. D. and Nelson, P. I.},
	month = jul,
	year = {1986},
	note = {Publisher: IOP
ADS Bibcode: 1986ApJ...306..490I},
	pages = {490},
}

@ARTICLE{2020SciPy-NMeth,
  author  = {Virtanen, Pauli and Gommers, Ralf and Oliphant, Travis E. and
            Haberland, Matt and Reddy, Tyler and Cournapeau, David and
            Burovski, Evgeni and Peterson, Pearu and Weckesser, Warren and
            Bright, Jonathan and {van der Walt}, St{\'e}fan J. and
            Brett, Matthew and Wilson, Joshua and Millman, K. Jarrod and
            Mayorov, Nikolay and Nelson, Andrew R. J. and Jones, Eric and
            Kern, Robert and Larson, Eric and Carey, C J and
            Polat, {\.I}lhan and Feng, Yu and Moore, Eric W. and
            {VanderPlas}, Jake and Laxalde, Denis and Perktold, Josef and
            Cimrman, Robert and Henriksen, Ian and Quintero, E. A. and
            Harris, Charles R. and Archibald, Anne M. and
            Ribeiro, Ant{\^o}nio H. and Pedregosa, Fabian and
            {van Mulbregt}, Paul and {SciPy 1.0 Contributors}},
  title   = {{{SciPy} 1.0: Fundamental Algorithms for Scientific
            Computing in Python}},
  journal = {Nature Methods},
  year    = {2020},
  volume  = {17},
  pages   = {261--272},
  adsurl  = {https://rdcu.be/b08Wh},
  doi     = {10.1038/s41592-019-0686-2},
}

@Article{         harris2020array,
 title         = {Array programming with {NumPy}},
 author        = {Charles R. Harris and K. Jarrod Millman and St{\'{e}}fan J.
                 van der Walt and Ralf Gommers and Pauli Virtanen and David
                 Cournapeau and Eric Wieser and Julian Taylor and Sebastian
                 Berg and Nathaniel J. Smith and Robert Kern and Matti Picus
                 and Stephan Hoyer and Marten H. van Kerkwijk and Matthew
                 Brett and Allan Haldane and Jaime Fern{\'{a}}ndez del
                 R{\'{i}}o and Mark Wiebe and Pearu Peterson and Pierre
                 G{\'{e}}rard-Marchant and Kevin Sheppard and Tyler Reddy and
                 Warren Weckesser and Hameer Abbasi and Christoph Gohlke and
                 Travis E. Oliphant},
 year          = {2020},
 month         = sep,
 journal       = {Nature},
 volume        = {585},
 number        = {7825},
 pages         = {357--362},
 doi           = {10.1038/s41586-020-2649-2},
 publisher     = {Springer Science and Business Media {LLC}},
 url           = {https://doi.org/10.1038/s41586-020-2649-2}
}

@ARTICLE{serafinelli24,
       author = {{Serafinelli}, Roberto and {De Rosa}, Alessandra and {Tortosa}, Alessia and {Stella}, Luigi and {Vagnetti}, Fausto and {Bianchi}, Stefano and {Ricci}, Claudio and {Kammoun}, Elias and {Petrucci}, Pierre-Olivier and {Middei}, Riccardo and {Lanzuisi}, Giorgio and {Marinucci}, Andrea and {Ursini}, Francesco and {Matt}, Giorgio},
        title = "{Investigating the interplay between the coronal properties and the hard X-ray variability of active galactic nuclei with NuSTAR}",
      journal = {\aap},
         year = 2024,
        month = oct,
       volume = {690},
          eid = {A145},
        pages = {A145},
          doi = {10.1051/0004-6361/202450777},
archivePrefix = {arXiv},
       eprint = {2407.06769},
 primaryClass = {astro-ph.HE},
       adsurl = {https://ui.adsabs.harvard.edu/abs/2024A&A...690A.145S}
}

@ARTICLE{evans09,
       author = {{Evans}, P.~A. and {Beardmore}, A.~P. and {Page}, K.~L. and {Osborne}, J.~P. and {O'Brien}, P.~T. and {Willingale}, R. and {Starling}, R.~L.~C. and {Burrows}, D.~N. and {Godet}, O. and {Vetere}, L. and {Racusin}, J. and {Goad}, M.~R. and {Wiersema}, K. and {Angelini}, L. and {Capalbi}, M. and {Chincarini}, G. and {Gehrels}, N. and {Kennea}, J.~A. and {Margutti}, R. and {Morris}, D.~C. and {Mountford}, C.~J. and {Pagani}, C. and {Perri}, M. and {Romano}, P. and {Tanvir}, N.},
        title = "{Methods and results of an automatic analysis of a complete sample of Swift-XRT observations of GRBs}",
      journal = {\mnras},
         year = 2009,
        month = aug,
       volume = {397},
       number = {3},
        pages = {1177-1201},
          doi = {10.1111/j.1365-2966.2009.14913.x},
archivePrefix = {arXiv},
       eprint = {0812.3662},
 primaryClass = {astro-ph},
       adsurl = {https://ui.adsabs.harvard.edu/abs/2009MNRAS.397.1177E}
}

@ARTICLE{hi4pi16,
       author = {{HI4PI Collaboration} and {Ben Bekhti}, N. and {Fl{\"o}er}, L. and {Keller}, R. and {Kerp}, J. and {Lenz}, D. and {Winkel}, B. and {Bailin}, J. and {Calabretta}, M.~R. and {Dedes}, L. and {Ford}, H.~A. and {Gibson}, B.~K. and {Haud}, U. and {Janowiecki}, S. and {Kalberla}, P.~M.~W. and {Lockman}, F.~J. and {McClure-Griffiths}, N.~M. and {Murphy}, T. and {Nakanishi}, H. and {Pisano}, D.~J. and {Staveley-Smith}, L.},
        title = "{HI4PI: A full-sky H I survey based on EBHIS and GASS}",
      journal = {\aap},
         year = 2016,
        month = oct,
       volume = {594},
          eid = {A116},
        pages = {A116},
          doi = {10.1051/0004-6361/201629178},
archivePrefix = {arXiv},
       eprint = {1610.06175},
 primaryClass = {astro-ph.GA},
       adsurl = {https://ui.adsabs.harvard.edu/abs/2016A&A...594A.116H}
}

@ARTICLE{halpern84,
       author = {{Halpern}, J.~P.},
        title = "{Variable X-ray absorption in the QSO MR 2251-178.}",
      journal = {\apj},
         year = 1984,
        month = jun,
       volume = {281},
        pages = {90-94},
          doi = {10.1086/162077},
       adsurl = {https://ui.adsabs.harvard.edu/abs/1984ApJ...281...90H}
}

@ARTICLE{blustin05,
       author = {{Blustin}, A.~J. and {Page}, M.~J. and {Fuerst}, S.~V. and {Branduardi-Raymont}, G. and {Ashton}, C.~E.},
        title = "{The nature and origin of Seyfert warm absorbers}",
      journal = {\aap},
         year = 2005,
        month = feb,
       volume = {431},
        pages = {111-125},
          doi = {10.1051/0004-6361:20041775},
archivePrefix = {arXiv},
       eprint = {astro-ph/0411297},
 primaryClass = {astro-ph},
       adsurl = {https://ui.adsabs.harvard.edu/abs/2005A&A...431..111B}
}

@ARTICLE{laha14,
       author = {{Laha}, Sibasish and {Guainazzi}, Matteo and {Dewangan}, Gulab C. and {Chakravorty}, Susmita and {Kembhavi}, Ajit K.},
        title = "{Warm absorbers in X-rays (WAX), a comprehensive high-resolution grating spectral study of a sample of Seyfert galaxies - I. A global view and frequency of occurrence of warm absorbers.}",
      journal = {\mnras},
         year = 2014,
        month = jul,
       volume = {441},
       number = {3},
        pages = {2613-2643},
          doi = {10.1093/mnras/stu669},
archivePrefix = {arXiv},
       eprint = {1404.0899},
 primaryClass = {astro-ph.HE},
       adsurl = {https://ui.adsabs.harvard.edu/abs/2014MNRAS.441.2613L}
}

@ARTICLE{behar03,
       author = {{Behar}, Ehud and {Rasmussen}, Andrew P. and {Blustin}, Alexander J. and {Sako}, Masao and {Kahn}, Steven M. and {Kaastra}, Jelle S. and {Branduardi-Raymont}, Graziella and {Steenbrugge}, Katrien C.},
        title = "{A Long Look at NGC 3783 with the XMM-Newton Reflection Grating Spectrometer}",
      journal = {\apj},
         year = 2003,
        month = nov,
       volume = {598},
       number = {1},
        pages = {232-241},
          doi = {10.1086/378853},
archivePrefix = {arXiv},
       eprint = {astro-ph/0307467},
 primaryClass = {astro-ph},
       adsurl = {https://ui.adsabs.harvard.edu/abs/2003ApJ...598..232B}
}

@ARTICLE{kallman01,
       author = {{Kallman}, T. and {Bautista}, M.},
        title = "{Photoionization and High-Density Gas}",
      journal = {\apjs},
         year = 2001,
        month = mar,
       volume = {133},
       number = {1},
        pages = {221-253},
          doi = {10.1086/319184},
       adsurl = {https://ui.adsabs.harvard.edu/abs/2001ApJS..133..221K}
}

@ARTICLE{serafinelli19,
       author = {{Serafinelli}, Roberto and {Tombesi}, Francesco and {Vagnetti}, Fausto and {Piconcelli}, Enrico and {Gaspari}, Massimo and {Saturni}, Francesco G.},
        title = "{Multiphase quasar-driven outflows in PG 1114+445. I. Entrained ultra-fast outflows}",
      journal = {\aap},
         year = 2019,
        month = jul,
       volume = {627},
          eid = {A121},
        pages = {A121},
          doi = {10.1051/0004-6361/201935275},
archivePrefix = {arXiv},
       eprint = {1906.02765},
 primaryClass = {astro-ph.GA},
       adsurl = {https://ui.adsabs.harvard.edu/abs/2019A&A...627A.121S}
}

@ARTICLE{tombesi11,
       author = {{Tombesi}, F. and {Cappi}, M. and {Reeves}, J.~N. and {Palumbo}, G.~G.~C. and {Braito}, V. and {Dadina}, M.},
        title = "{Evidence for Ultra-fast Outflows in Radio-quiet Active Galactic Nuclei. II. Detailed Photoionization Modeling of Fe K-shell Absorption Lines}",
      journal = {\apj},
         year = 2011,
        month = nov,
       volume = {742},
       number = {1},
          eid = {44},
        pages = {44},
          doi = {10.1088/0004-637X/742/1/44},
archivePrefix = {arXiv},
       eprint = {1109.2882},
 primaryClass = {astro-ph.HE},
       adsurl = {https://ui.adsabs.harvard.edu/abs/2011ApJ...742...44T}
}

@ARTICLE{asplund09,
       author = {{Asplund}, Martin and {Grevesse}, Nicolas and {Sauval}, A. Jacques and {Scott}, Pat},
        title = "{The Chemical Composition of the Sun}",
      journal = {\araa},
         year = 2009,
        month = sep,
       volume = {47},
       number = {1},
        pages = {481-522},
          doi = {10.1146/annurev.astro.46.060407.145222},
archivePrefix = {arXiv},
       eprint = {0909.0948},
 primaryClass = {astro-ph.SR},
       adsurl = {https://ui.adsabs.harvard.edu/abs/2009ARA&A..47..481A}
}

@ARTICLE{garcia10,
       author = {{Garc{\'\i}a}, J. and {Kallman}, T.~R.},
        title = "{X-ray Reflected Spectra from Accretion Disk Models. I. Constant Density Atmospheres}",
      journal = {\apj},
         year = 2010,
        month = aug,
       volume = {718},
       number = {2},
        pages = {695-706},
          doi = {10.1088/0004-637X/718/2/695},
archivePrefix = {arXiv},
       eprint = {1006.0485},
 primaryClass = {astro-ph.HE},
       adsurl = {https://ui.adsabs.harvard.edu/abs/2010ApJ...718..695G}
}

@ARTICLE{matzeu23,
       author = {{Matzeu}, G.~A. and {Brusa}, M. and {Lanzuisi}, G. and {Dadina}, M. and {Bianchi}, S. and {Kriss}, G. and {Mehdipour}, M. and {Nardini}, E. and {Chartas}, G. and {Middei}, R. and {Piconcelli}, E. and {Gianolli}, V. and {Comastri}, A. and {Longinotti}, A.~L. and {Krongold}, Y. and {Ricci}, F. and {Petrucci}, P.~O. and {Tombesi}, F. and {Luminari}, A. and {Zappacosta}, L. and {Miniutti}, G. and {Gaspari}, M. and {Behar}, E. and {Bischetti}, M. and {Mathur}, S. and {Perna}, M. and {Giustini}, M. and {Grandi}, P. and {Torresi}, E. and {Vignali}, C. and {Bruni}, G. and {Cappi}, M. and {Costantini}, E. and {Cresci}, G. and {De Marco}, B. and {De Rosa}, A. and {Gilli}, R. and {Guainazzi}, M. and {Kaastra}, J. and {Kraemer}, S. and {La Franca}, F. and {Marconi}, A. and {Panessa}, F. and {Ponti}, G. and {Proga}, D. and {Ursini}, F. and {Baldini}, P. and {Fiore}, F. and {King}, A.~R. and {Maiolino}, R. and {Matt}, G. and {Merloni}, A.},
        title = "{Supermassive Black Hole Winds in X-rays: SUBWAYS. I. Ultra-fast outflows in quasars beyond the local Universe}",
      journal = {\aap},
         year = 2023,
        month = feb,
       volume = {670},
          eid = {A182},
        pages = {A182},
          doi = {10.1051/0004-6361/202245036},
archivePrefix = {arXiv},
       eprint = {2212.02960},
 primaryClass = {astro-ph.HE},
       adsurl = {https://ui.adsabs.harvard.edu/abs/2023A&A...670A.182M}
}

@ARTICLE{Liu2025NatAs,
       author = {{Liu}, Jun-Rong and {Wang}, Jian-Min and {Fermi-LAT Collaboration} and {Abdollahi}, S. and {Ajello}, M. and {Batista}, R. Alves and {Baldini}, L. and {Bartolini}, C. and {Bastieri}, D. and {Gonzalez}, J. Becerra and {Bellazzini}, R. and {Berenji}, B. and {Bissaldi}, E. and {Blandford}, R.~D. and {Bonino}, R. and {Bruel}, P. and {Buson}, S. and {Cameron}, R.~A. and {Caraveo}, P.~A. and {Cavazzuti}, E. and {Chiaro}, G. and {Cibrario}, N. and {Ciprini}, S. and {Cristarella Orestano}, P. and {Cutini}, S. and {D'Ammando}, F. and {Di Lalla}, N. and {Dinesh}, A. and {Di Venere}, L. and {Dom{\'\i}nguez}, A. and {Fegan}, S.~J. and {Fiori}, A. and {Franckowiak}, A. and {Fukazawa}, Y. and {Funk}, S. and {Fusco}, P. and {Gargano}, F. and {Gasbarra}, C. and {Gasparrini}, D. and {Germani}, S. and {Giglietto}, N. and {Giliberti}, M. and {Giordano}, F. and {Giroletti}, M. and {Green}, D. and {Grenier}, I.~A. and {Guiriec}, S. and {Hashizume}, M. and {Hays}, E. and {Hewitt}, J.~W. and {Horan}, D. and {Hou}, Xian and {Karwin}, C. and {Kayanoki}, T. and {Kuss}, M. and {Laviron}, A. and {Lemoine-Goumard}, M. and {Li}, Jian and {Liodakis}, I. and {Longo}, F. and {Loparco}, F. and {Lorusso}, L. and {Lubrano}, P. and {Maldera}, S. and {Marcotulli}, L. and {Mart{\'\i}-Devesa}, G. and {Mazziotta}, M.~N. and {Mereu}, I. and {Michelson}, P.~F. and {Mirabal}, N. and {Mitthumsiri}, W. and {Mizuno}, T. and {Monzani}, M.~E. and {Morishita}, T. and {Morselli}, A. and {Moskalenko}, I.~V. and {Negro}, M. and {Niwa}, R. and {Omodei}, N. and {Orienti}, M. and {Orlando}, E. and {Ormes}, J.~F. and {Paneque}, D. and {Panzarini}, G. and {Persic}, M. and {Pesce-Rollins}, M. and {Pillera}, R. and {Porter}, T.~A. and {Principe}, G. and {Rain{\`o}}, S. and {Rando}, R. and {Rani}, B. and {Razzano}, M. and {Reimer}, A. and {Reimer}, O. and {S{\'a}nchez-Conde}, M. and {Saz Parkinson}, P.~M. and {Serini}, D. and {Sgr{\`o}}, C. and {Siskind}, E.~J. and {Spandre}, G. and {Spinelli}, P. and {Suson}, D.~J. and {Tajima}, H. and {Thayer}, J.~B. and {Torres}, D.~F. and {Zhao}, Zi-Hao},
        title = "{Fermi detection of gamma-ray emission from the hot coronae of radio-quiet active galactic nuclei}",
      journal = {Nature Astronomy},
         year = 2025,
        month = jul,
       volume = {9},
        pages = {1086-1097},
          doi = {10.1038/s41550-025-02538-2},
archivePrefix = {arXiv},
       eprint = {2502.19189},
 primaryClass = {astro-ph.HE},
       adsurl = {https://ui.adsabs.harvard.edu/abs/2025NatAs...9.1086L}
}

@ARTICLE{Inoue2024,
       author = {{Inoue}, Yoshiyuki and {Takasao}, Shinsuke and {Khangulyan}, Dmitry},
        title = "{Upper limit on the coronal cosmic ray energy budget in Seyfert galaxies}",
      journal = {\pasj},
         year = 2024,
        month = oct,
       volume = {76},
       number = {5},
        pages = {996-1001},
          doi = {10.1093/pasj/psae065},
archivePrefix = {arXiv},
       eprint = {2401.07580},
 primaryClass = {astro-ph.HE},
       adsurl = {https://ui.adsabs.harvard.edu/abs/2024PASJ...76..996I}
}

@article{michiyama_alma_2024,
	title = {{ALMA} {Confirmation} of {Millimeter} {Time} {Variability} in the {Gamma}-{Ray} {Detected} {Seyfert} {Galaxy} {GRS} 1734-292},
	volume = {965},
	issn = {0004-637X},
	url = {https://ui.adsabs.harvard.edu/abs/2024ApJ...965...68M},
	doi = {10.3847/1538-4357/ad2fae},
	urldate = {2025-08-28},
	journal = {\apj},
	author = {Michiyama, Tomonari and Inoue, Yoshiyuki and Doi, Akihiro and Yamada, Tomoya and Fukazawa, Yasushi and Kubo, Hidetoshi and Barnier, Samuel},
	month = apr,
	year = {2024},
	note = {ADS Bibcode: 2024ApJ...965...68M},
	pages = {68},
}

@article{tananbaum_x-ray_1979,
	title = {X-ray studies of quasars with the {Einstein} {Observatory}.},
	volume = {234},
	issn = {0004-637X},
	url = {https://ui.adsabs.harvard.edu/abs/1979ApJ...234L...9T},
	doi = {10.1086/183100},
	urldate = {2025-09-01},
	journal = {\apj},
	author = {Tananbaum, H. and Avni, Y. and Branduardi, G. and Elvis, M. and Fabbiano, G. and Feigelson, E. and Giacconi, R. and Henry, J. P. and Pye, J. P. and Soltan, A. and Zamorani, G.},
	month = nov,
	year = {1979},
	note = {ADS Bibcode: 1979ApJ...234L...9T},
	pages = {L9--L13},
}

@article{avni_cosmological_1982,
	title = {On the cosmological evolution of the {X}-ray emission from quasars},
	volume = {262},
	issn = {0004-637X},
	url = {https://ui.adsabs.harvard.edu/abs/1982ApJ...262L..17A},
	doi = {10.1086/183903},
	urldate = {2025-09-01},
	journal = {\apj},
	author = {Avni, Y. and Tananbaum, H.},
	month = nov,
	year = {1982},
	note = {ADS Bibcode: 1982ApJ...262L..17A},
	pages = {L17--L21},
}

@article{avni_x-ray_1986,
	title = {X-{Ray} {Properties} of {Optically} {Selected} {QSOs}},
	volume = {305},
	issn = {0004-637X},
	url = {https://ui.adsabs.harvard.edu/abs/1986ApJ...305...83A},
	doi = {10.1086/164230},
	urldate = {2025-09-01},
	journal = {\apj},
	author = {Avni, Y. and Tananbaum, H.},
	month = jun,
	year = {1986},
	note = {ADS Bibcode: 1986ApJ...305...83A},
	pages = {83},
}

@article{wilkes_einstein_1994,
	title = {The {Einstein} {Database} of {IPC} {X}-{Ray} {Observations} of {Optically} {Selected} and {Radio}-selected {Quasars}. {I}.},
	volume = {92},
	issn = {0067-0049},
	url = {https://ui.adsabs.harvard.edu/abs/1994ApJS...92...53W},
	doi = {10.1086/191959},
	urldate = {2025-09-01},
	journal = {\apj\,Supplement Series},
	author = {Wilkes, Belinda J. and Tananbaum, Harvey and Worrall, D. M. and Avni, Yoram and Oey, M. S. and Flanagan, Joan},
	month = may,
	year = {1994},
	note = {ADS Bibcode: 1994ApJS...92...53W},
	pages = {53},
}

@ARTICLE{rybak_detection_2025,
       author = {{Rybak}, M. and {Sluse}, D. and {Gupta}, K.~K. and {Millon}, M. and {Behar}, E. and {Courbin}, F. and {McKean}, J.~P. and {Stacey}, H.~R.},
        title = "{Detection of millimetre-wave coronal emission in a quasar at cosmological distance using microlensing}",
      journal = {\aap},
         year = 2025,
        month = sep,
       volume = {701},
          eid = {A215},
        pages = {A215},
          doi = {10.1051/0004-6361/202554595},
archivePrefix = {arXiv},
       eprint = {2503.13313},
 primaryClass = {astro-ph.GA},
       adsurl = {https://ui.adsabs.harvard.edu/abs/2025A&A...701A.215R}
}

@article{peng_detailed_2002,
	title = {Detailed {Structural} {Decomposition} of {Galaxy} {Images}},
	volume = {124},
	issn = {0004-6256},
	url = {https://ui.adsabs.harvard.edu/abs/2002AJ....124..266P},
	doi = {10.1086/340952},
	urldate = {2025-09-02},
	journal = {The Astronomical Journal},
	author = {Peng, Chien Y. and Ho, Luis C. and Impey, Chris D. and Rix, Hans-Walter},
	month = jul,
	year = {2002},
	note = {ADS Bibcode: 2002AJ....124..266P},
	pages = {266--293},
}

@article{peng_detailed_2010,
	title = {Detailed {Decomposition} of {Galaxy} {Images}. {II}. {Beyond} {Axisymmetric} {Models}},
	volume = {139},
	issn = {0004-6256},
	url = {https://ui.adsabs.harvard.edu/abs/2010AJ....139.2097P},
	doi = {10.1088/0004-6256/139/6/2097},
	urldate = {2025-09-02},
	journal = {The Astronomical Journal},
	author = {Peng, Chien Y. and Ho, Luis C. and Impey, Chris D. and Rix, Hans-Walter},
	month = jun,
	year = {2010},
	note = {ADS Bibcode: 2010AJ....139.2097P},
	pages = {2097--2129},
}

@article{condon_nrao_1998,
	title = {The {NRAO} {VLA} {Sky} {Survey}},
	volume = {115},
	issn = {0004-6256},
	url = {https://ui.adsabs.harvard.edu/abs/1998AJ....115.1693C},
	doi = {10.1086/300337},
	urldate = {2025-09-04},
	journal = {The Astronomical Journal},
	author = {Condon, J. J. and Cotton, W. D. and Greisen, E. W. and Yin, Q. F. and Perley, R. A. and Taylor, G. B. and Broderick, J. J.},
	month = may,
	year = {1998},
	note = {ADS Bibcode: 1998AJ....115.1693C},
	pages = {1693--1716},
}

@article{mullaney_defining_2011,
	title = {Defining the intrinsic {AGN} infrared spectral energy distribution and measuring its contribution to the infrared output of composite galaxies},
	volume = {414},
	issn = {0035-8711},
	url = {https://ui.adsabs.harvard.edu/abs/2011MNRAS.414.1082M},
	doi = {10.1111/j.1365-2966.2011.18448.x},
	urldate = {2025-09-04},
	journal = {\mnras},
	author = {Mullaney, J. R. and Alexander, D. M. and Goulding, A. D. and Hickox, R. C.},
	month = jun,
	year = {2011},
	note = {ADS Bibcode: 2011MNRAS.414.1082M},
	pages = {1082--1110},
}

@article{den_brok_muse_2020,
	title = {The {MUSE} {Atlas} of {Discs} ({MAD}): {Ionized} gas kinematic maps and an application to diffuse ionized gas},
	volume = {491},
	issn = {0035-8711},
	shorttitle = {The {MUSE} {Atlas} of {Discs} ({MAD})},
	url = {https://ui.adsabs.harvard.edu/abs/2020MNRAS.491.4089D},
	doi = {10.1093/mnras/stz3184},
	urldate = {2025-09-05},
	journal = {\mnras},
	author = {den Brok, Mark and Carollo, C. Marcella and Erroz-Ferrer, Santiago and Fagioli, Martina and Brinchmann, Jarle and Emsellem, Eric and Krajnović, Davor and Marino, Raffaella A. and Onodera, Masato and Tacchella, Sandro and Weilbacher, Peter M. and Woo, Joanna},
	month = jan,
	year = {2020},
	note = {ADS Bibcode: 2020MNRAS.491.4089D},
	pages = {4089--4107},
}

@article{gravity_collaboration_central_2021,
	title = {The central parsec of {NGC} 3783: a rotating broad emission line region, asymmetric hot dust structure, and compact coronal line region},
	volume = {648},
	copyright = {https://creativecommons.org/licenses/by/4.0},
	issn = {0004-6361, 1432-0746},
	shorttitle = {The central parsec of {NGC} 3783},
	url = {https://www.aanda.org/10.1051/0004-6361/202040061},
	doi = {10.1051/0004-6361/202040061},
	urldate = {2025-09-05},
	journal = {A\&A},
	author = {{GRAVITY Collaboration} and Amorim, A. and Bauböck, M. and Brandner, W. and Bolzer, M. and Clénet, Y. and Davies, R. and De Zeeuw, P. T. and Dexter, J. and Drescher, A. and Eckart, A. and Eisenhauer, F. and Förster Schreiber, N. M. and Gao, F. and Garcia, P. J. V. and Genzel, R. and Gillessen, S. and Gratadour, D. and Hönig, S. and Kaltenbrunner, D. and Kishimoto, M. and Lacour, S. and Lutz, D. and Millour, F. and Netzer, H. and Ott, T. and Paumard, T. and Perraut, K. and Perrin, G. and Peterson, B. M. and Petrucci, P. O. and Pfuhl, O. and Prieto, M. A. and Rouan, D. and Sanchez-Bermudez, J. and Shangguan, J. and Shimizu, T. and Schartmann, M. and Stadler, J. and Sternberg, A. and Straub, O. and Straubmeier, C. and Sturm, E. and Tacconi, L. J. and Tristram, K. R. W. and Vermot, P. and Von Fellenberg, S. and Waisberg, I. and Widmann, F. and Woillez, J.},
	month = apr,
	year = {2021},
	pages = {A117},
}

@article{teng_fermilat_2011,
	title = {Fermi/{LAT} {Observations} of {Swift}/{BAT} {Seyfert} {Galaxies}: {On} the {Contribution} of {Radio}-quiet {Active} {Galactic} {Nuclei} to the {Extragalactic} γ-{Ray} {Background}},
	volume = {742},
	issn = {0004-637X},
	shorttitle = {Fermi/{LAT} {Observations} of {Swift}/{BAT} {Seyfert} {Galaxies}},
	url = {https://ui.adsabs.harvard.edu/abs/2011ApJ...742...66T},
	doi = {10.1088/0004-637X/742/2/66},
	urldate = {2025-09-16},
	journal = {\apj},
	author = {Teng, Stacy H. and Mushotzky, Richard F. and Sambruna, Rita M. and Davis, David S. and Reynolds, Christopher S.},
	month = dec,
	year = {2011},
	note = {ADS Bibcode: 2011ApJ...742...66T},
	pages = {66},
}

@article{baldi_pg-rqs_2022,
	title = {The {PG}-{RQS} survey. {Building} the radio spectral distribution of radio-quiet quasars. {I}. {The} 45-{GHz} data},
	volume = {510},
	issn = {0035-8711},
	url = {https://ui.adsabs.harvard.edu/abs/2022MNRAS.510.1043B},
	doi = {10.1093/mnras/stab3445},
	urldate = {2025-09-16},
	journal = {\mnras},
	author = {Baldi, R. D. and Laor, A. and Behar, E. and Horesh, A. and Panessa, F. and McHardy, I. and Kimball, A.},
	month = feb,
	year = {2022},
	note = {ADS Bibcode: 2022MNRAS.510.1043B},
	pages = {1043--1058},
}

@misc{Virtanen2020,
       author = {{Virtanen}, Pauli and {Gommers}, Ralf and {Burovski}, Evgeni and {Oliphant}, Travis E. and {Weckesser}, Warren and {Cournapeau}, David and {Alexbrc} and {Peterson}, Pearu and {Reddy}, Tyler and {Haberland}, Matt and {Wilson}, Josh and {Nelson}, Andrew and {Endolith} and {Mayorov}, Nikolay and {Van Der Walt}, Stefan and {Laxalde}, Denis and {Polat}, Ilhan and {Brett}, Matthew and {Larson}, Eric and {Millman}, Jarrod and {Lars} and {Van Mulbregt}, Paul and {Eric-Jones} and {Carey}, CJ and {Moore}, Eric and {Kern}, Robert and {Leslie}, Tim and {Perktold}, Josef and {Striega}, Kai and {Feng}, Yu},
        title = "{scipy/scipy: SciPy 1.6.0}",
         year = 2020,
        month = dec,
          eid = {10.5281/zenodo.4406806},
          doi = {10.5281/zenodo.4406806},
      version = {v1.6.0},
    publisher = {Zenodo},
       adsurl = {https://ui.adsabs.harvard.edu/abs/2020zndo...4406806V}
}

@ARTICLE{reeves08,
       author = {{Reeves}, James and {Done}, Chris and {Pounds}, Ken and {Terashima}, Yuichi and {Hayashida}, Kiyoshi and {Anabuki}, Naohisa and {Uchino}, Masahiro and {Turner}, Martin},
        title = "{On why the iron K-shell absorption in AGN is not a signature of the local warm/hot intergalactic medium}",
      journal = {\mnras},
         year = 2008,
        month = mar,
       volume = {385},
       number = {1},
        pages = {L108-L112},
          doi = {10.1111/j.1745-3933.2008.00443.x},
archivePrefix = {arXiv},
       eprint = {0801.1587},
 primaryClass = {astro-ph},
       adsurl = {https://ui.adsabs.harvard.edu/abs/2008MNRAS.385L.108R}
}

@ARTICLE{gupta_bass_2025,
       author = {{Gupta}, Kriti Kamal and {Ricci}, Claudio and {Tortosa}, Alessia and {Temple}, Matthew J. and {Koss}, Michael J. and {Trakhtenbrot}, Benny and {Bauer}, Franz E. and {Treister}, Ezequiel and {Mushotzky}, Richard and {Kammoun}, Elias and {Papadakis}, Iossif and {Oh}, Kyuseok and {Rojas}, Alejandra and {Chang}, Chin-Shin and {Diaz}, Yaherlyn and {Jana}, Arghajit and {Kakkad}, Darshan and {del Moral-Castro}, Ignacio and {Peca}, Alessandro and {Powell}, Meredith C. and {Stern}, Daniel and {Urry}, C. Megan and {Harrison}, Fiona},
        title = "{BASS. LIII. The Eddington Ratio as the Primary Regulator of the Fraction of X-Ray Emission in Active Galactic Nuclei}",
      journal = {\apj},
         year = 2025,
        month = sep,
       volume = {990},
       number = {1},
          eid = {86},
        pages = {86},
          doi = {10.3847/1538-4357/adf0f8},
archivePrefix = {arXiv},
       eprint = {2507.12541},
 primaryClass = {astro-ph.GA},
       adsurl = {https://ui.adsabs.harvard.edu/abs/2025ApJ...990...86G}
}

@ARTICLE{Tortosa2018,
       author = {{Tortosa}, A. and {Bianchi}, S. and {Marinucci}, A. and {Matt}, G. and {Petrucci}, P.~O.},
        title = "{A NuSTAR census of coronal parameters in Seyfert galaxies}",
      journal = {\aap},
         year = 2018,
        month = jun,
       volume = {614},
          eid = {A37},
        pages = {A37},
          doi = {10.1051/0004-6361/201732382},
archivePrefix = {arXiv},
       eprint = {1801.04456},
 primaryClass = {astro-ph.GA},
       adsurl = {https://ui.adsabs.harvard.edu/abs/2018A&A...614A..37T}
}

@article{piconcelli_x-ray_2011,
	title = {X-ray spectroscopy of the {Compton}-thick {Seyfert} 2 {ESO} 138 - {G1}},
	volume = {534},
	issn = {0004-6361},
	url = {https://ui.adsabs.harvard.edu/abs/2011A&A...534A.126P},
	doi = {10.1051/0004-6361/201117462},
	urldate = {2025-12-18},
	journal = {A\&A},
	author = {Piconcelli, E. and Bianchi, S. and Vignali, C. and Jiménez-Bailón, E. and Fiore, F.},
	month = oct,
	year = {2011},
	note = {Publisher: EDP
ADS Bibcode: 2011A\&A...534A.126P},
	pages = {A126},
}

@article{rodriguez-ardila_narrow-line_2024,
	title = {The narrow-line region properties of {ESO} 138-{G001} unveiled by {SOAR}/{SIFS} observations},
	volume = {527},
	issn = {0035-8711},
	url = {https://ui.adsabs.harvard.edu/abs/2024MNRAS.52710649R},
	doi = {10.1093/mnras/stad3872},
	urldate = {2025-12-18},
	journal = {\mnras},
	author = {Rodríguez-Ardila, A. and May, D. and Panda, S. and Fonseca-Faria, M. A. and Fraga, L.},
	month = feb,
	year = {2024},
	note = {Publisher: OUP
ADS Bibcode: 2024MNRAS.52710649R},
	pages = {10649--10667},
}

@article{cerqueira-campos_coronal-line_2021,
	title = {Coronal-line forest active galactic nuclei - {I}. {Physical} properties of the emission-line regions},
	volume = {500},
	issn = {0035-8711},
	url = {https://ui.adsabs.harvard.edu/abs/2021MNRAS.500.2666C},
	doi = {10.1093/mnras/staa3320},
	urldate = {2026-01-08},
	journal = {\mnras},
	author = {Cerqueira-Campos, F. C. and Rodríguez-Ardila, A. and Riffel, R. and Marinello, M. and Prieto, A. and Dahmer-Hahn, L. G.},
	month = jan,
	year = {2021},
	note = {Publisher: OUP
ADS Bibcode: 2021MNRAS.500.2666C},
	pages = {2666--2684},
}

@ARTICLE{Brinchmann2004,
       author = {{Brinchmann}, J. and {Charlot}, S. and {White}, S.~D.~M. and {Tremonti}, C. and {Kauffmann}, G. and {Heckman}, T. and {Brinkmann}, J.},
        title = "{The physical properties of star-forming galaxies in the low-redshift Universe}",
      journal = {\mnras},
         year = 2004,
        month = jul,
       volume = {351},
       number = {4},
        pages = {1151-1179},
          doi = {10.1111/j.1365-2966.2004.07881.x},
archivePrefix = {arXiv},
       eprint = {astro-ph/0311060},
 primaryClass = {astro-ph},
       adsurl = {https://ui.adsabs.harvard.edu/abs/2004MNRAS.351.1151B}
}

@ARTICLE{Noeske2007,
       author = {{Noeske}, K.~G. and {Faber}, S.~M. and {Weiner}, B.~J. and {Koo}, D.~C. and {Primack}, J.~R. and {Dekel}, A. and {Papovich}, C. and {Conselice}, C.~J. and {Le Floc'h}, E. and {Rieke}, G.~H. and {Coil}, A.~L. and {Lotz}, J.~M. and {Somerville}, R.~S. and {Bundy}, K.},
        title = "{Star Formation in AEGIS Field Galaxies since z=1.1: Staged Galaxy Formation and a Model of Mass-dependent Gas Exhaustion}",
      journal = {\apjl},
         year = 2007,
        month = may,
       volume = {660},
       number = {1},
        pages = {L47-L50},
          doi = {10.1086/517927},
archivePrefix = {arXiv},
       eprint = {astro-ph/0703056},
 primaryClass = {astro-ph},
       adsurl = {https://ui.adsabs.harvard.edu/abs/2007ApJ...660L..47N}
}

@ARTICLE{Hankla2026,
       author = {{Hankla}, Amelia M. and {Philippov}, Alexander and {Mbarek}, Rostom and {Mushotzky}, Richard F. and {Musoke}, G. and {Gro{\v{s}}elj}, Daniel and {Liska}, Matthew},
        title = "{An Outflow from the X-Ray Corona as the Origin of Millimeter Emission from Radio-quiet AGNs}",
      journal = {\apj},
         year = 2026,
        month = feb,
       volume = {997},
       number = {2},
          eid = {224},
        pages = {224},
          doi = {10.3847/1538-4357/ae2478},
archivePrefix = {arXiv},
       eprint = {2512.01662},
 primaryClass = {astro-ph.HE},
       adsurl = {https://ui.adsabs.harvard.edu/abs/2026ApJ...997..224H}
}

@ARTICLE{Fabian2017,
       author = {{Fabian}, A.~C. and {Lohfink}, A. and {Belmont}, R. and {Malzac}, J. and {Coppi}, P.},
        title = "{Properties of AGN coronae in the NuSTAR era - II. Hybrid plasma}",
      journal = {\mnras},
         year = 2017,
        month = may,
       volume = {467},
       number = {3},
        pages = {2566-2570},
          doi = {10.1093/mnras/stx221},
archivePrefix = {arXiv},
       eprint = {1701.06774},
 primaryClass = {astro-ph.HE},
       adsurl = {https://ui.adsabs.harvard.edu/abs/2017MNRAS.467.2566F}
}

@ARTICLE{Yamada2024,
       author = {{Yamada}, Tomoya and {Sakai}, Nobuyuki and {Inoue}, Yoshiyuki and {Michiyama}, Tomonari},
        title = "{Deciphering Radio Emissions from Accretion Disk Winds in Radio-quiet Active Galactic Nuclei}",
      journal = {\apj},
         year = 2024,
        month = jun,
       volume = {968},
       number = {2},
          eid = {116},
        pages = {116},
          doi = {10.3847/1538-4357/ad3a63},
archivePrefix = {arXiv},
       eprint = {2404.04632},
 primaryClass = {astro-ph.HE},
       adsurl = {https://ui.adsabs.harvard.edu/abs/2024ApJ...968..116Y}
}

@ARTICLE{Paul2026,
       author = {{Paul}, Jeremiah D. and {Plotkin}, Richard M.},
        title = "{Ruling Out Compact Jets as the Dominant Source of Radio Emission in Radio-quiet, High-Eddington-ratio Active Galactic Nuclei}",
      journal = {\apj},
         year = 2026,
        month = feb,
       volume = {998},
       number = {2},
          eid = {296},
        pages = {296},
          doi = {10.3847/1538-4357/ae36a6},
archivePrefix = {arXiv},
       eprint = {2601.06480},
 primaryClass = {astro-ph.HE},
       adsurl = {https://ui.adsabs.harvard.edu/abs/2026ApJ...998..296P}
}

@ARTICLE{NhatLy2026,
       author = {{Nhat Ly}, Minh and {Inoue}, Yoshiyuki and {Sentoku}, Yasuhiko and {Sano}, Takayoshi},
        title = "{Proton Acceleration by Collisionless Shocks in Supermassive Black Hole Coronae: Implications for High-Energy Neutrinos}",
      journal = {arXiv e-prints},
         year = 2026,
        month = jan,
          eid = {arXiv:2601.01999},
        pages = {arXiv:2601.01999},
          doi = {10.48550/arXiv.2601.01999},
archivePrefix = {arXiv},
       eprint = {2601.01999},
 primaryClass = {astro-ph.HE},
       adsurl = {https://ui.adsabs.harvard.edu/abs/2026arXiv260101999N}
}

@ARTICLE{Haardt1991,
       author = {{Haardt}, F. and {Maraschi}, L.},
        title = "{A Two-Phase Model for the X-Ray Emission from Seyfert Galaxies}",
      journal = {\apjl},
         year = 1991,
        month = oct,
       volume = {380},
        pages = {L51},
          doi = {10.1086/186171},
       adsurl = {https://ui.adsabs.harvard.edu/abs/1991ApJ...380L..51H}
}

@ARTICLE{Markoff2005,
       author = {{Markoff}, Sera and {Nowak}, Michael A. and {Wilms}, J{\"o}rn},
        title = "{Going with the Flow: Can the Base of Jets Subsume the Role of Compact Accretion Disk Coronae?}",
      journal = {\apj},
         year = 2005,
        month = dec,
       volume = {635},
       number = {2},
        pages = {1203-1216},
          doi = {10.1086/497628},
archivePrefix = {arXiv},
       eprint = {astro-ph/0509028},
 primaryClass = {astro-ph},
       adsurl = {https://ui.adsabs.harvard.edu/abs/2005ApJ...635.1203M}
}

@ARTICLE{Field93,
       author = {{Field}, G.~B. and {Rogers}, R.~D.},
        title = "{Radiation from Magnetized Accretion Disks in Active Galactic Nuclei}",
      journal = {\apj},
         year = 1993,
        month = jan,
       volume = {403},
        pages = {94},
          doi = {10.1086/172185},
       adsurl = {https://ui.adsabs.harvard.edu/abs/1993ApJ...403...94F}
}

@ARTICLE{Hickox18,
       author = {{Hickox}, Ryan C. and {Alexander}, David M.},
        title = "{Obscured Active Galactic Nuclei}",
      journal = {\araa},
         year = 2018,
        month = sep,
       volume = {56},
        pages = {625-671},
          doi = {10.1146/annurev-astro-081817-051803},
archivePrefix = {arXiv},
       eprint = {1806.04680},
 primaryClass = {astro-ph.GA},
       adsurl = {https://ui.adsabs.harvard.edu/abs/2018ARA&A..56..625H}
}

@ARTICLE{scholz_anderson_1987,
       author = {{Scholz}, F.~W. and {Stephens}, M.~A.},
        title = "{K-Sample Anderson-Darling Tests}",
      journal = {Journal of the American Statistical Association},
         year = 1987,
        month = sep,
       volume = {82},
       number = {399},
        pages = {918-924},
          doi = {10.2307/2288805},
       adsurl = {https://www.jstor.org/stable/2288805}
}

@ARTICLE{Chen2025,
       author = {{Chen}, Sina and {Laor}, Ari and {Behar}, Ehud and {Baldi}, Ranieri D. and {Gelfand}, Joseph D. and {Kimball}, Amy E.},
        title = "{A Dichotomy in the 1─24 GHz Parsec-scale Radio Spectra of Radio-quiet Quasars}",
      journal = {\apj},
         year = 2025,
        month = feb,
       volume = {979},
       number = {2},
          eid = {241},
        pages = {241},
          doi = {10.3847/1538-4357/ada142},
archivePrefix = {arXiv},
       eprint = {2410.07889},
 primaryClass = {astro-ph.GA},
       adsurl = {https://ui.adsabs.harvard.edu/abs/2025ApJ...979..241C}
}

@ARTICLE{Droguett-Callejas26,
       author = {{Droguett-Callejas}, Macarena and {Treister}, Ezequiel and {Barcos-Mu{\~n}oz}, Loreto and {Johnstone}, Makoto and {Bauer}, Franz E. and {Kawamuro}, Taiki and {Torres-Alb{\`a}}, N{\'u}ria and {Ricci}, Claudio and {Koss}, Michael and {Song}, Yiqing and {Peca}, Alessandro and {Evans}, Aaron and {Gonz{\'a}lez}, Jorge},
        title = "{Probing Heavily Obscured Active Galactic Nuclei in Major Galaxy Mergers Using the Millimeter─X-Ray Correlation}",
      journal = {\apjl},
         year = 2026,
        month = mar,
       volume = {999},
       number = {1},
          eid = {L11},
        pages = {L11},
          doi = {10.3847/2041-8213/ae4008},
archivePrefix = {arXiv},
       eprint = {2601.15186},
 primaryClass = {astro-ph.GA},
       adsurl = {https://ui.adsabs.harvard.edu/abs/2026ApJ...999L..11D}
}

@ARTICLE{Koss23,
       author = {{Koss}, Michael J. and {Treister}, Ezequiel and {Kakkad}, Darshan and {Casey-Clyde}, J. Andrew and {Kawamuro}, Taiki and {Williams}, Jonathan and {Foord}, Adi and {Trakhtenbrot}, Benny and {Bauer}, Franz E. and {Privon}, George C. and {Ricci}, Claudio and {Mushotzky}, Richard and {Barcos-Munoz}, Loreto and {Blecha}, Laura and {Connor}, Thomas and {Harrison}, Fiona and {Liu}, Tingting and {Magno}, Macon and {Mingarelli}, Chiara M.~F. and {Muller-Sanchez}, Francisco and {Oh}, Kyuseok and {Shimizu}, T. Taro and {Smith}, Krista Lynne and {Stern}, Daniel and {Tello}, Miguel Parra and {Urry}, C. Megan},
        title = "{UGC 4211: A Confirmed Dual Active Galactic Nucleus in the Local Universe at 230 pc Nuclear Separation}",
      journal = {\apjl},
         year = 2023,
        month = jan,
       volume = {942},
       number = {1},
          eid = {L24},
        pages = {L24},
          doi = {10.3847/2041-8213/aca8f0},
archivePrefix = {arXiv},
       eprint = {2301.03609},
 primaryClass = {astro-ph.GA},
       adsurl = {https://ui.adsabs.harvard.edu/abs/2023ApJ...942L..24K}
}

@ARTICLE{Serafinelli25,
       author = {{Serafinelli}, Roberto and {Nicastro}, Fabrizio and {Luminari}, Alfredo and {Krongold}, Yair and {Camilloni}, Francesco and {Kammoun}, Elias and {Middei}, Riccardo and {Piconcelli}, Enrico and {Piro}, Luigi},
        title = "{Time-evolving Diagnostic of the Ionized Absorbers in NGC 4051. I. High-resolution Time-averaged Spectroscopy}",
      journal = {\apj},
         year = 2025,
        month = dec,
       volume = {995},
       number = {1},
          eid = {6},
        pages = {6},
          doi = {10.3847/1538-4357/ae1614},
archivePrefix = {arXiv},
       eprint = {2510.18069},
 primaryClass = {astro-ph.GA},
       adsurl = {https://ui.adsabs.harvard.edu/abs/2025ApJ...995....6S}
}

@ARTICLE{Shablovinskaia25,
       author = {{Shablovinskaya}, E. and {Ricci}, C. and {Paladino}, R. and {Laor}, A. and {Chang}, C.-S. and {Belfiori}, D. and {Kawamuro}, T. and {Lopez-Rodriguez}, E. and {Rosario}, D.~J. and {Aalto}, S. and {Koss}, M. and {Mushotzky}, R. and {Privon}, G.~C.},
        title = "{ALMA 3 mm polarimetry of radio-quiet active galactic nuclei}",
      journal = {\aap},
         year = 2025,
        month = nov,
       volume = {703},
          eid = {A82},
        pages = {A82},
          doi = {10.1051/0004-6361/202555796},
archivePrefix = {arXiv},
       eprint = {2506.05973},
 primaryClass = {astro-ph.HE},
       adsurl = {https://ui.adsabs.harvard.edu/abs/2025A&A...703A..82S}
}
\bibliographystyle{aasjournal}

\appendix
\renewcommand{\thefigure}{\arabic{figure}} 
\setcounter{figure}{4}

\vspace{-1\baselineskip}
\section{ALMA data}\label{AppendixALMA}

\subsection{Continuum observations}\label{AppendixALMAcont}

\begin{figure}[th!]
    \centering
    \begin{minipage}[t]{0.33\textwidth}
        \centering
        \includegraphics[width=\linewidth]{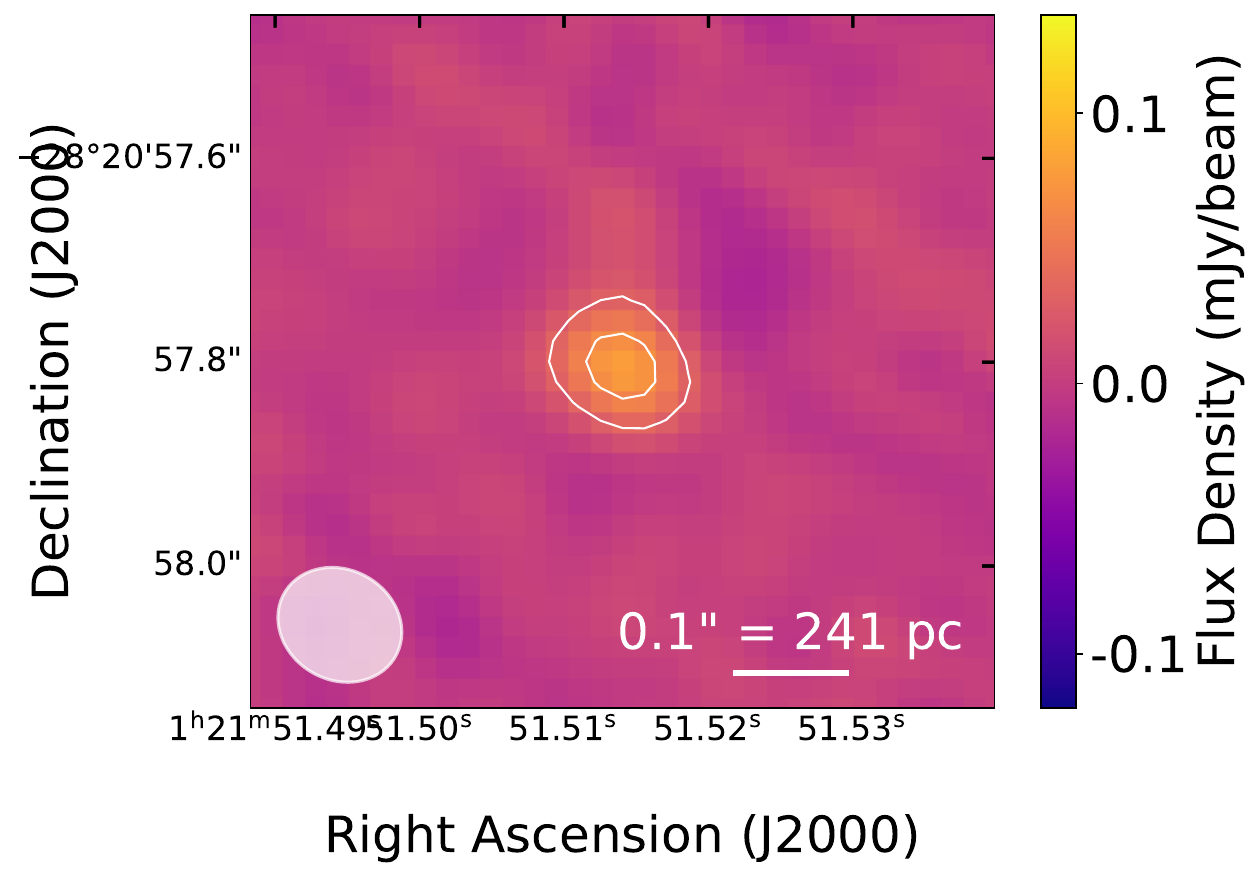}
        \subcaption{\parbox[t][1.5cm][t]{\linewidth}{\textbf{Q\,0119--286}\newline Contours at 5$\sigma$ and 10$\sigma$.}}
    \end{minipage}\hfill
    \begin{minipage}[t]{0.33\textwidth}
        \centering
        \includegraphics[width=\linewidth]{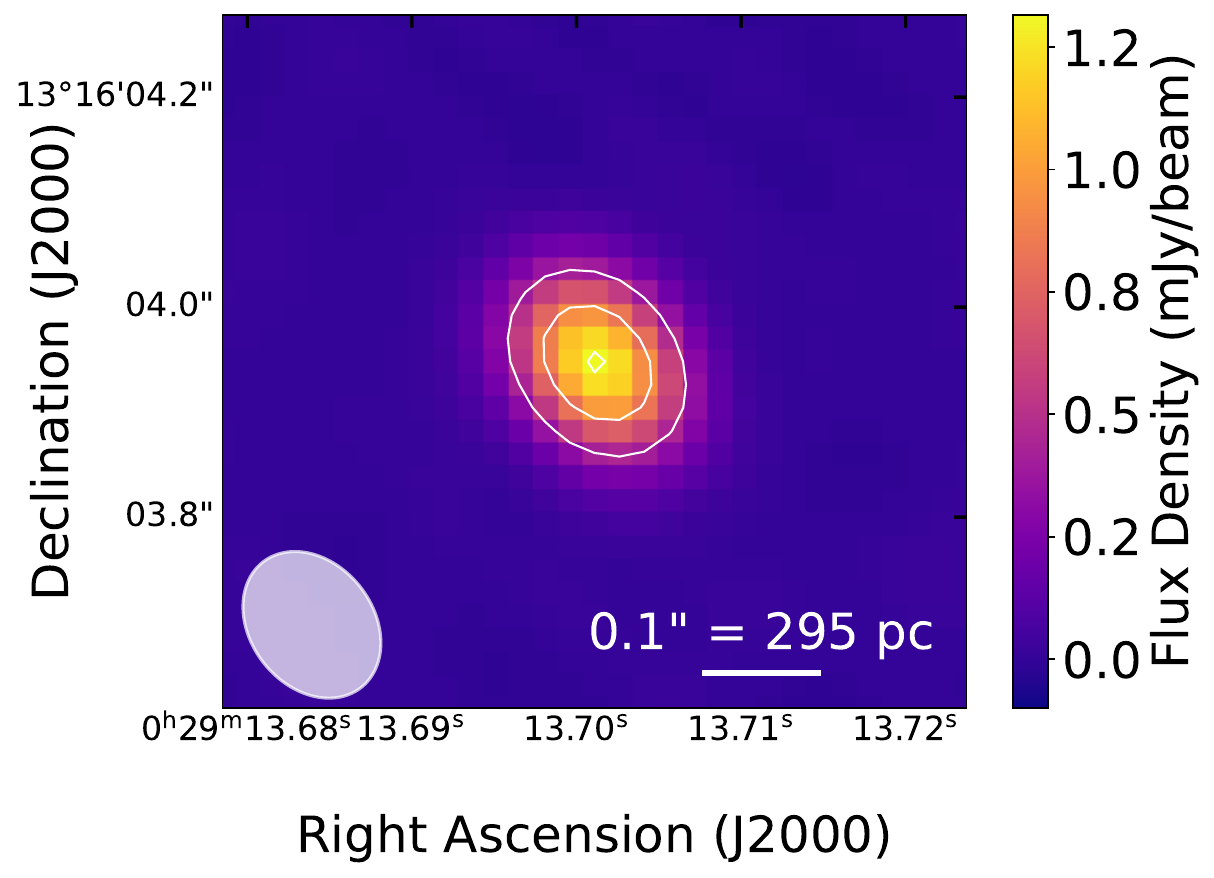}
        \subcaption{\parbox[t][1.5cm][t]{\linewidth}{\textbf{PG\,0026+129}\newline Contours at 50$\sigma$, 100$\sigma$, 150$\sigma$.}}
    \end{minipage}\hfill
    \begin{minipage}[t]{0.33\textwidth}
        \centering
        \includegraphics[width=\linewidth]{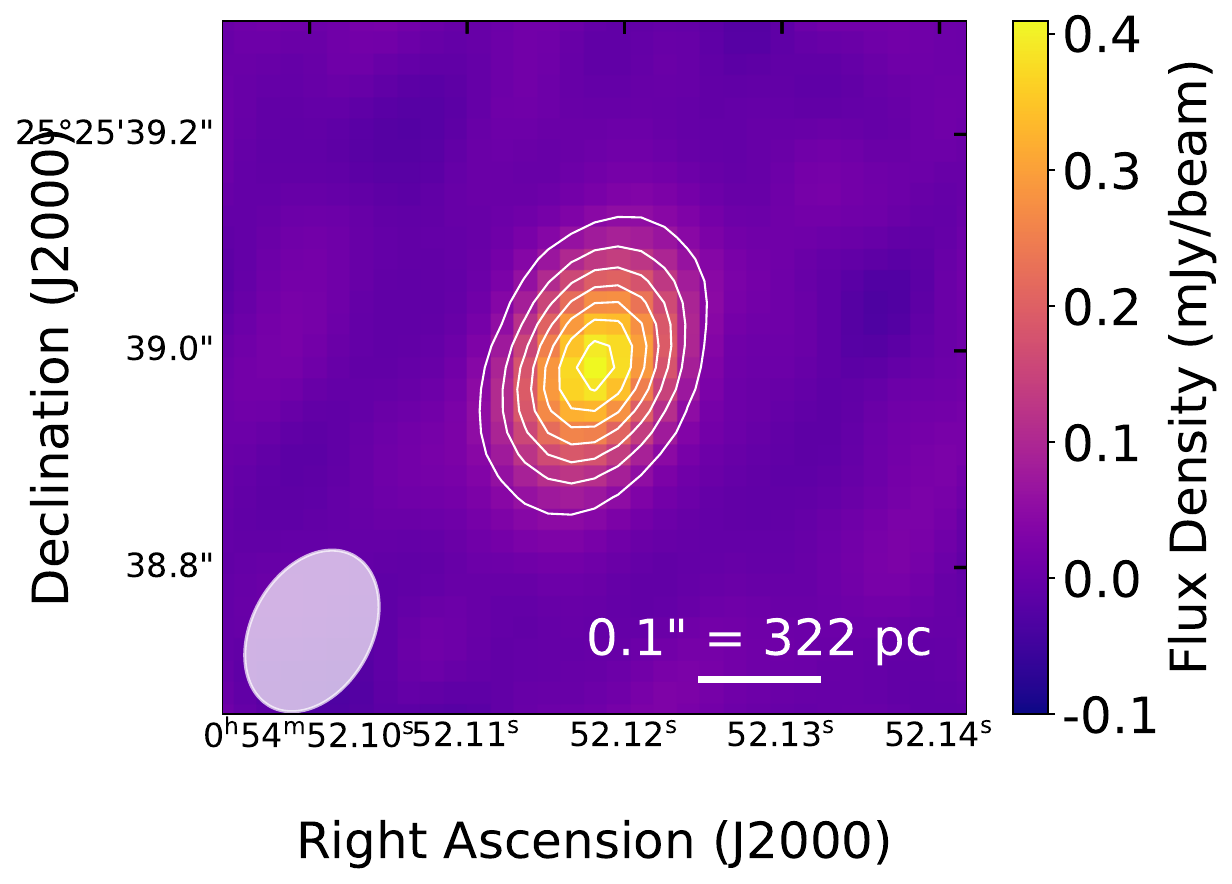}
        \subcaption{\parbox[t][1.5cm][t]{\linewidth}{\textbf{PG\,0052+251}\newline Contours from 5$\sigma$ to 35$\sigma$.}}
    \end{minipage} 

    \vspace{-8mm} 

    \begin{minipage}[t]{0.33\textwidth}
        \centering
        \includegraphics[width=\linewidth]{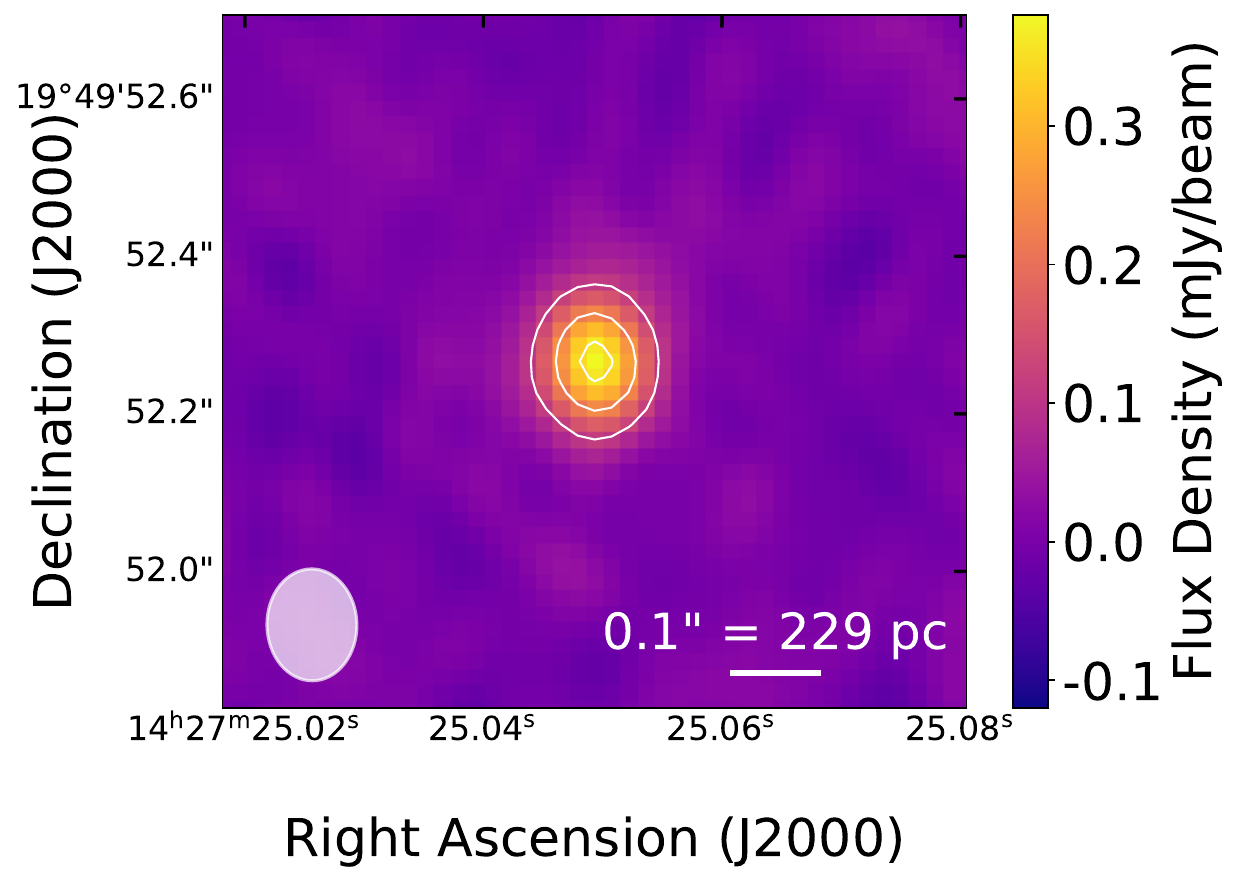}
        \subcaption{\parbox[t][1.5cm][t]{\linewidth}{\textbf{Mrk\,813}\newline Contours at 5$\sigma$, 10$\sigma$, and 15$\sigma$.}}
    \end{minipage}\hfill
    \begin{minipage}[t]{0.33\textwidth}
        \centering
        \includegraphics[width=\linewidth]{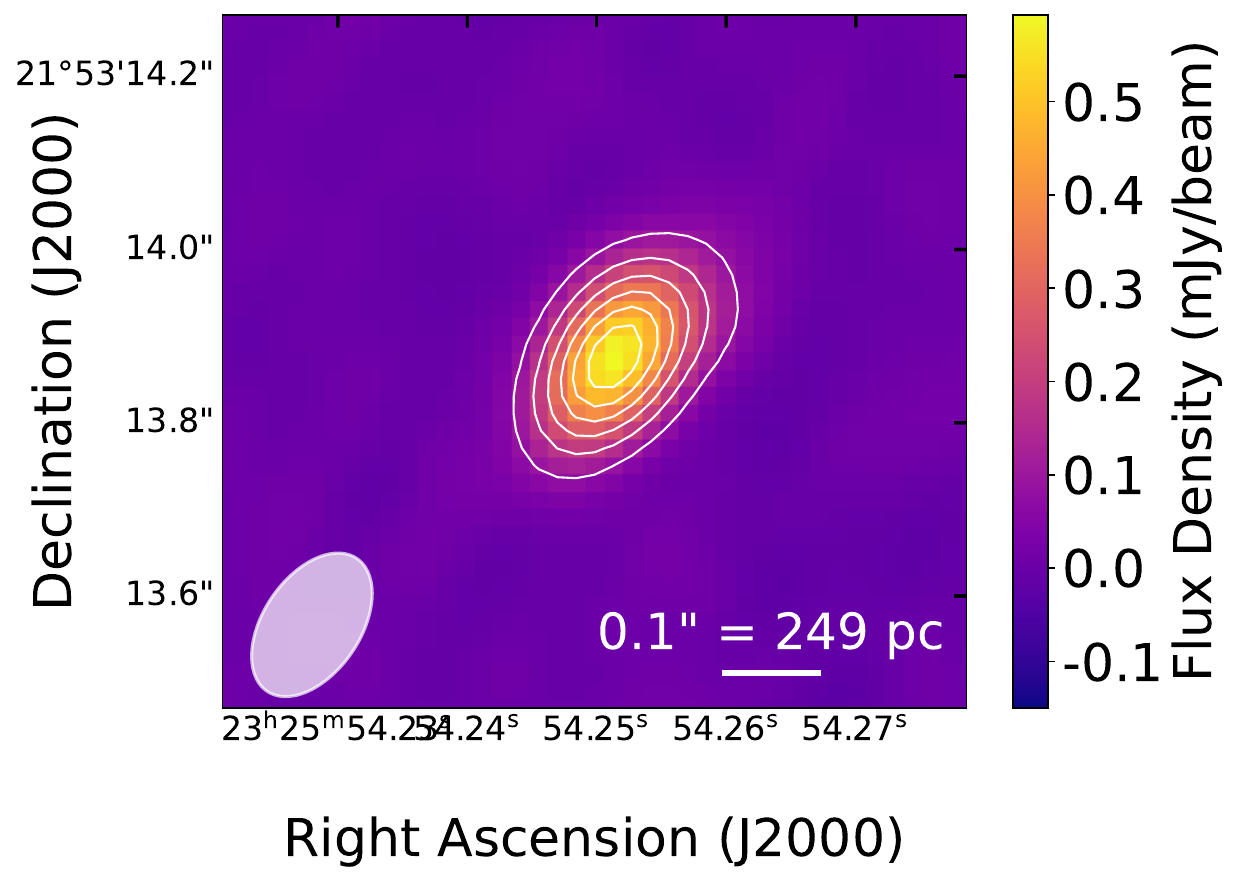}
        \subcaption{\parbox[t][1.5cm][t]{\linewidth}{\textbf{RHS\,61}\newline Contours from 10$\sigma$ to 70$\sigma$.}}
    \end{minipage}\hfill
    \begin{minipage}[t]{0.33\textwidth}
        \centering
        \includegraphics[width=\linewidth]{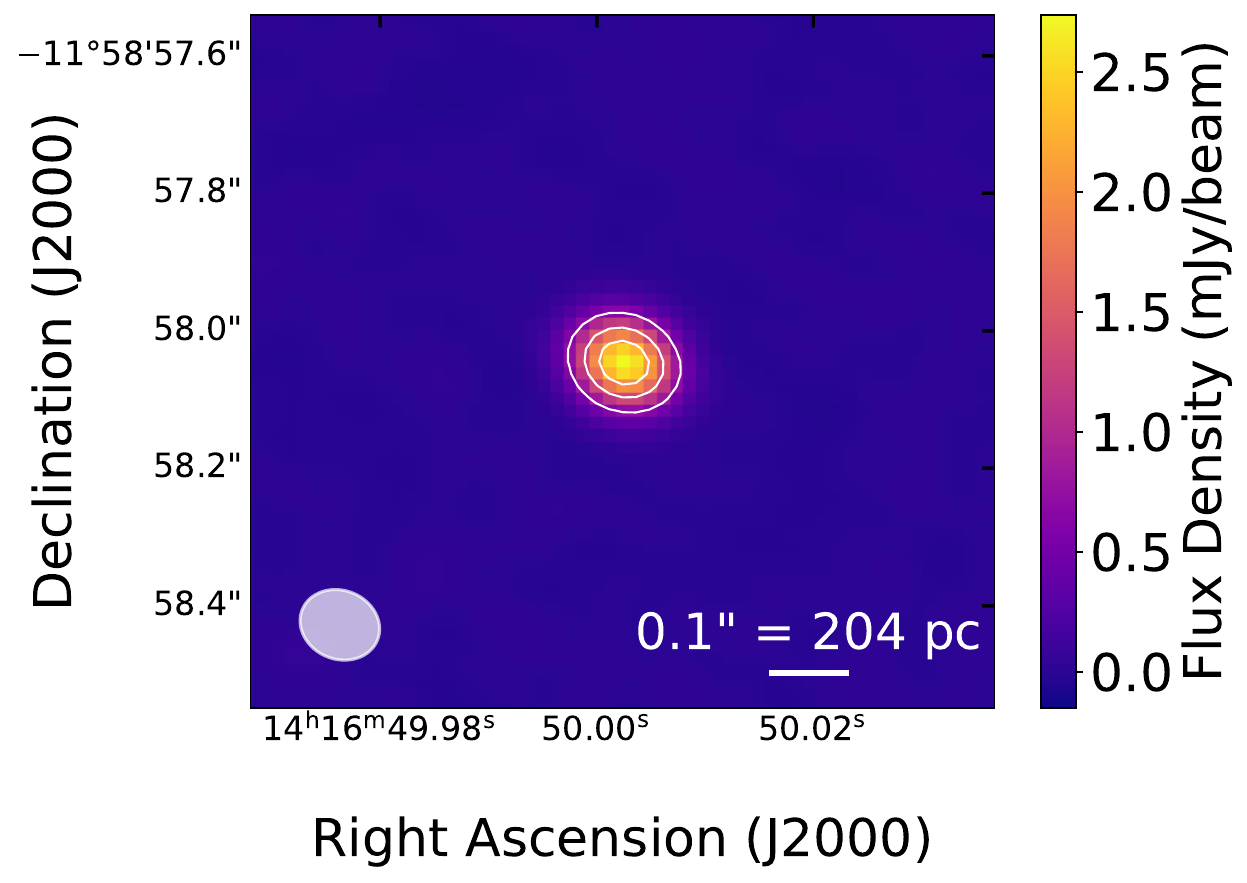}
        \subcaption{\parbox[t][1.5cm][t]{\linewidth}{\textbf{LEDA\,12622}\newline Contours at 50$\sigma$, 100$\sigma$, and 150$\sigma$.}}
    \end{minipage} 

    \vspace{-8mm} 

    \begin{minipage}[t]{0.33\textwidth}
        \centering
        \includegraphics[width=\linewidth]{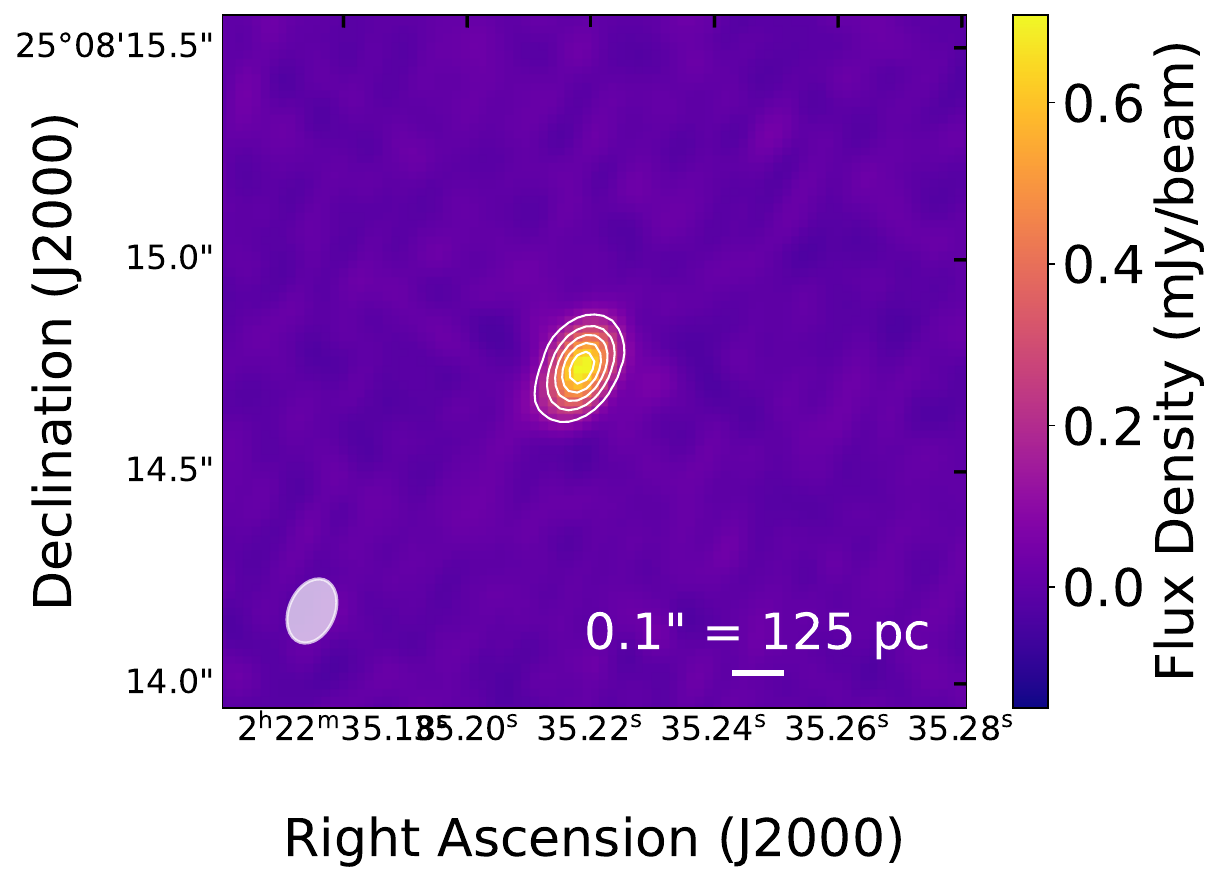}
        \subcaption{\parbox[t][1.5cm][t]{\linewidth}{\textbf{2MASX\,J02223523+2508143}\newline Contours from 10$\sigma$ to 50$\sigma$.}}
    \end{minipage}\hfill
    \begin{minipage}[t]{0.33\textwidth}
        \centering
        \includegraphics[width=\linewidth]{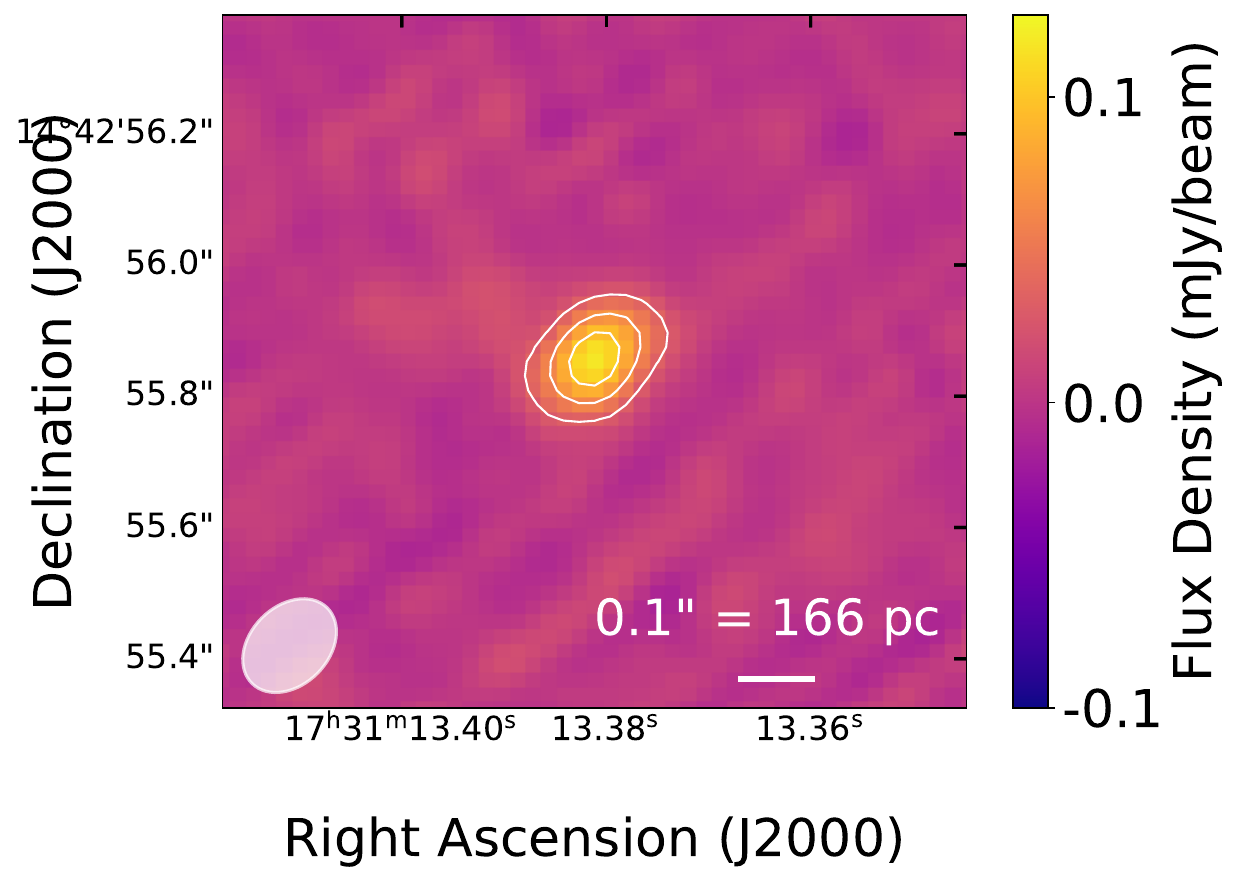}
        \subcaption{\parbox[t][1.5cm][t]{\linewidth}{\textbf{2MASX\,J17311341+1442561}\newline Contours at 5$\sigma$, 10$\sigma$, and 15$\sigma$.}}
    \end{minipage}\hfill
    \begin{minipage}[t]{0.33\textwidth}
        \centering
        \includegraphics[width=\linewidth]{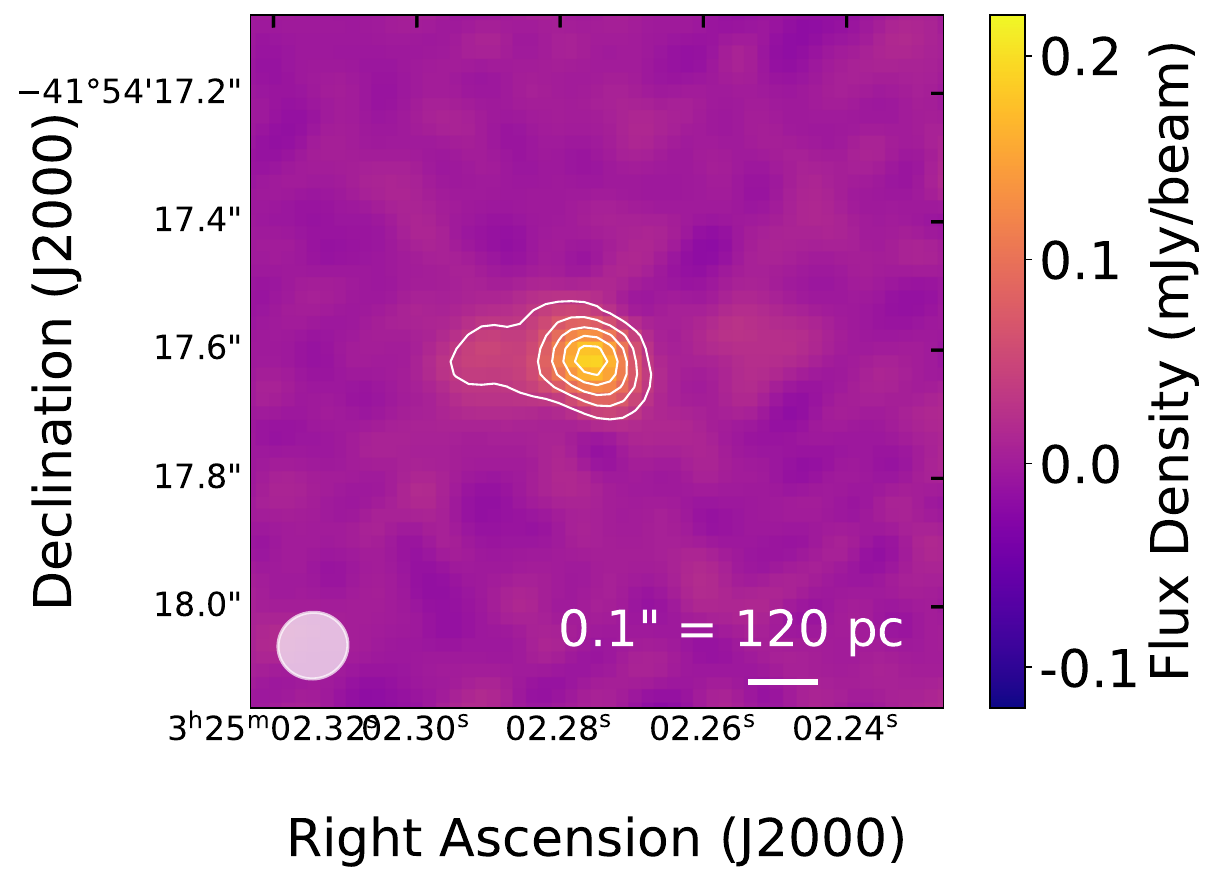}
        \subcaption{\parbox[t][1.5cm][t]{\linewidth}{\textbf{LEDA\,12773}\newline Contours from 5$\sigma$ to 25$\sigma$. The extra structure to the left peaks at 7$\sigma$.}}
        \label{ALMAimages-LEDA127}
    \end{minipage} 

\caption{ALMA observations of our nine high-luminosity AGN. The beam sizes are displayed in the bottom-left corner. Furthermore, the bar in the bottom-right corner represents a scale of 0.1\arcsec\ and the corresponding size in pc. Contours are displayed at multiples of $\sigma$ as specified for each source. The respective values of $\sigma$ and the beam size ($\theta$) are listed in Table\,\ref{tab:ObsALMA}.}
\label{fig:ALMAimages}
\end{figure}

We have obtained new ALMA observations of our sample of nine RQ AGN. The observations and imaging of the observations are discussed in Section\,\ref{almadata} and Section\,\ref{Datamm}, respectively.
The ALMA images obtained by our observations are displayed in Figure\,\ref{fig:ALMAimages}. Each image shows the beam in the bottom-left corner and the physical size of 0\farcs1 in pc in the bottom-right corner.

\subsection{Spectral index determination}\label{ApSpectralindex}

\begin{figure}[th!]
    \centering
    \begin{minipage}[b]{0.31\textwidth}
        \centering
        \includegraphics[width=1.1\textwidth]{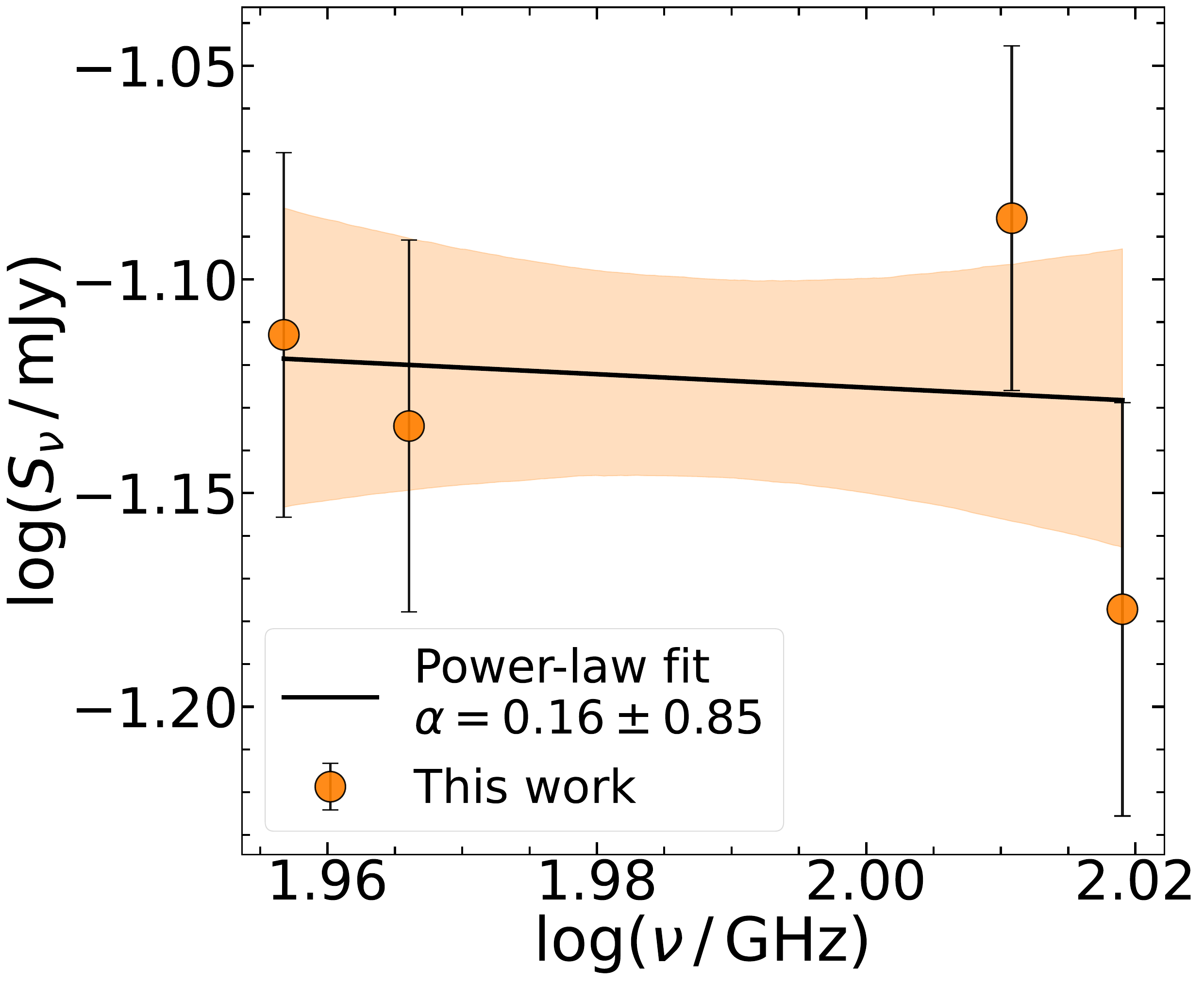}
        \subcaption{\textbf{Q\,0119--286}}
    \end{minipage}\hfill
    \begin{minipage}[b]{0.31\textwidth}
        \centering
        \includegraphics[width=1.1\textwidth]{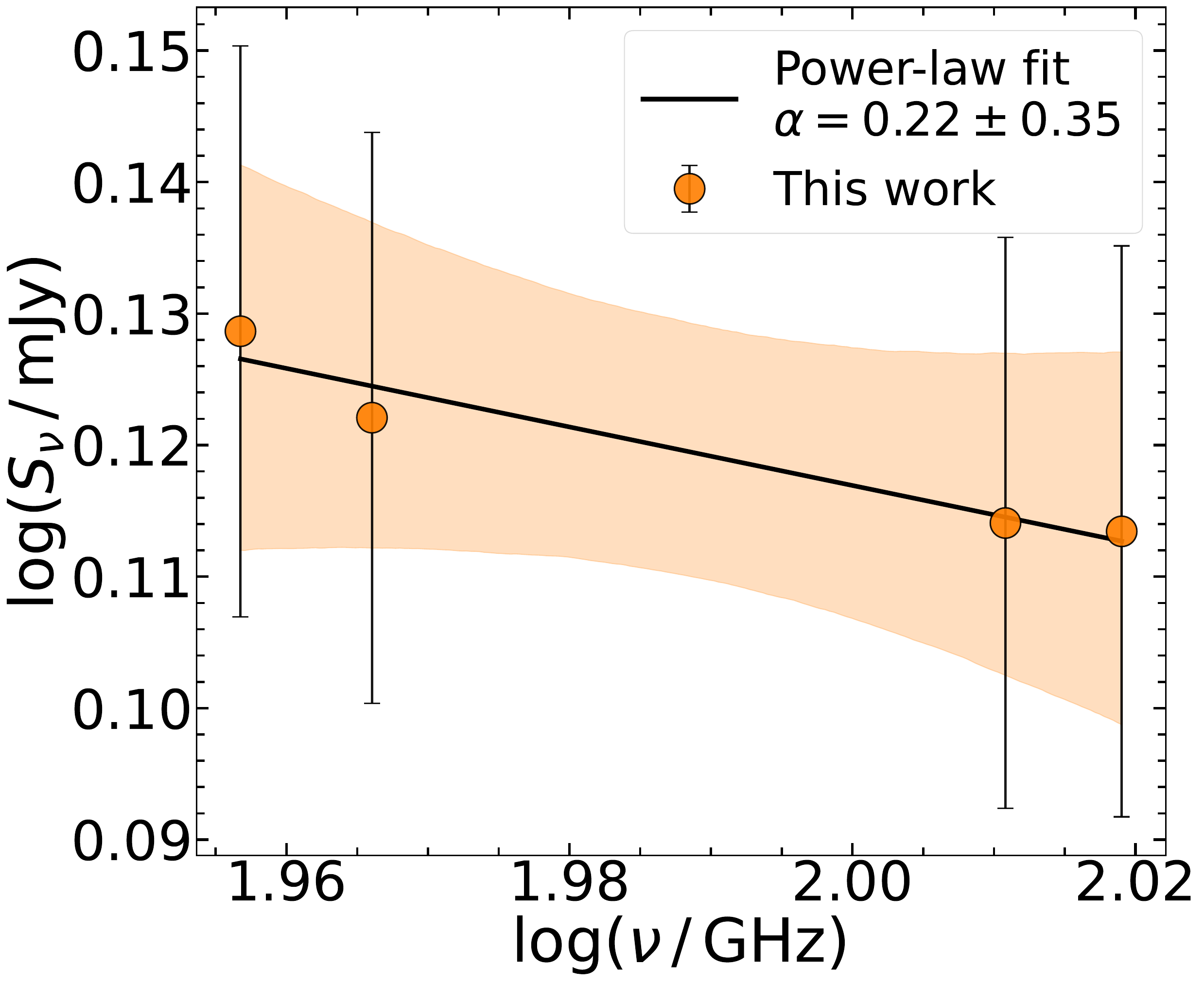}
        \subcaption{\textbf{PG\,0026+129}}
    \end{minipage}\hfill
    \begin{minipage}[b]{0.31\textwidth}
        \centering
        \includegraphics[width=1.1\textwidth]{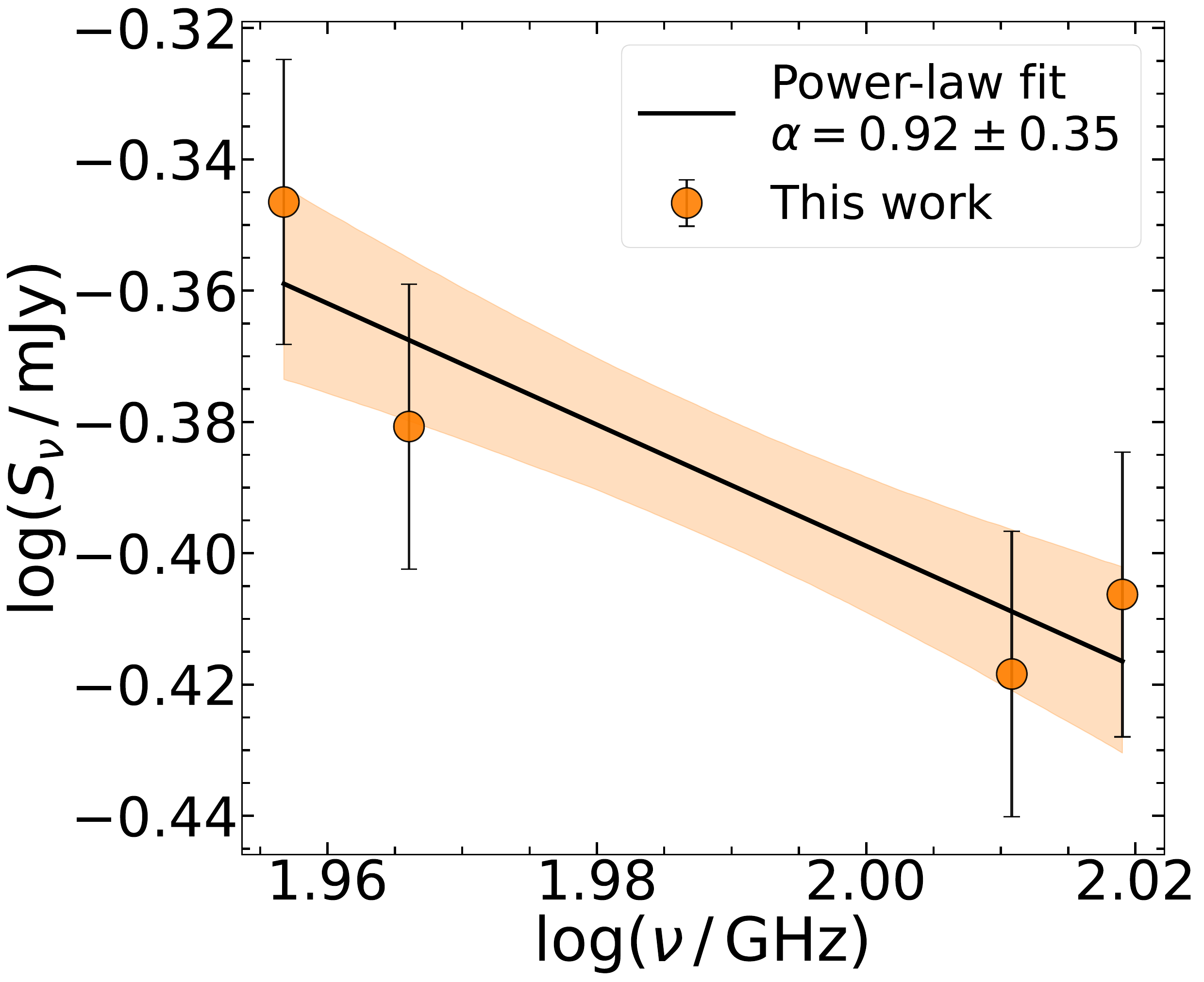}
        \subcaption{\textbf{PG\,0052+251}}
    \end{minipage}  
    \begin{minipage}[b]{0.31\textwidth}
        \centering
        \includegraphics[width=1.1\textwidth]{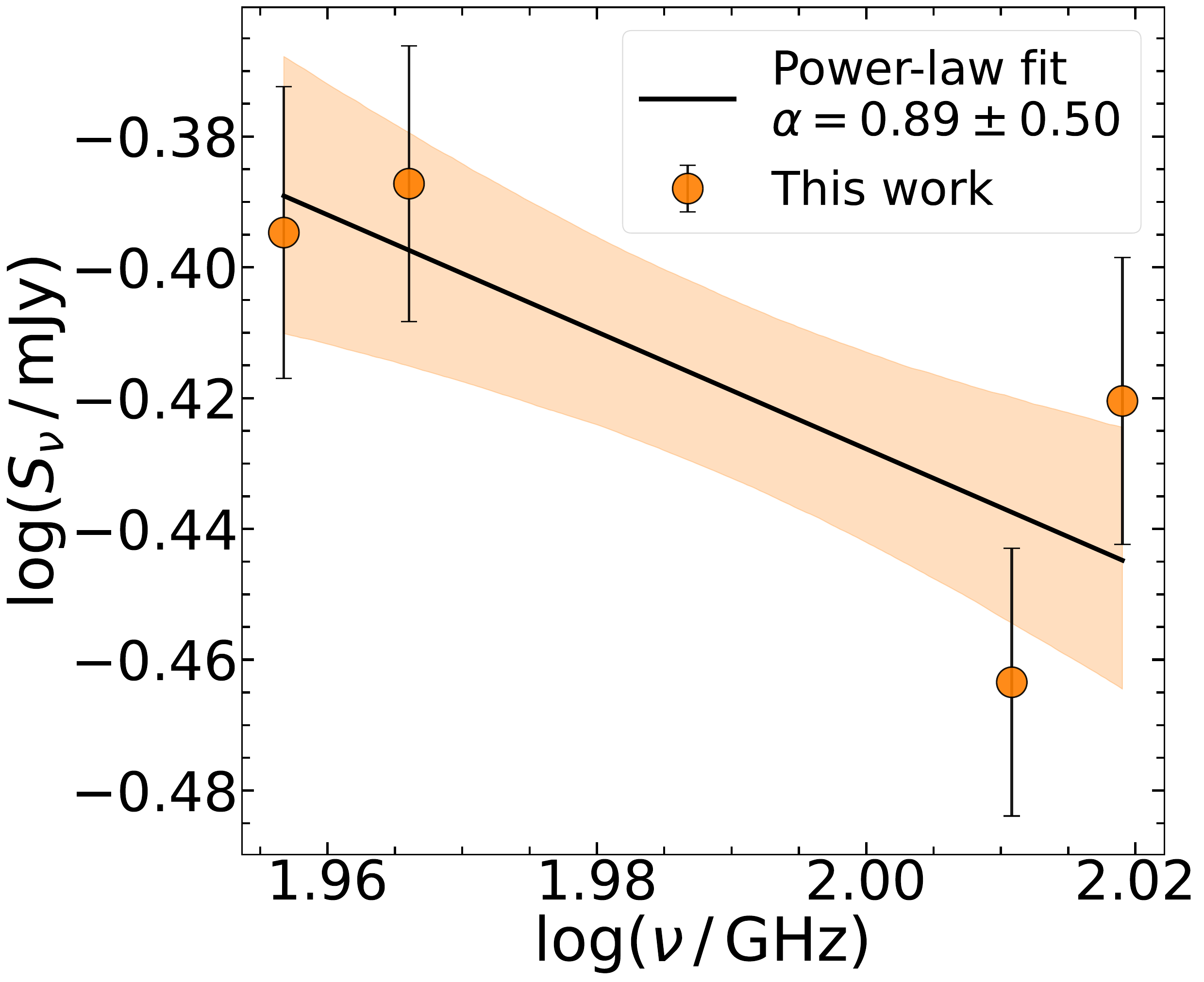}
        \subcaption{\textbf{Mrk\,813}}
    \end{minipage}\hfill
    \begin{minipage}[b]{0.31\textwidth}
        \centering
        \includegraphics[width=1.1\textwidth]{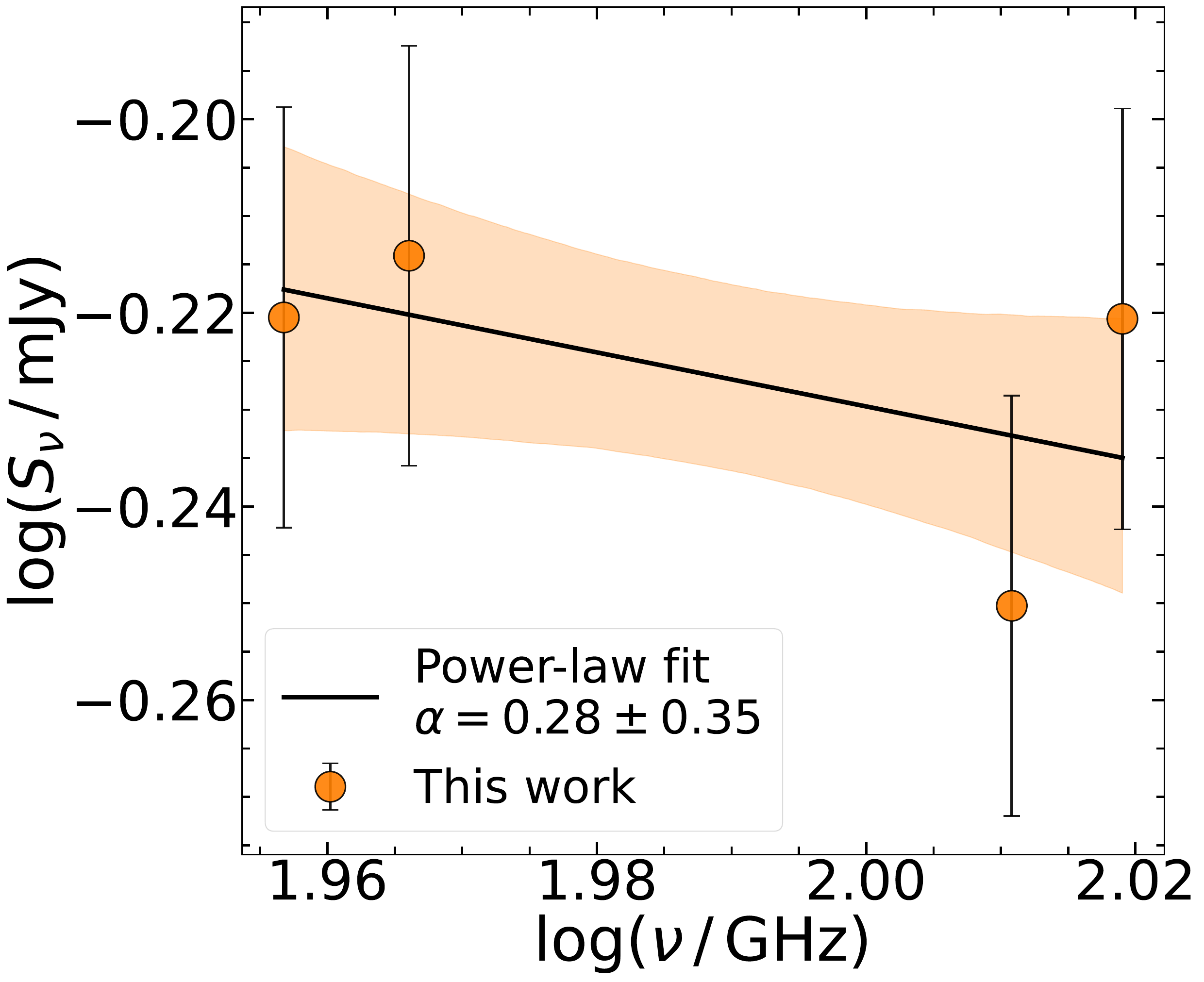}
        \subcaption{\textbf{RHS\,61}}
    \end{minipage}\hfill
    \begin{minipage}[b]{0.31\textwidth}
        \centering
        \includegraphics[width=1.1\textwidth]{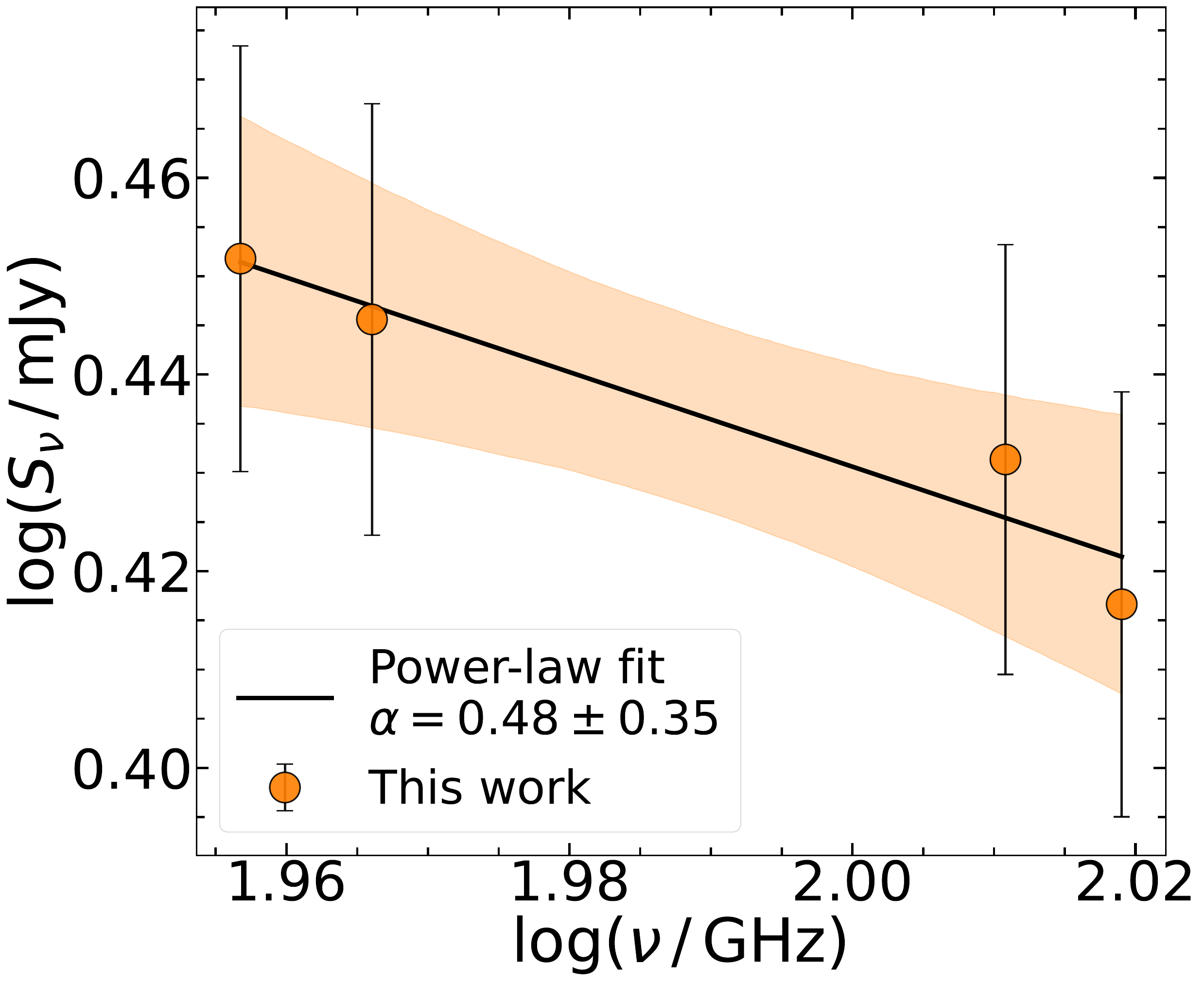}
        \subcaption{\textbf{LEDA\,12622}}
    \end{minipage}
    \begin{minipage}[b]{0.31\textwidth}
        \centering
        \includegraphics[width=1.1\textwidth]{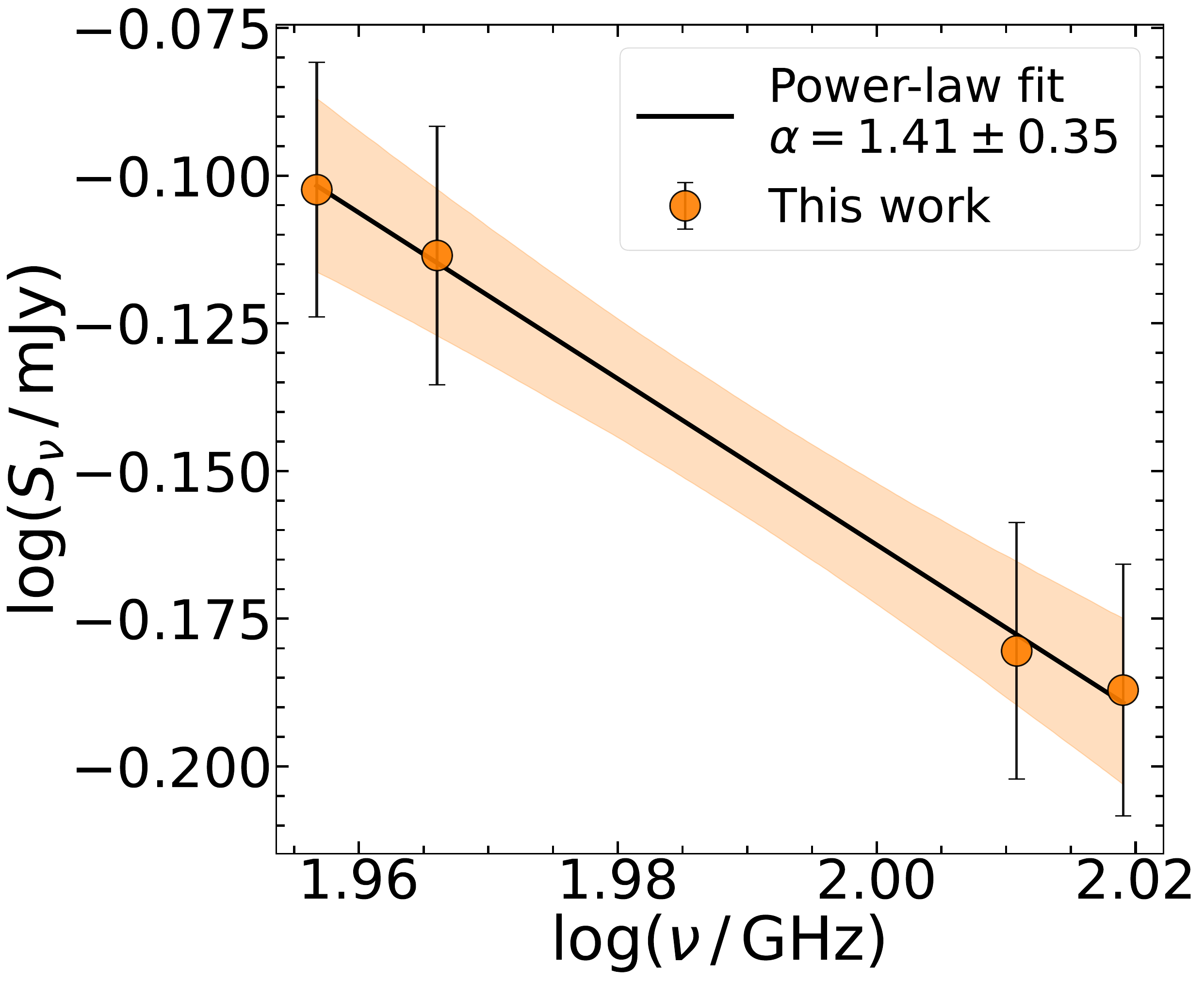}
        \subcaption{\textbf{2MASX\,J02223523+2508143}}
    \end{minipage}\hfill
    \begin{minipage}[b]{0.31\textwidth}
        \centering
        \includegraphics[width=1.1\textwidth]{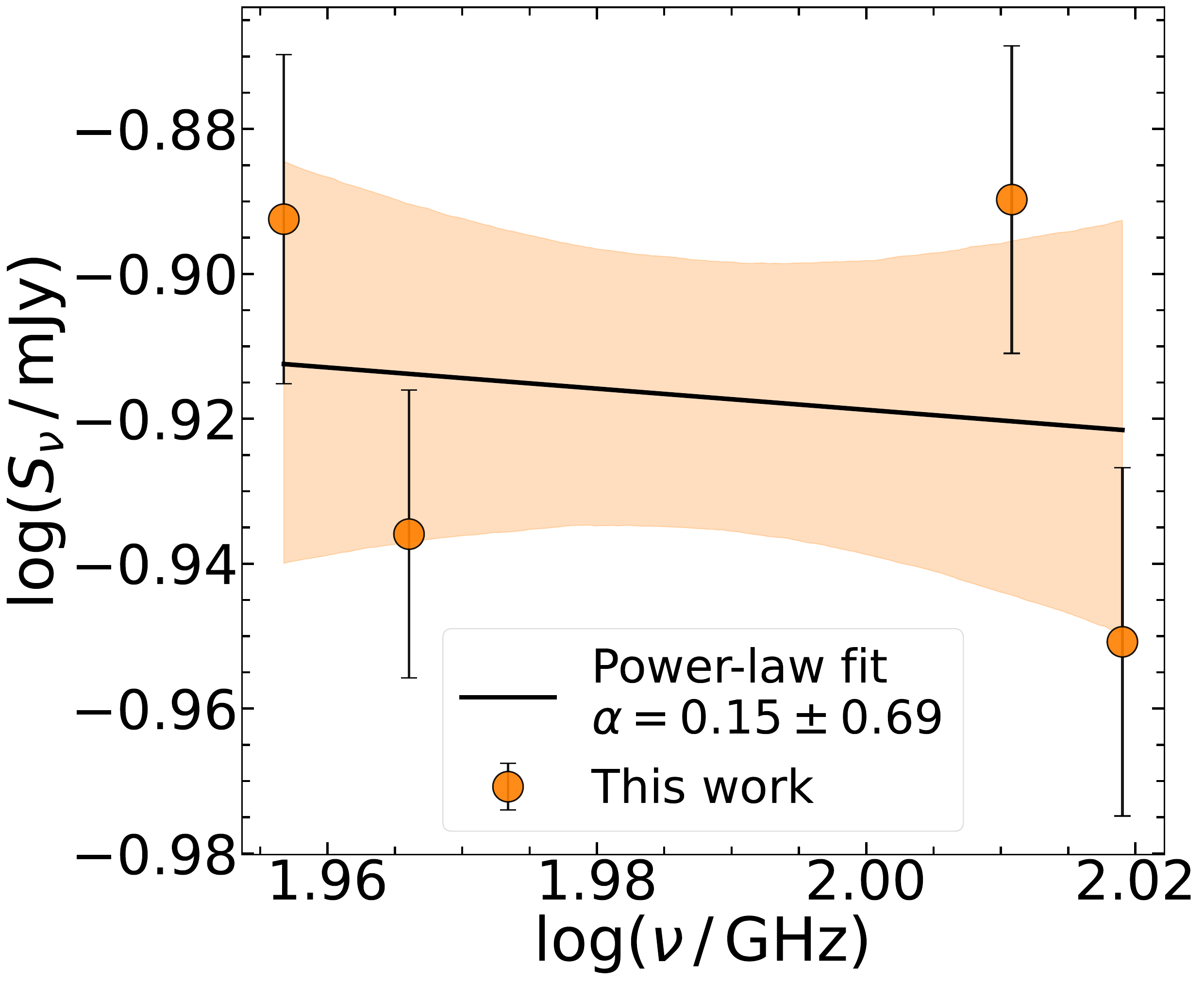}
        \subcaption{\textbf{2MASX\,J17311341+1442561}}
    \end{minipage}\hfill
    \begin{minipage}[b]{0.31\textwidth}
        \centering
        \includegraphics[width=1.1\textwidth]{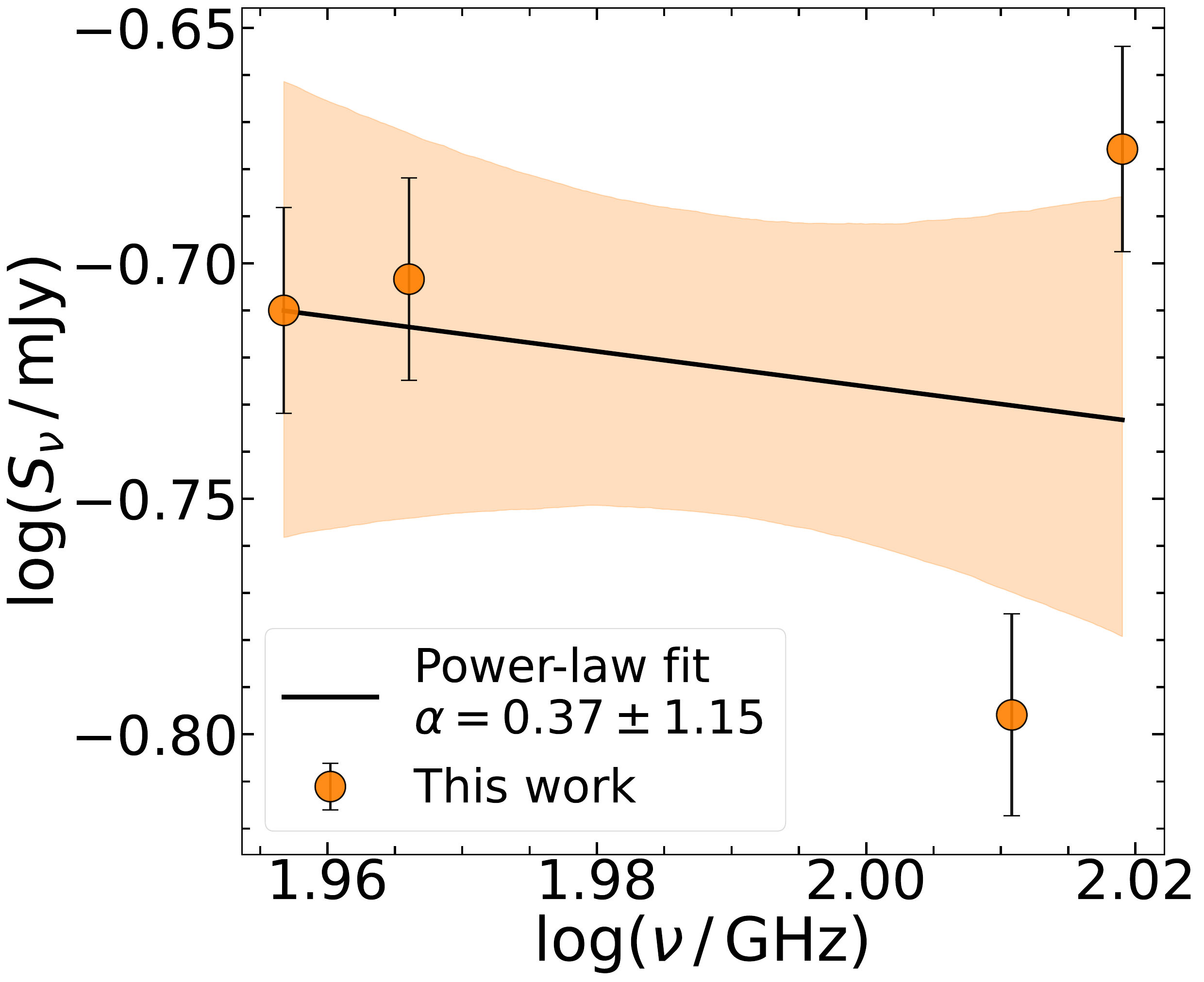}
        \subcaption{\textbf{LEDA\,12773}}
    \end{minipage}
    \caption{The fit of $ S_{\rm mm} \propto \nu ^{-\alpha_{\rm mm}}$ to the 100\,GHz fluxes in mJy from the four separate SPWs for each of our sources. Systematic uncertainties on the 100\,GHz fluxes are $\sigma_{\rm S_{100GHz}^{peak}}=\sqrt{(\rm rms)^2+(0.05\times S_{\rm 100GHz}^{\rm peak})^2}$ (see Section\,\ref{Datamm}). The $ \alpha_{\rm mm}$ values resulting from the fit, including the errors, are indicated in each figure.}
    \label{fig:alpha}
\end{figure}

Figure \ref{fig:alpha} shows for each of our nine sources the resulting fit of $S_{\nu}\propto\nu^{-\alpha_{\rm mm}}$ to the peak fluxes in the four SPWs. The $ \alpha_{\rm mm}$ values resulting from the fit are displayed in the figures as well.

\section{\textit{Swift} data}\label{AppendixXray}

We have obtained new \textit{Swift} observations of our sample of nine RQ AGN as well. The observations and images are discussed in Section\,\ref{swiftdata} and Section\,\ref{sect:XrayDataReduction}, respectively.

Table \ref{tab:AdditionalSwiftData} displays additional details on the XRT and UVOT data. We present for each source the \textit{Swift} ID, the observation ID, the Galactic absorption column densities\footnote{Obtained from the HEASARC $N_{\rm H, Gal}$ calculator \url{https://heasarc.gsfc.nasa.gov/cgi-bin/Tools/w3nh/w3nh.pl}} ($ N_{\rm H, Gal}$) as used in our X-ray spectral modeling and the photon index $\Gamma$ derived from our spectral fits. We also provide the UV magnitudes, both including contributions from the host galaxy and AGN, and those corresponding to the host galaxy alone. As described in Section\,\ref{sect:UVDATARED}, we apply a correction factor for dust extinction. 

We note that the source 2MASX\,J02223523+2508143 has a high correction factor of 45.0. 

\begin{table*}[ht!]
\setlength{\tabcolsep}{4.5pt}
\centering
\caption{Additional information on the \textit{Swift} data}
\begin{tabularx}{\textwidth}{lll|ccc|ccccccccc}
\hline
&&&XRT&&&UVOT\\
\hline
\hline
(1) & (2) & (3) & (4) & (5) & (6) & (7) & (8) & (9) \\
Source & SWIFT ID & Observation &  $ \log(N_{\rm H, Gal})$ & $\Gamma$ & C-stat/ & $ m_{\rm galaxy+AGN}$ & $ m_{\rm galaxy}$     & Correction \\
       &          &         ID       & (cm$^{-2}$)         &          & dof     &                       &                       & Factor     \\
\hline
Q\,0119$-$286           &  SWIFT\,J0122.0$-$2818  & 00045919010  & 20.2 & 1.7$\pm$0.4 & 20/35  & 15.4 & 16.8 & 0.50 \\
        
PG\,0026+129            &   SWIFT\,J0029.2+1319   & 00037603004  & 20.7 & 2.0$\pm$0.2 & 106/120  & 15.8 & 17.3 & 0.73 \\

PG\,0052+251            &   SWIFT\,J0054.9+2524   & 00037604006  & 20.6 & 1.8$\pm$0.2 & 158/176& 16.1 & 18.0 & 0.66 \\

Mrk\,813                &   SWIFT\,J1427.5+1949   & 00035307048  & 20.4 & 1.7$\pm$0.2 & 133/163  & 16.0 & 16.8 & 0.59 \\

RHS\,61                 &   SWIFT\,J2325.6+2157   & 00031323002  & 20.6 & 2.0$\pm$0.2 & 165/161  & \nodata      & \nodata      & \nodata      \\

LEDA\,126226            &   SWIFT\,J1416.9$-$1158 & 00040708005  & 20.7 & 2.2$\pm$0.2 & 92/118  & 15.5 & 18.4 & 0.42 \\

2MASX\,J02223523+2508143&   SWIFT\,J0222.3+2509   & 00032246003  & 20.8 & 1.7$\pm$0.8 & 81/95  & 18.9 & 19.6 & 45.0 \\

2MASX\,J17311341+1442561&   SWIFT\,J1731.3+1442   & 00041781004  & 20.8 & 2.3$\pm$0.2 & 112/145  & 16.6 & 19.9 & 0.77 \\

LEDA\,12773             &   SWIFT\,J0325.0$-$4154 & 00037556005  & 20.0 & 2.0$\pm$0.2 & 122/153  & 16.6 & 18.6 & 0.69 \\    
\hline
\end{tabularx}
\par\vspace{1ex}
\parbox{0.95\textwidth}{\small
\textbf{Note:} (1) Source names, (2) ID of the \textit{Swift} observations, (3) Observation ID of the \textit{Swift} observations, (4) Galactic absorption $N_{\rm H, Gal}$ in cm$^{-2}$, (5) Photon index obtained from the X-ray fits with their 90\% confidence uncertainties, (6) Quality of the fit of the X-ray model expressed as the total fit statistics divided by the degrees of freedom, (7) The UV magnitude of the host galaxy and AGN combined, (8) UV magnitude of host galaxy, and (9) Correction factor for dust $\rm 10^{0.4 R_{V} E(B-V) k(\lambda)}$.}

\label{tab:AdditionalSwiftData}
\end{table*}

\subsection{X-ray spectra}\label{AppendixXrayspectra}

The X-ray spectra of our nine sources in the 2--10\,keV range are displayed in Figure\,\ref{fig:Xray_spectra}. 
The fitted model is described in Section\,\ref{sect:XrayDataReduction} and the fitted parameters are presented in Table\,\ref{tab:AdditionalSwiftData}.

\begin{figure}[h!]
    \centering
    \begin{minipage}[b]{0.32\textwidth}
        \centering
        \includegraphics[width=1.1\textwidth]{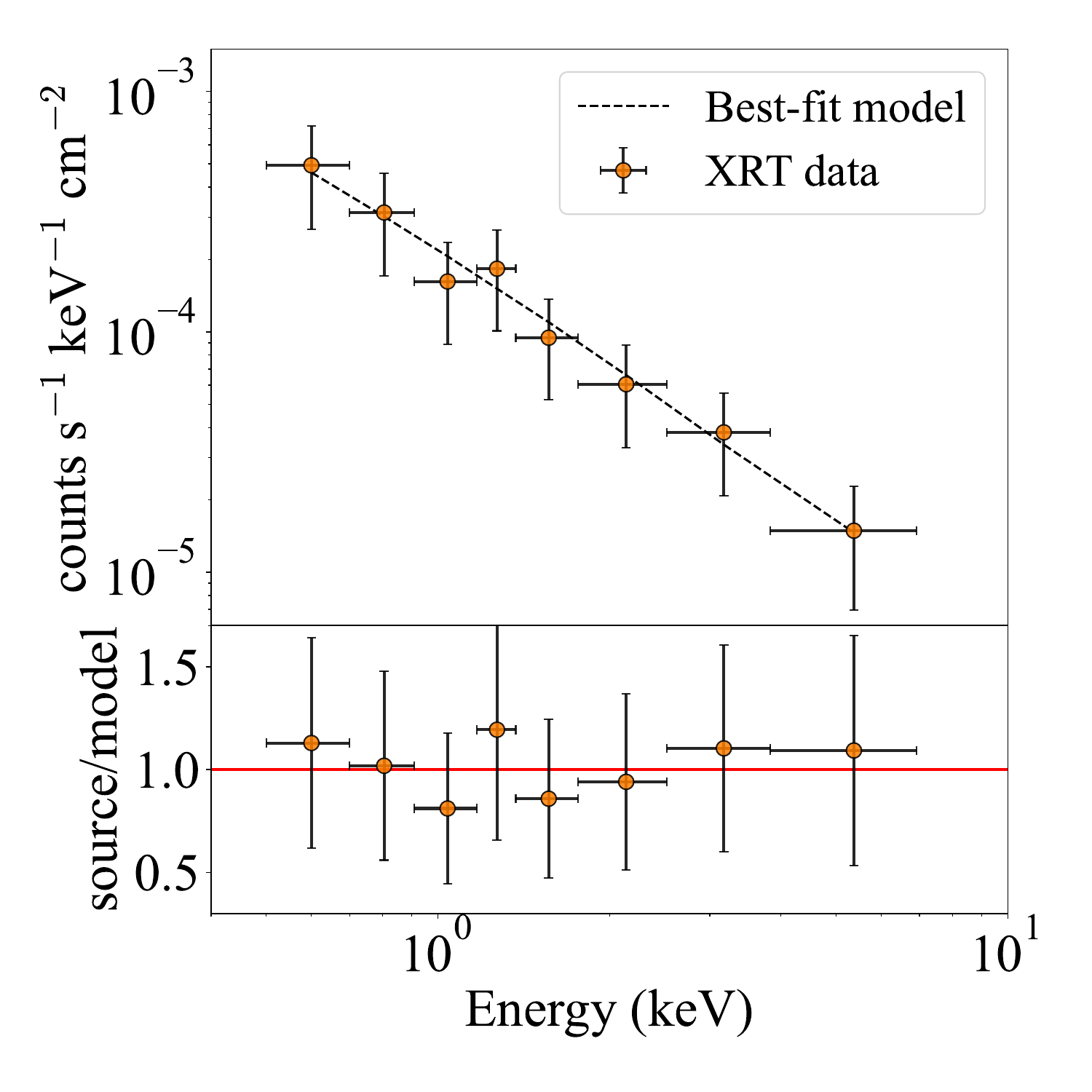}
        \subcaption{\textbf{Q\,0119--286}}
    \end{minipage}\hfill
    \begin{minipage}[b]{0.32\textwidth}
        \centering
        \includegraphics[width=1.1\textwidth]{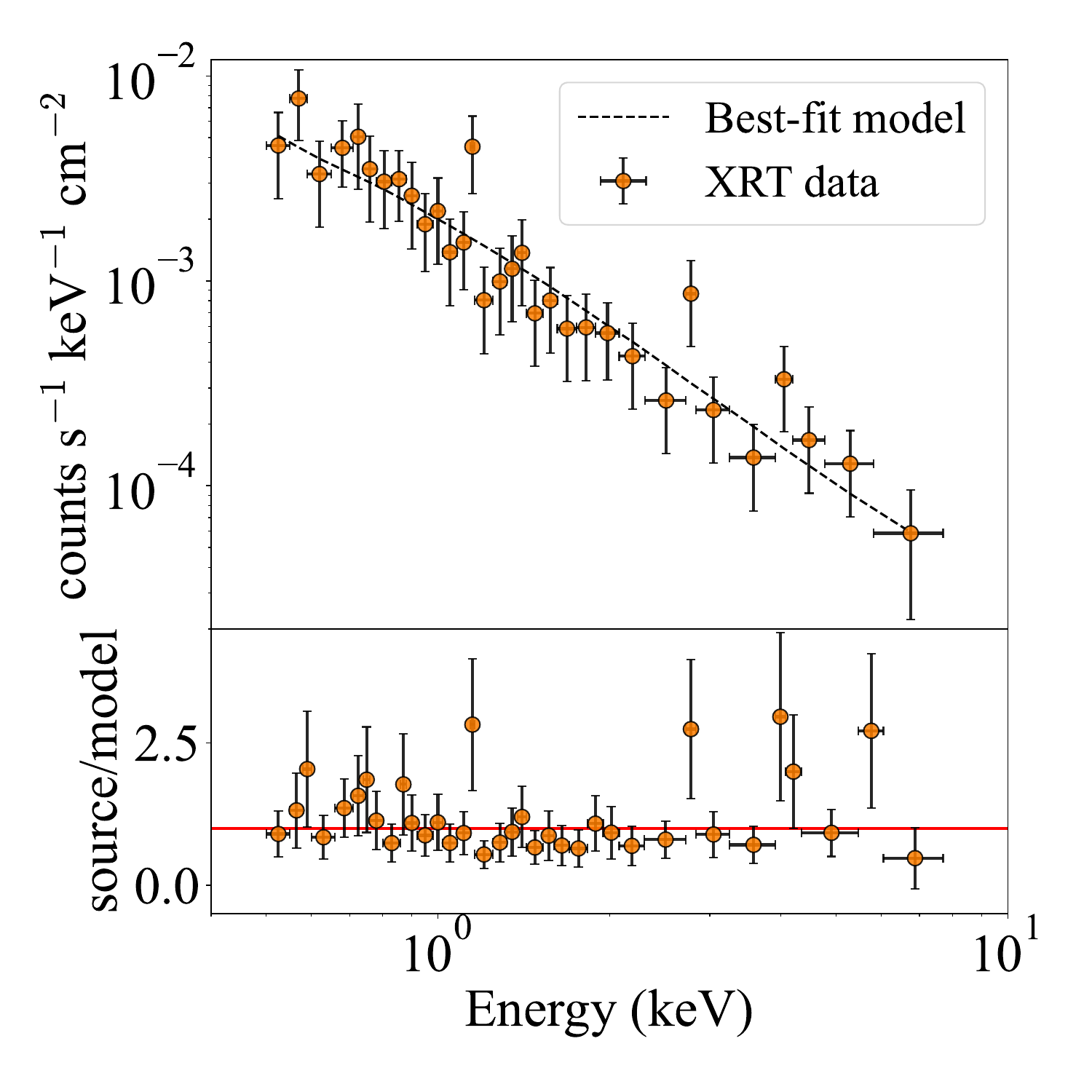}
        \subcaption{\textbf{PG\,0026+129}}
    \end{minipage}\hfill
    \begin{minipage}[b]{0.32\textwidth}
        \centering
        \includegraphics[width=1.1\textwidth]{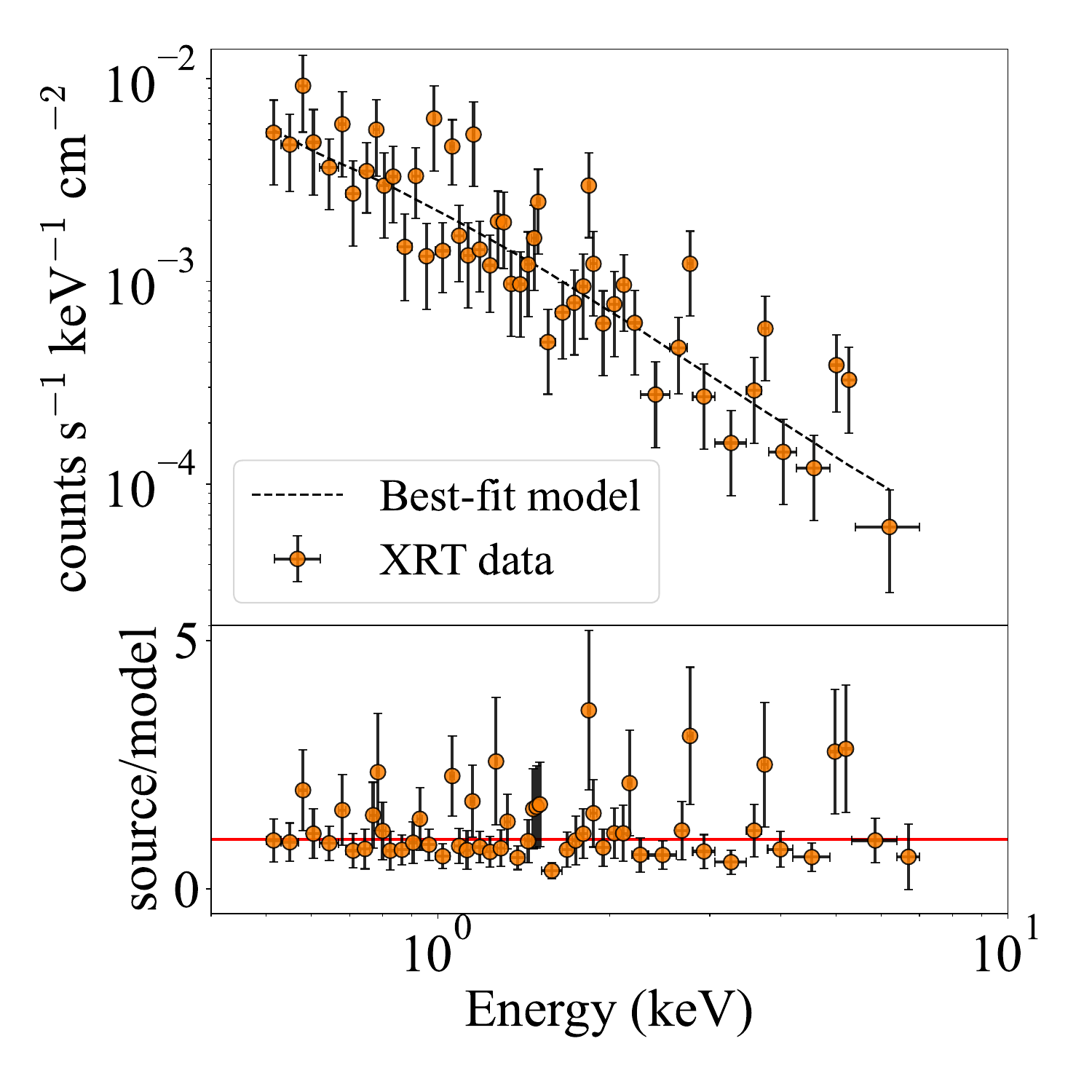}
        \subcaption{\textbf{PG\,0052+251}}
    \end{minipage} 
\caption{(Part 1 of 2) X-ray spectra of our nine sources at 2--10\,keV. The top panel shows the spectra in counts. The bottom panel shows the ratio of the data and the fitted model.}
\end{figure}
\begin{figure}[ht!]
\ContinuedFloat
\centering
    \begin{minipage}[b]{0.32\textwidth}
        \centering
        \includegraphics[width=1.1\textwidth]{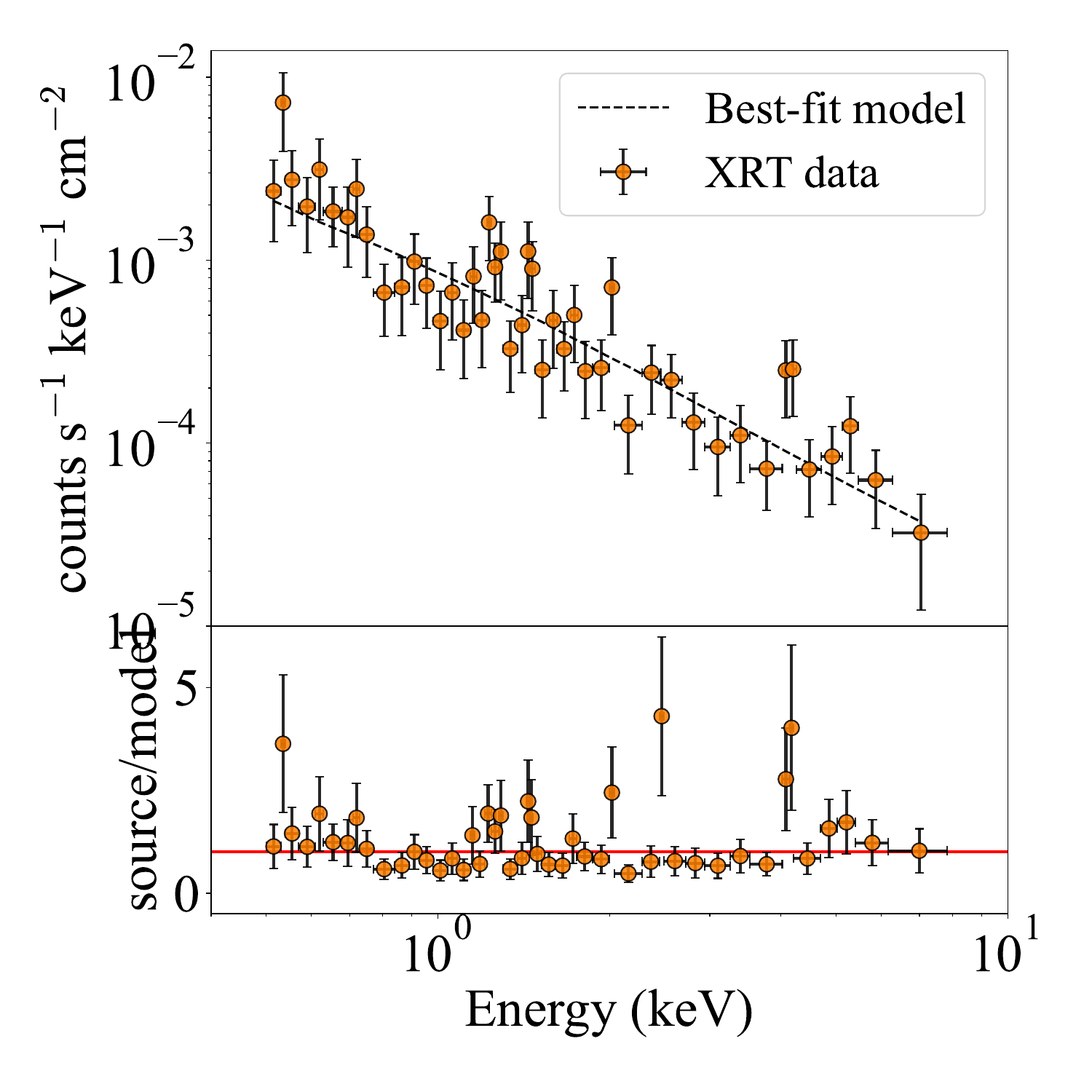}
        \subcaption{\textbf{Mrk\,813}}
    \end{minipage}\hfill
    \begin{minipage}[b]{0.32\textwidth}
        \centering
        \includegraphics[width=1.1\textwidth]{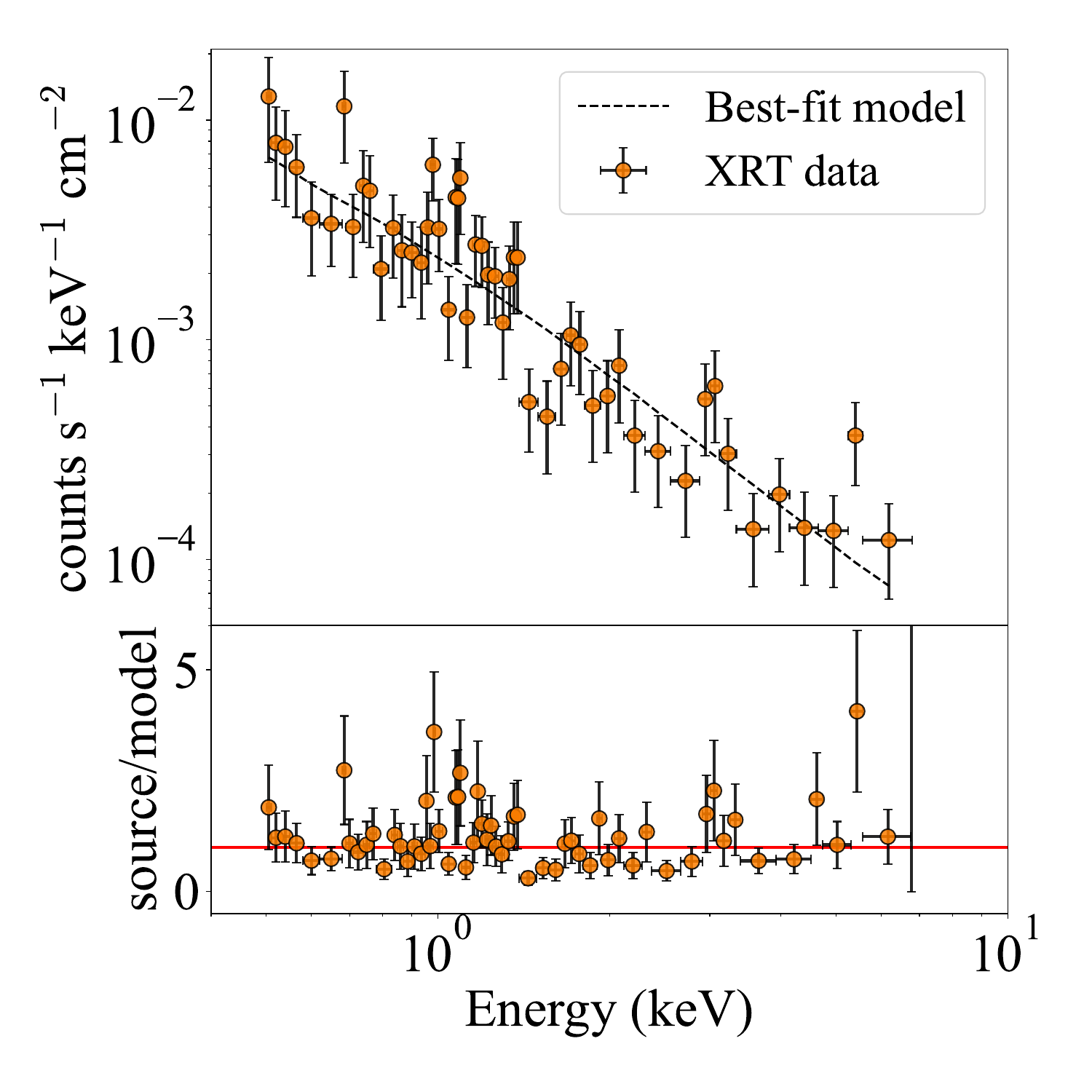}
        \subcaption{\textbf{RHS\,61}}
    \end{minipage}\hfill
    \begin{minipage}[b]{0.32\textwidth}
        \centering
        \includegraphics[width=1.1\textwidth]{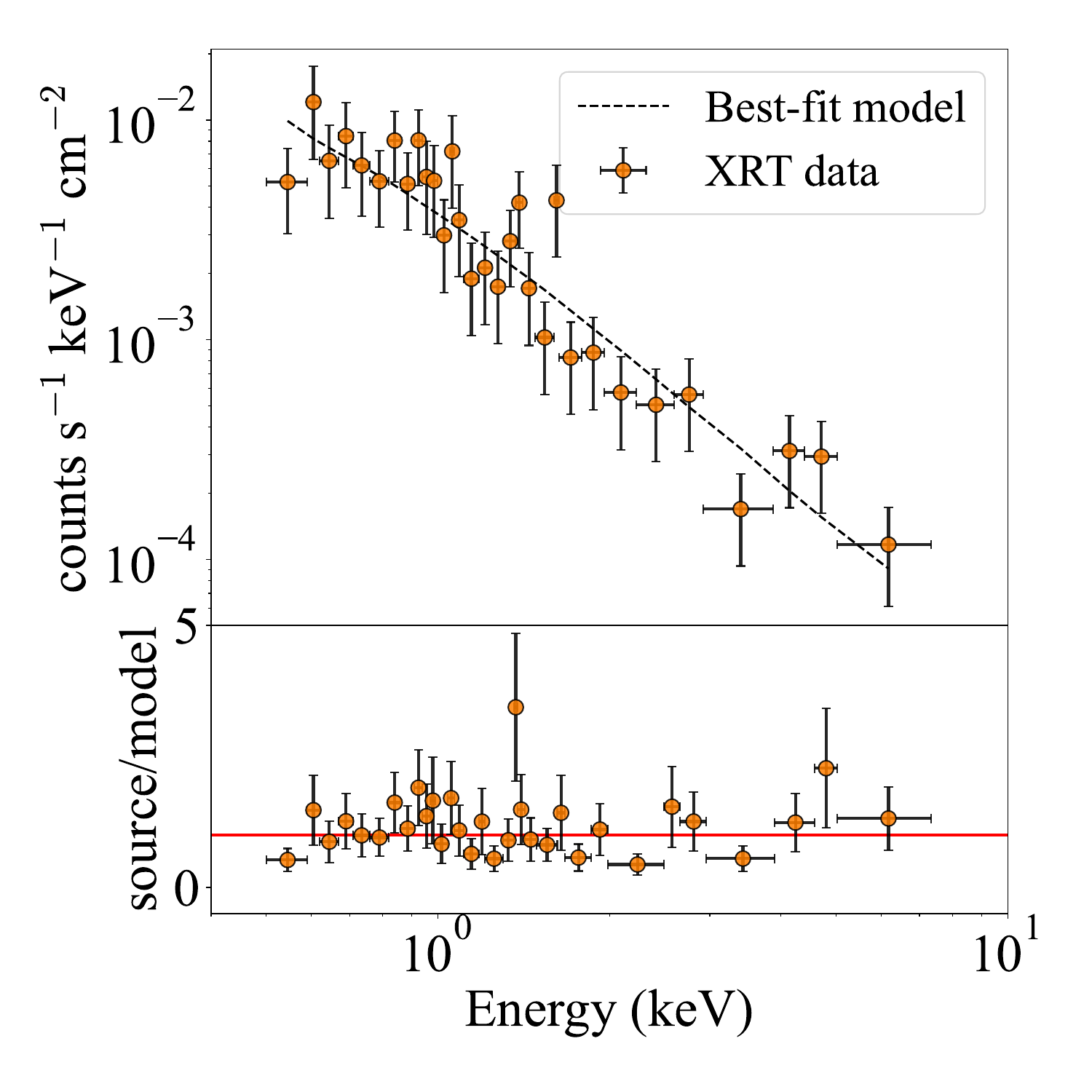}
        \subcaption{\textbf{LEDA\,12622}}
    \end{minipage}
    \begin{minipage}[b]{0.32\textwidth}
        \centering
        \includegraphics[width=1.1\textwidth]{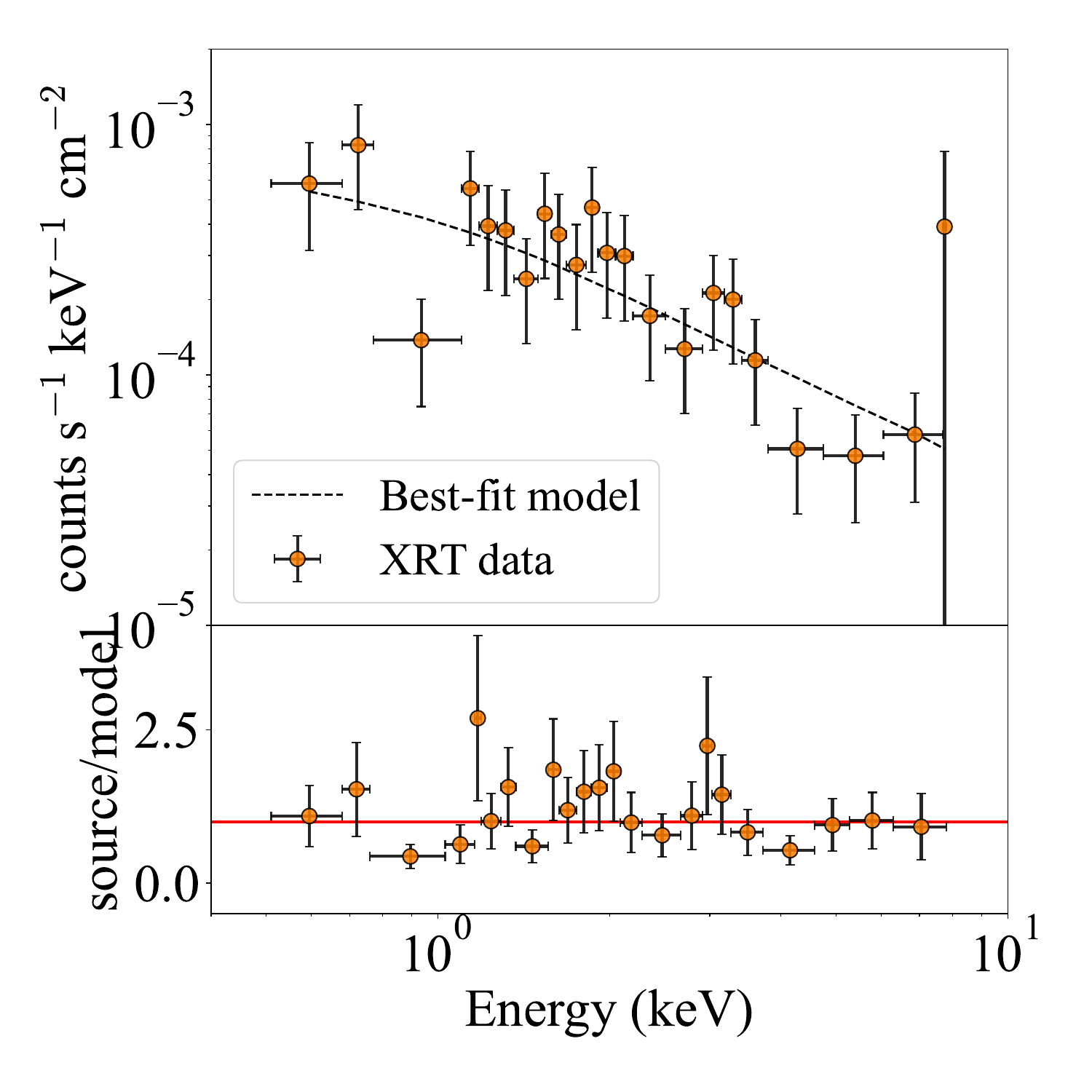}
        \subcaption{\textbf{2MASX\,J02223523+2508143}}
    \end{minipage}\hfill
    \begin{minipage}[b]{0.32\textwidth}
        \centering
        \includegraphics[width=1.1\textwidth]{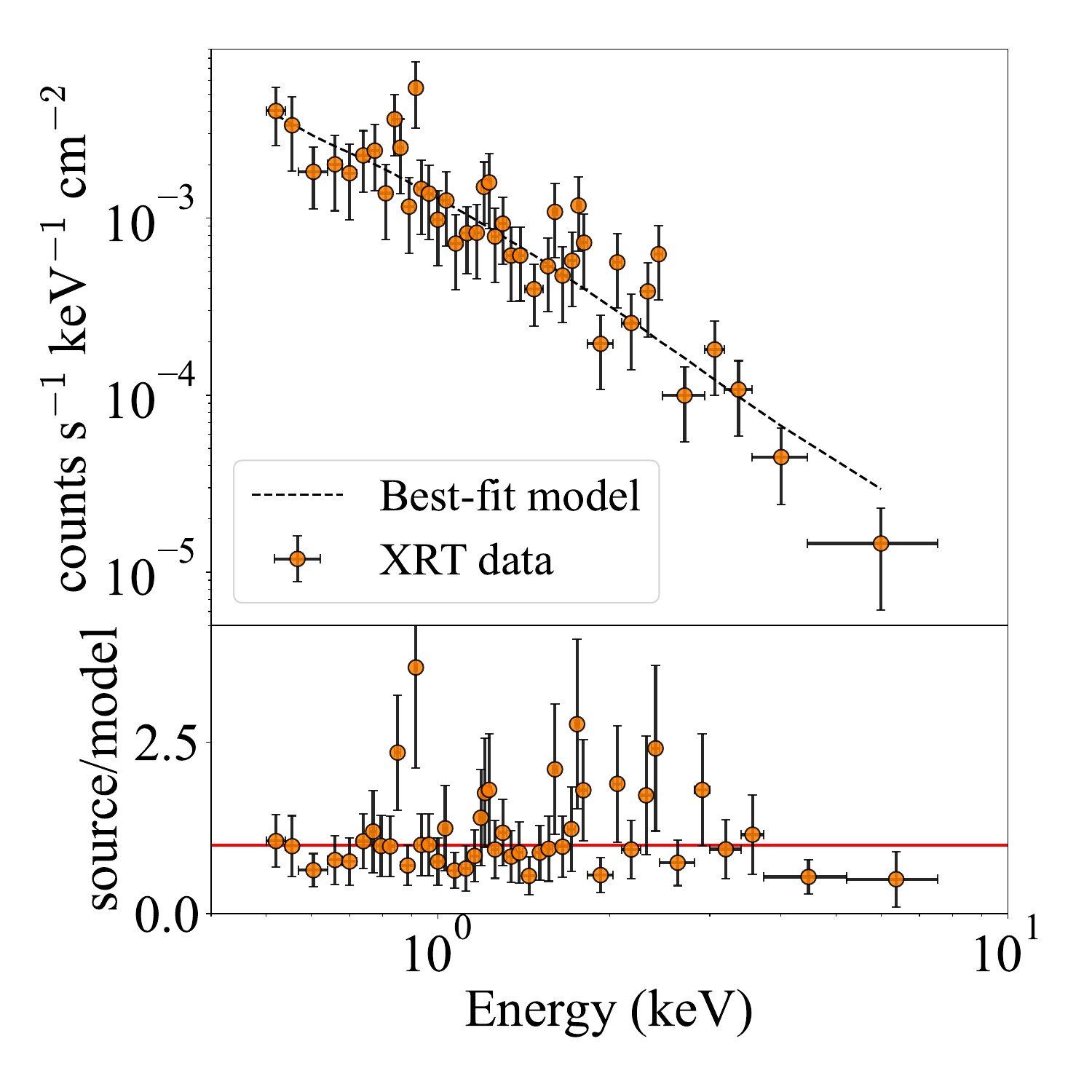}
        \subcaption{\textbf{2MASX\,J17311341+1442561}}
    \end{minipage}\hfill
    \begin{minipage}[b]{0.32\textwidth}
        \centering
        \includegraphics[width=1.1\textwidth]{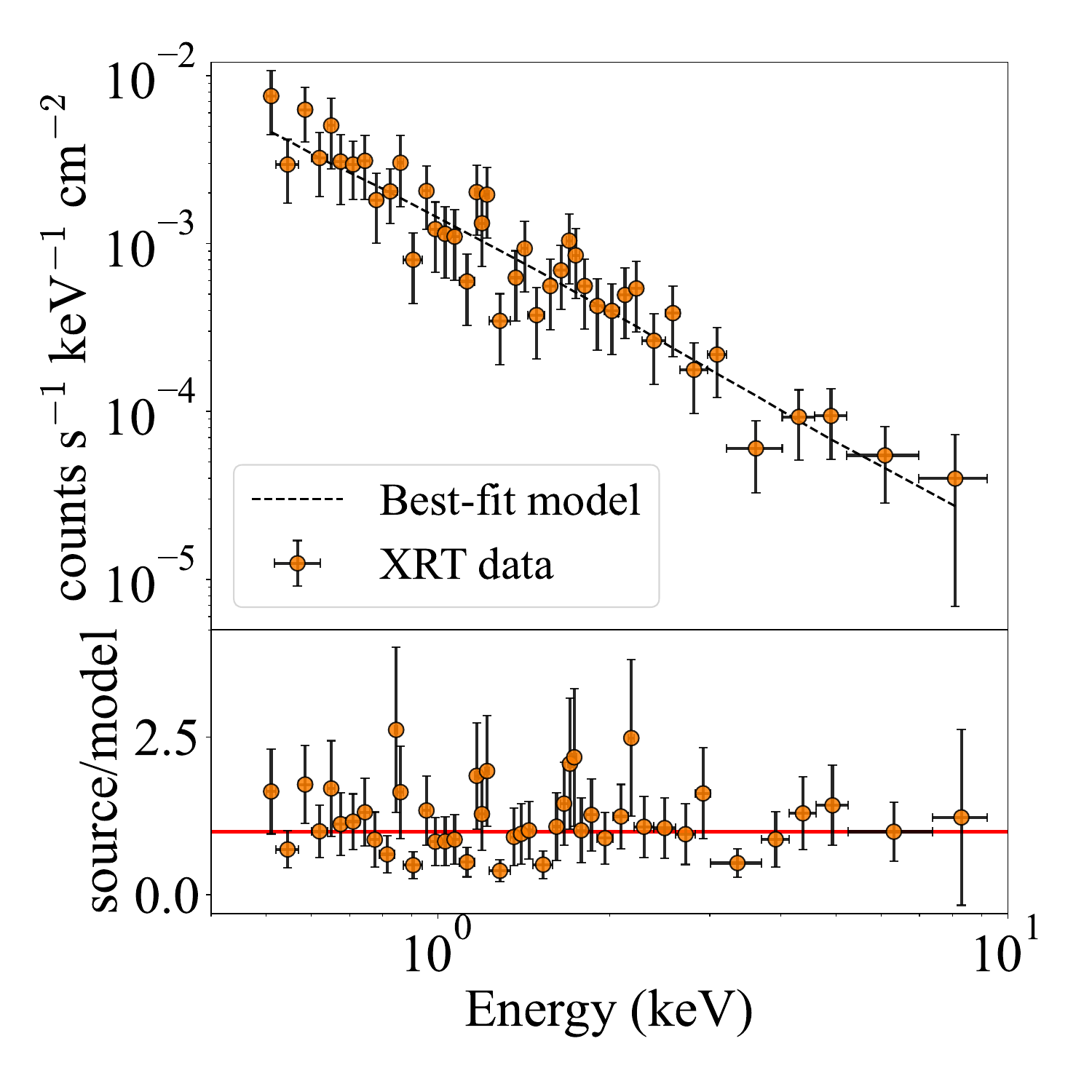}
        \subcaption{\textbf{LEDA\,12773}}
    \end{minipage}
    \caption{(Part 2 of 2) X-ray spectra of our nine sources at 2--10\,keV. The top panel shows the spectra in counts. The bottom panel shows the ratio of the data and the fitted model.} 
    \label{fig:Xray_spectra}
\end{figure}

\newpage

\section{Additional tests on the millimeter/X-ray relation}\label{AppendixAdditional}

\subsection{The 0.3--2\,keV energy range}\label{soft}

\begin{figure*}[t]
\centering
\includegraphics[width=0.48\textwidth]{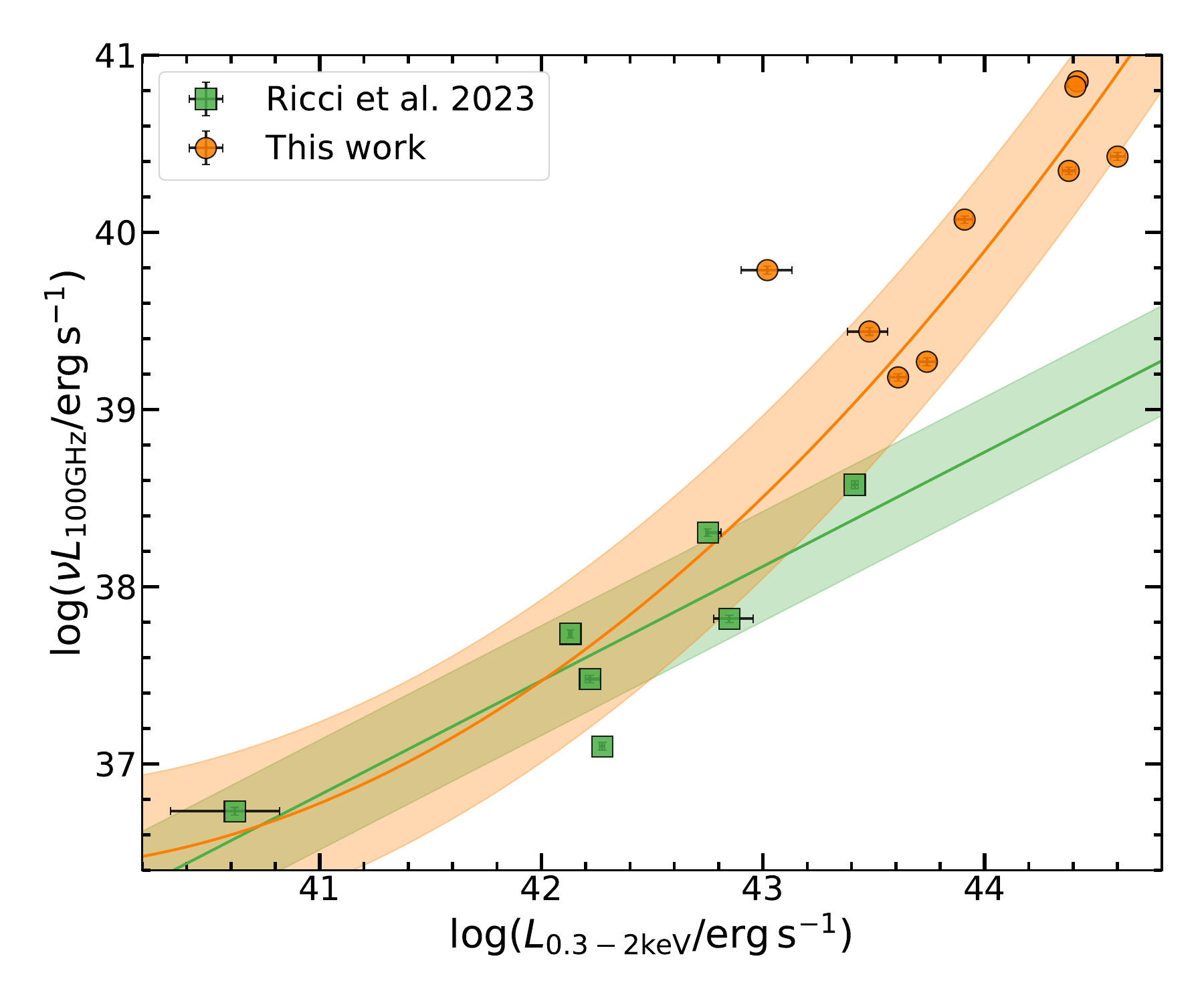}
\hfill
\includegraphics[width=0.48\textwidth]{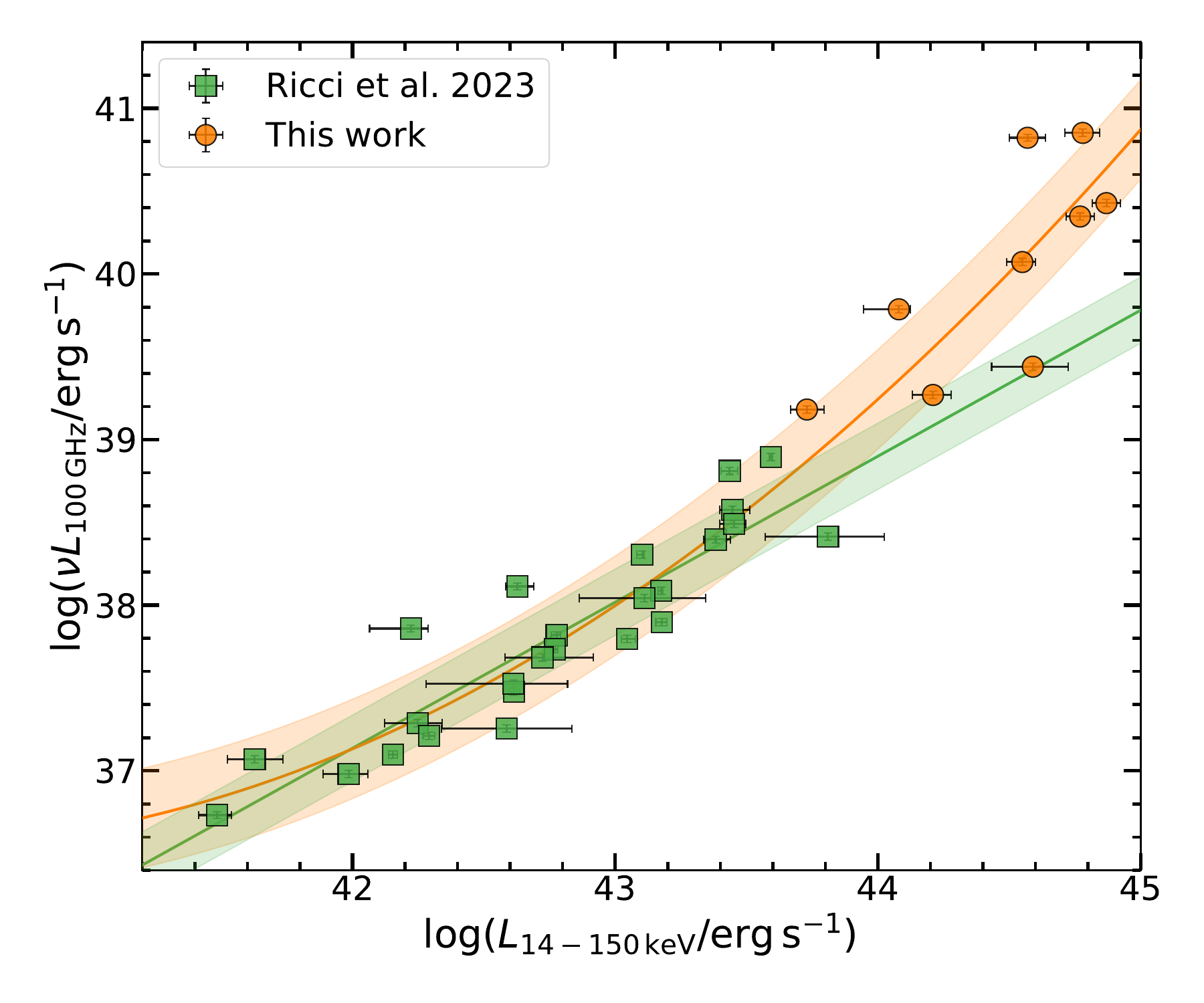}
\caption{ 
\textbf{Left:} The 100\,GHz luminosity $ \nu L_{\rm 100GHz}$ in erg\,s$^{-1}$ vs. the intrinsic 0.3--2\,keV luminosity $ L_{\rm 0.3-2keV}$ in erg\,s$^{-1}$ for sources by \citetalias{ricci_tight_2023} and this work. We have fitted a second-degree polynomial relation to the millimeter and X-ray emission, presented in Equation\,\ref{eq:mmXray_032}, which has an intrinsic scatter of 0.46\,dex.
\textbf{Right:} The 100\,GHz luminosity $ \nu L_{\rm 100GHz}$ in erg\,s$^{-1}$ vs. the 14--150\,keV luminosity $ L_{\rm 14-150keV}$ in erg\,s$^{-1}$. We have fitted a second-degree polynomial, presented in Equation\,\ref{eq:mmXray_14150}, with an intrinsic scatter of 0.30\,dex. The ALMA and \textit{Swift} observations in this figure were not obtained simultaneously.}
\label{fig:corrFL032}
\end{figure*}

In addition to investigating the relation between the 2--10\,keV and millimeter emission, we investigated the relation between 0.3--2\,keV and millimeter emission, which is where the soft excess is located.
Here, we have only adopted the sources from \citetalias{ricci_tight_2023} that are unobscured.
As can be seen in Figure\,\ref{fig:corrFL032}, we observe a similar trend between the millimeter and 0.3--2\,keV emission as with the millimeter and 2--10\,keV emission. 
Again, we fitted a second-degree polynomial and obtained the relation:
\begin{equation}
\log \left( \dfrac{L_\mathrm{100\,GHz}}{10^{38}\,\mathrm{erg\,s}^{-1}} \right) 
  = (0.17\pm0.09) \, \log \left( \dfrac{L_\mathrm{0.3-2\,keV}}{10^{43}\,\mathrm{erg\,s}^{-1}} \right)^{2} 
   + (1.21\pm0.12) \, \log \left( \dfrac{L_\mathrm{0.3-2\,keV}}{10^{43}\,\mathrm{erg\,s}^{-1}} \right) + (0.51\pm0.17)
 \label{eq:mmXray_032}
\end{equation}
which has a large intrinsic scatter of 0.46\,dex.

\subsection{The 14-150\,keV energy range}\label{App14150}

As discussed in Section\,\ref{millimeter/Xray}, we adopt the 2--10\,keV band rather than the 14--150\,keV band also presented by \citetalias{ricci_tight_2023}, since the former was observed quasi-simultaneously with the ALMA data.
Nevertheless, 14--150\,keV luminosities are available from \cite{ricci_bat_2017} (also see Section\,\ref{sampleselection}), which are time-averaged over several years.
Given the tight correlation reported by \citetalias{ricci_tight_2023} between $L_{\rm 14-150keV}$ and $L_{\rm 100GHz}$, we show the corresponding relation for our sample in Figure\,\ref{fig:corrFL032}, although with nonsimultaneous observations.
We obtain the following second-degree polynomial relation:
\begin{equation}
    \log \left( \dfrac{L_\mathrm{100\,GHz}}{10^{38}\,\mathrm{erg\,s}^{-1}} \right) 
  = (0.19\pm0.06) \, \log \left( \dfrac{L_\mathrm{14-150\,keV}}{10^{43}\,\mathrm{erg\,s}^{-1}} \right)^{2} 
   + (1.06\pm0.07) \, \log \left( \dfrac{L_\mathrm{14-150\,keV}}{10^{43}\,\mathrm{erg\,s}^{-1}} \right) - (0.003\pm0.07)
 \label{eq:mmXray_14150}
\end{equation}
with an intrinsic scatter of 0.30\,dex for the full sample and a scatter of 0.44\,dex for the high-luminosity sources.


\subsection{Physical scale}\label{physicalscale}

\begin{figure*}[ht]
    \centering
    \includegraphics[width=0.42\textwidth]{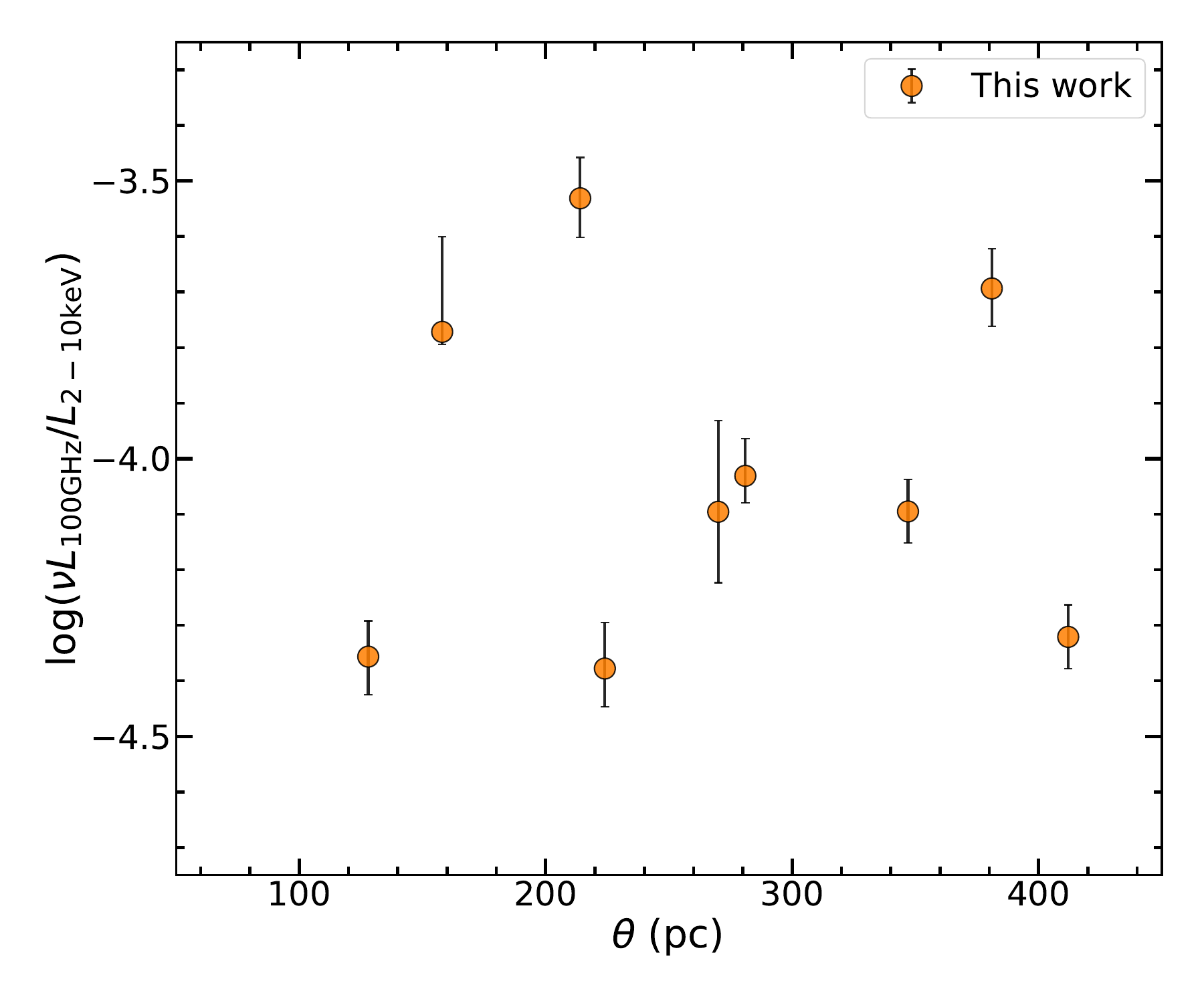}
    \caption{The millimeter/X-ray luminosity ratio $ \log(\nu L_{\rm 100GHz}/L_{\rm 2-10keV})$ vs. the physical beam size $\theta$ of our ALMA observations in pc. We obtain a $p$-value of 0.93, which indicates that there is no correlation between the two parameters.} 
    \label{fig:ratioLbeam}
\end{figure*}

Figure \ref{fig:ratioLbeam} shows the millimeter/X-ray luminosity ratio versus the physical beam scale of the 100\,GHz ALMA observations. We concluded there is no correlation between the ratio and increasing the physical beam size, obtaining a $p$-value of 0.93. Therefore, we suggest that the excess in millimeter emission observed in our luminous sources is likely not primarily caused by the larger beam size compared to \citetalias{ricci_tight_2023}. However, future high-resolution observations will confirm this.

\newpage

\subsection{$L_{\rm 100GHz}/L_{\rm 2-10keV}$ dependence on $M_{\rm BH}$ and SFR}\label{AppendixMSFR}

\begin{figure*}[h]
\centering
\includegraphics[width=0.95\textwidth]{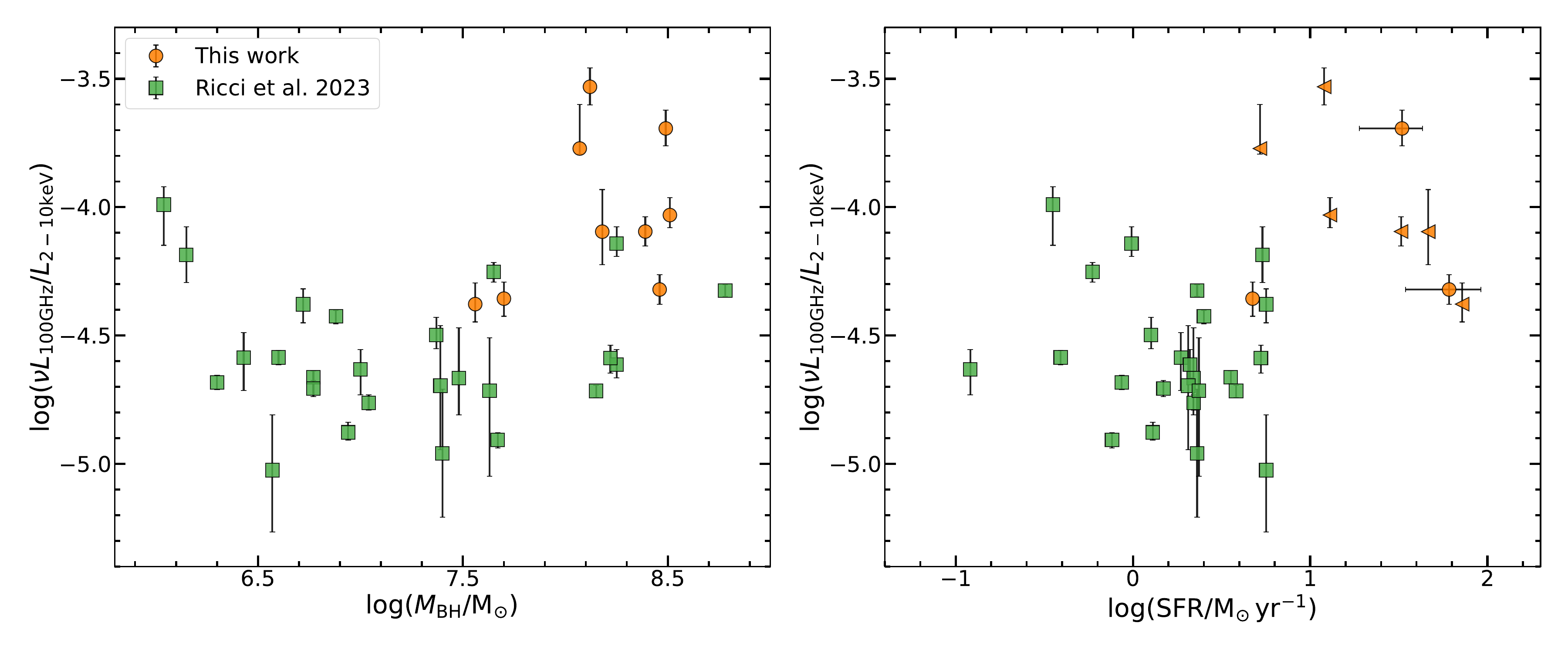}
\caption{\textbf{Left:} The millimeter/X-ray luminosity ratio vs. the black hole mass $ M_{\rm BH}$ of the sources from this work and \citetalias{ricci_tight_2023}. We find no correlation between the two, obtaining a $p$-value of 0.03. 
\textbf{Right:} The millimeter/X-ray luminosity ratio vs. the star-formation rate (SFR). Eight out of nine sources in this work only have upper limits on SFR. No correlation is found, obtaining a $p$-value of 0.22.}
\label{fig:ratio_vs_MBH_SFR}
\end{figure*}

Figure\,\ref{fig:ratio_vs_MBH_SFR} displays the millimeter/X-ray luminosity ratio $\log(\nu L_{\rm 100GHz}/L_{\rm 2-10keV})$ as a function of $ M_{\rm BH}$ and SFR. The lack of correlation we observe between the millimeter/X-ray luminosity ratio and $M_{\rm BH}$ and SFR and its implications are described in Section\,\ref{MbhSFR}.
The SFR values for seven out of nine sources, consisting of six upper limits and one detection, are adopted from \cite{ichikawa_complete_2017,ichikawa_bat_2019} and were determined through IR SED decomposition. Finally, the SFR values for PG\,0026+129 and PG\,0052+251 were determined by Y. D\'iaz et al. (2026, in preparation) from the strength of the PAH feature.

\subsection{X-ray outflows}\label{AppXrayWinds}

To further investigate the potential connection between the observed millimeter emission and the presence of outflows, we analyzed the available high-quality X-ray spectra for the sources in our sample. 
Among the nine AGN, five have archival X-ray observations with sufficiently high signal-to-noise ratios to enable a meaningful spectroscopic analysis aimed at identifying signatures of ionized absorption. In particular, we searched for evidence of absorbers or outflowing gas that could produce shocks and contribute to the observed millimeter continuum emission. 

\indent Q\,0119--286 was observed by a short ($\sim9$\,ks) pointing by XMM-Newton. We extract the EPIC-pn spectrum using {\footnotesize SASv20.0.0}, from a circular region of $40^"$ that includes the source, and the background from a source-free region of the same size, after applying filters to remove bad time intervals with particle flaring and the processing command {\footnotesize EPPROC}. We fit the source with a model that includes neutral Galactic absorption \citep[$\rm N_{\rm H}=1.5\times10^{20}$\,cm$^{-2}$][]{hi4pi16}, a blackbody to model a weak soft excess, a power--law, and a Gaussian line for the Fe K$\alpha$ emission line. We find a statistic of $\chi^2/{\rm dof}=92/72$. The quality of this spectrum is not enough to use a photoionized grid; however, an absorption edge, typical of absorbers, is clearly present in a visual check of the residuals. Adding an edge model, we obtain an energy for the edge of $E=620\pm35$ eV, marginally consistent with the oxygen K-edge at $540$ eV. The statistic increases by $\Delta\chi^2/\Delta{\rm dof}=14/2$ to $\chi^2/{\rm dof}=78/70$, corresponding to a $p$-value of $p_{\rm val}=0.0032$ ($\lesssim3\sigma$). The blackbody temperature and the power-law photon index are $kT=130\pm10$ eV and $\Gamma=2.33\pm0.09$, respectively.\\
\indent PG\,0026+129 was observed for 13 ks by XMM-Newton and for 150 ks by {\it NuSTAR}. We model the spectra with an absorbed \citep[$N_{\rm H}=4.7\times10^{20}$\,cm$^{-2}$][]{hi4pi16} power-law plus blackbody and ionized reflection using {\footnotesize XILLVER} \citep{garcia10}, prominent at $E>10$\,keV. Since a detailed fit is beyond the scope of this paper, we freeze the iron abundance and inclination of the disk in the model. The best-fit model has a statistic of $\chi^2/{\rm dof}=1043/913$. Negative residuals are present in the soft X-ray band ($E=0.5-2$\,keV), strongly suggesting the presence of a absorber. We multiply the above model by our photoionized grid and the statistic improves to $\chi^2/{\rm dof}=998/911$ ($\Delta\chi^2/\Delta{\rm dof}=45/2$). This ionized gas is characterized by column density $ N_{\rm H}=(2\pm1)\times10^{21}$\,cm$^{-2}$ and an ionization parameter $\log\xi/({\rm erg\;cm\;s^{-1}})=1.25^{+0.25}_{-0.50}$. We also retrieve a blackbody temperature $ kT=179\pm16$ eV, a photon index $\Gamma=1.92^{+0.03}_{-0.05}$ and a cutoff energy $ E_{\rm cut}=94_{-31}^{+82}$ keV.\\
\indent One source, PG\,0052+251, was already studied by \cite{matzeu23} and a very low-ionization absorber ($\rm \log\xi/({\rm erg\;cm\;s^{-1}})\sim-1.1$) with column density $ N_{\rm H}\sim6\times10^{21}$\,cm$^{-2}$ was found.\\
\indent Mrk\,813 was observed 73 times with \textit{Swift}-XRT and once by NuSTAR (OBSID 60160583002). All the available XRT observations of Mrk\,813 were combined into a single spectrum of $\sim65$~ks using the online tool\footnote{\url{https://www.swift.ac.uk/user_objects/}} described in \cite{evans09}. In contrast, the FPMA and FPMB spectra ($\sim25$~ks each) of the NuSTAR observations were extracted using {\footnotesize NUSTARDAS}, following the same procedure described by \citet{serafinelli24}. The three spectra were fitted with a power-law absorbed by Galactic absorption \citep[$N_{\rm H}=2.3\times10^{20}$][]{hi4pi16}, plus a blackbody to phenomenologically model a moderate soft excess. We obtain a statistic of $\rm \chi^2/{\rm dof}=710/552$, where dof is the number of degrees of freedom of the fit. 
Clear residuals are present at $\sim1$\,keV, which might indicate the presence of an absorber through the presence of an Fe UTA absorption complex \citep[e.g.,][]{halpern84,behar03,blustin05,laha14,Serafinelli25}. 
We create a photoionization grid using {\footnotesize XSTAR} \citep{kallman01}. Following e.g., \citet{tombesi11} and \citet{serafinelli19} we assume a SED with $\Gamma=2$, $ E_{\rm cut}=100$~keV, solar abundances \citep{asplund09}, a fully covering medium (covering factor $C_f=1$) and a turbulent velocity of $v_{\rm turb}=100$\,km\,s$^{-1}$ \citep[e.g.][]{laha14}. We apply the photoionization grid and we find $\rm N_{\rm H,ion}=3.4^{+0.8}_{-0.6}\times10^{21}$\,cm$^{-2}$ and a ionization parameter upper limit of $\rm \log\xi/({\rm erg\;cm\;s^{-1}})<0.8$. The continuum photon index is $\Gamma=1.87\pm0.03$, while the blackbody temperature is $kT=84_{-6}^{+22}$ eV. The statistic improves by $\Delta\chi^2/\Delta{\rm dof}=101/2$ to a final $\chi^2/{\rm dof}=609/500$, indicating that the absorber is extremely significant. We note that although our best-fit is consistent with neutral absorption since we found an upper limit to the ionization parameter, the adoption of a neutral absorption model is disfavored by a much smaller improvement of the fit statistic ($\Delta\chi^2/\Delta{\rm dof}=13/1$).\\
\indent As discussed in Section\,\ref{sect:XrayDataReduction}, we found evidence for the presence of an absorber in 2MASX\,J02223523+2508143. We analyzed the archival \textit{Swift}/XRT observations of this source, combining all available exposures to obtain a spectrum with $\sim 3500$ counts. The spectral fitting was performed in \textsc{XSPEC} following the same procedure adopted for the other sources. The spectrum clearly reveals the presence of ionized absorption. We modeled this component with our standard \textsc{xstar} table. The best-fit absorber has a column density of $N_{\rm H} = (8.8 \pm 2.2) \times 10^{21}$ cm$^{-2}$, and ionization parameter $\log \xi/({\rm erg cm s}^{-1}) = 0.9 \pm 0.2$. In addition to the continuum, two emission features were required by the fit: a soft line at $\sim0.7$\,keV, and the neutral Fe\,K$\alpha$ line at 6.4\,keV. The primary X-ray continuum was well described by a power--law with photon index $\Gamma = 1.94 \pm 0.15$. The overall fit is statistically acceptable ($\chi^2/{\rm dof}=103/124$).\\
\indent Finally, the remaining source with X-ray observations of sufficient quality for spectral analysis is LEDA\,12773, for which 30 \textit{Swift}-XRT exposures were combined into a total effective exposure time of 65 ks, following the procedure described by \citet{evans09}. For this source, we do not detect any evidence of ionized X-ray absorption. The spectrum is well fitted by a model consisting of a power--law plus blackbody component, absorbed by Galactic neutral hydrogen \citep[$ N_{\rm H} = 10^{20}$\,cm$^{-2}$;][]{hi4pi16}, yielding a fit statistic of $\chi^2/{\rm dof} = 547/665$.

\subsection{Spectral index dependence on AGN parameters}\label{Appendix_alpha_dep}

\begin{figure*}[h!]
\centering
\includegraphics[width=1\textwidth]{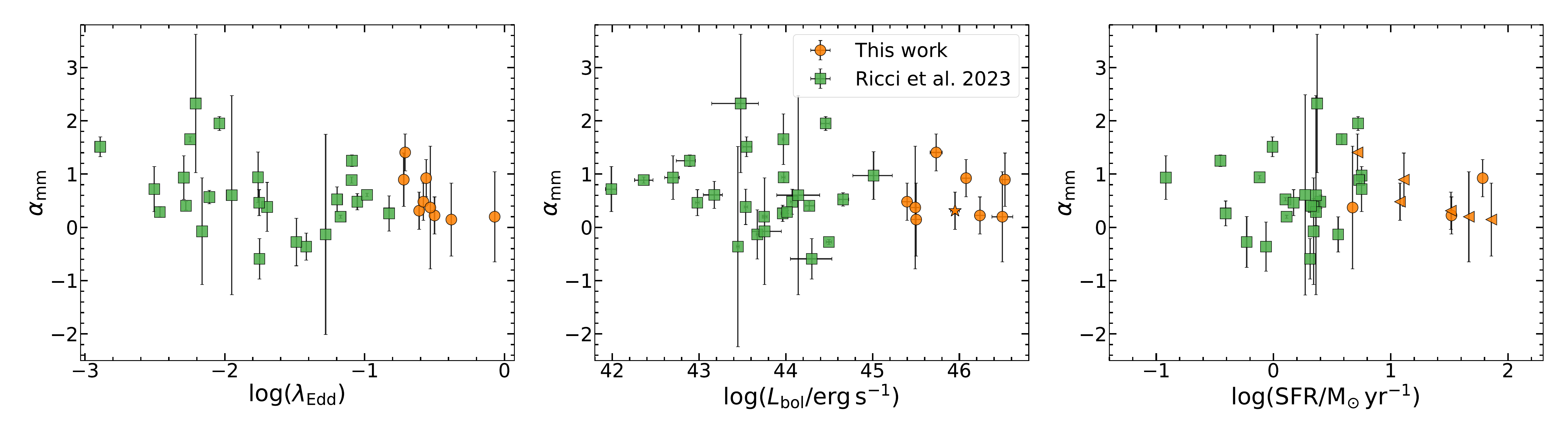}
\caption{\textbf{Left:} Spectral index ($\rm \alpha_{mm}$) of the 100\,GHz observations for this work, as obtained from fitting $ S_{\rm mm} \propto \nu ^{-\alpha_{\rm mm}}$ to the 100\,GHz fluxes in mJy in the four separate SPWs, and \citetalias{ricci_tight_2023} vs. the Eddington ratio $\lambda_{\rm Edd}$. We do not observe a correlation, obtaining a $p$-value of 0.17. \textbf{Middle:} $\rm \alpha_{mm}$ vs. the bolometric luminosity $L_{\rm bol}$. Here, no correlation is observed either since we obtain a $p$-value of 0.37. \textbf{Right:} $\rm \alpha_{mm}$ vs. SFR. We obtain a $p$-value of 0.85, observing no correlation.}
\label{fig:alpha_vs_x}
\end{figure*}

Figure\,\ref{fig:alpha_vs_x} displays the millimeter spectral index $\rm \alpha_{mm}$ as a function of the Eddington ratio $\rm \lambda_{Edd}$. We obtain a $p$-value of 0.17 and, therefore, determine no correlation.
Figure \ref{fig:alpha_vs_x} also displays the bolometric luminosity $L_{\rm bol}$ as a function of the Eddington ratio $\rm \lambda_{Edd}$. We obtain a $p$-value of 0.37, which indicates no correlation between the two parameters.
Finally, we present $\rm \alpha_{mm}$ as a function of SFR in Figure\,\ref{fig:alpha_vs_x}, where no correlation is observed either, obtaining a $p$-value of 0.85.

\subsection{Radio-to-submillimeter SEDs}\label{AppSED}

To determine whether the observed millimeter emission in our high-luminosity AGN can be fully attributed to the X-ray corona, we performed an SED fitting procedure following the work by \cite{del_palacio_millimeter_2025}.
We present below the available ancillary data in addition to our ALMA Band\,3 observations, along with the SED fitting results for the sources in our sample. The best-constrained SEDs are shown in Figure\,\ref{fig:SEDs} in Section\,\ref{SEDmod}. 

\indent For Q\,0019--286, the SED is very poorly constrained, with only upper limits from the Rapid ASKAP Continuum Survey (RACS) and the Very Large Array Sky Survey (VLASS). While the 100\,GHz ALMA flux could be dominated by free--free emission, a corona with $r_{\rm c} \sim 110$ and $\log \delta \sim -2.30$ 
could potentially be consistent with the SED fit.

\indent PG\,0026+129 has one of the best-constrained SEDs, thanks to VLA data in Bands C (4--8\,GHz) and Q (40--50\,GHz) and VLBA data at 1.5, 5.0, 8.4, and 23.6\,GHz. 
The fit (see Figure\,\ref{fig:SEDs}) indicates that the 100\,GHz flux is dominated by a corona with $r_\mathrm{c} = 217\pm13$ and $\log{\delta}=-1.06\pm0.11$. More details can be found in Section\,\ref{SEDmod}.

\indent PG\,0052+251 has a less constrained SED, presented in Figure\,\ref{fig:SEDs} as well, with VLA data in Bands C, X, and Q, VLBA data, and ALMA Band\,5 data (187--202\,GHz; Proposal ID 2023.1.01062.S; PI: F. Bauer). 
The 100\,GHz flux could be constrained by a corona with $r_{\rm c} = 131\pm66$ and $\log\delta = -1.68^{+0.31}_{-1.17}$, although this fit is more ambiguous compared to PG\,0026+129. More details on this fit can be found in Section\,\ref{SEDmod} as well.

\indent For Mrk\,813, the SED is poorly constrained, having only additional upper limits from RACS and VLASS, along with VLA C-band data. 
The 100\,GHz flux should arise exclusively from the corona, although the peak of the SED is highly uncertain. 
The derived parameters span a wide range: $r_{\rm c} \sim 380\pm200$ and $\log \delta \sim -2.3\pm0.3$.

\indent For RHS\,61, we combined the 100\,GHz data with upper limits from the low-resolution LoTSS, RACS, and VLASS surveys, as well as VLA C-band data. 
The corona is poorly constrained, and a significant fraction of the 100\,GHz flux could originate from free--free emission. 
The SED fit is consistent with a corona of size $r_{\rm c} \sim 200$ and $\log \delta \sim -2.0$. 

\indent LEDA\,126226 has a loosely constrained SED, with only low-resolution survey data available. The 100\,GHz flux could be dominated by a corona with $r_{\rm c} \sim 550$ and $\log\delta \sim -1.5$, but these values are highly uncertain, and free--free emission may contribute significantly.

\indent For 2MASX\,J02223523+2508143, the SED is also very loosely constrained, with only RACS and VLASS data. The 100\,GHz flux could primarily originate from the corona if $r_{\rm c} \sim 400$ and $\log\delta \sim -2.3$, but these parameters are poorly determined.

\indent Similarly, for 2MASX\,J17311341+1442561, the SED is very loosely constrained, with only RACS and VLASS data. The 100\,GHz flux may come mostly from a corona with $r_{\rm c} \sim 300$ and $\log \delta \sim -2.3$, although the constraints are weak.

\indent Finally, for LEDA\,12773, the SED is very poorly constrained, with only low-resolution RACS and VLASS data. The 100\,GHz flux could be dominated by a corona with $r_{\rm c} \sim 300$ and $\log\delta \sim -2.6$.


\newpage

\end{document}